\documentclass[11pt]{article}
\usepackage[a4paper,margin=1in]{geometry}
\usepackage{amsmath,amssymb,bbm,graphicx}
\usepackage{xcolor}
\usepackage{hyperref}
\usepackage{microtype}
\usepackage{braket}
\usepackage{subcaption}
\usepackage{doi}
\usepackage{float}
\usepackage{tikz}
\usepackage{pgfplots}
\usepgfplotslibrary{groupplots}
\usepackage{booktabs}
\usetikzlibrary{arrows.meta,positioning,shapes.geometric,calc}
\usetikzlibrary{arrows.meta, decorations.pathreplacing, decorations.markings}
\usepackage[
  backend=biber,
  style=numeric-comp,
  sorting=none,
  giveninits=true
]{biblatex}
\usepackage{booktabs}
\usepackage{multirow}

\usepackage{flafter}
\usepackage{placeins}
\usepackage{hyperref}
\usepackage[nameinlink,noabbrev]{cleveref}

\definecolor{ket}{RGB}{29,158,117}    
\definecolor{bra}{RGB}{127,119,221}   
\definecolor{op} {RGB}{186,117,23}    
\definecolor{ghost}{RGB}{180,175,220} 

\tikzset{
  myarrow/.style={-{Stealth[length=5pt,width=4pt]}},
  ket line/.style={ket,   line width=1.6pt, line cap=round},
  bra line/.style={bra,   line width=1.6pt, line cap=round},
  op arc/.style ={op,     line width=1.2pt, line cap=round},
  ghost line/.style={ghost, line width=0.9pt, dashed, dash pattern=on 4pt off 3pt},
  axis/.style={black!45, -{Stealth[length=4pt,width=3.5pt]}, thin},
  tick/.style={black!45, thin},
  label/.style={font=\small},
  sublabel/.style={font=\footnotesize},
}

\pgfplotsset{compat=1.18}
\definecolor{paperblue}{RGB}{36,92,150}
\definecolor{papergreen}{RGB}{43,130,103}
\definecolor{paperred}{RGB}{184,70,62}
\definecolor{paperamber}{RGB}{210,145,45}
\definecolor{sheetgreen}{RGB}{229,244,237}
\hypersetup{
  colorlinks=true,
  linkcolor=paperblue,
  urlcolor=paperblue,
  pdftitle={Asymptotic projection and the imaginary part}
}
\renewcommand{\arraystretch}{1.13}
\newcommand{\ii}{\mathrm{i}}
\newcommand{\e}{\mathrm{e}}
\newcommand{\dd}{\mathrm{d}}
\newcommand{\order}{\mathcal O}
\newcommand{\Binout}{B_{\mathrm{in\text{-}out}}}
\newcommand{\Binin}{B_{\mathrm{in\text{-}in}}}

\numberwithin{equation}{section}

\title{\bf Cosmological Correlators Using Tensor Networks}
\author{Ujjwal Basumatary$^{a,c}$, Aninda Sinha$^{a,b}$ and Xinan Zhou${^d}$ \footnote{ujjwalb.physics@gmail.com, asinha@iisc.ac.in, xinan.zhou@ucas.ac.cn}\\
\it ${^a}$Centre for High Energy Physics,
\it Indian Institute of Science, Bangalore, India.\\ \it ${^b}$Quantum Horizons Alberta, PHAS, University of Calgary, Canada.\\ \it $^{c}$ Raman Research Institute, Light and Matter Physics, Bangalore, India. \\ \it $^{d}$ Kavli Institute for Theoretical Sciences, University of Chinese Academy of Sciences,\\ \it Beijing 100190, China}

\begin{document}
\begingroup
\setlength{\tabcolsep}{0pt} 
\maketitle
\endgroup
\begin{abstract}
We develop tensor-network methods for computing cosmological correlators nonperturbatively in de Sitter space and use them to test the proposal that suitably defined in-in correlators can be obtained from an in-out formalism by gluing the expanding and contracting Poincar\'e patches. Focusing on interacting \(1+1\)-dimensional \(\phi^4\) theory, we formulate finite-time lattice observables using Matrix Product State (MPS) techniques and analyze the regulator subtleties associated with the singular behavior near the patching surface. \textcolor{black} {We consider two different contour prescriptions: a finite-time real-contour and a complex contour which properly accounts for the regulator. In the regulated light-field theory, exact-diagonalization and MPS checks support recovery of the full complex in--in/in--out equality at the tested finite cutoffs, suggesting a non-perturbative resolution of light field singularities in perturbation theory. 
} 
We briefly discuss how these methods extend to low-angular-momentum sectors in \(3+1\) dimensions, and why regimes of rapid entanglement growth may eventually motivate quantum-computing implementations.
\end{abstract}

\tableofcontents

\section{Introduction}
\label{sec:intro}

Cosmological correlators give us a direct theoretical probe of quantum fields in the early universe. In most analyses, the inflationary phase is approximated by a de Sitter spacetime \cite{cosmowhitepaper,desitterlectures} and the relevant observables are computed perturbatively in the so-called in-in formalism \cite{Schwinger1961Brownian,Keldysh1965Diagram,maldacena1,weinberg1}. In recent years, there have been several developments in understanding the structure of cosmological correlators from a systematic perturbative and effective-field-theory point of view, for example \cite{cosmoeft,ChenWangXianyu2017SK, Sleight:2019hfp, DiPietro:2021sjt, EmaMukaida2024Cutting,Goyal:2025hdm, cite1, cite2}, as well as from a bootstrap framework \cite{nima, cosmobootstrap,pierre, Werth:2024mjg, deRham:2025mjh}. While this approach has led to major advances, much less is known about cosmological correlators beyond perturbation theory. In this paper we address this gap using tensor-network methods.

One of the main questions that we study in this work is the applicability of the ``in-in\,=\,in-out'' proposal of \cite{Donath:2024}; closely related proposals include \cite{pimentel,Goodhew:2024eup}. In-in correlators rely on the Schwinger-Keldysh formalism, which doubles the types of propagators that appear and makes calculations tedious even at tree level \cite{ChenWangXianyu2017SK}. Donath and Pajer \cite{Donath:2024} proposed that, in close analogy with flat space, it is possible to use the standard in-out formalism to compute in-in correlators. The idea is to patch the de Sitter expanding Poincar\'e patch to a contracting Poincar\'e patch, defining the in-states in the expanding patch and the out-states in the contracting patch. This proposal was verified perturbatively in \cite{Donath:2024}. The main subtlety at leading order in perturbation theory is to establish conditions that ensure the convergence of the relevant integrals as one passes through the patching region. This leads to constraints on the dimensions of the external operators, as well as on the kinds of contact interactions that can be considered. Furthermore, a crucial assumption is that the Bunch-Davies vacuum \cite{BunchDavies1978DeSitter} remains invariant under the interacting unitary time-evolution operator up to an overall phase. The natural nonperturbative questions are therefore the following: (a) what happens to the perturbative constraints on the dimensions of the external operators, and (b) whether the assumption about the nature of the time evolution through the patched universe is actually true.

We will rely on Matrix Product State (MPS) techniques to probe these questions. More broadly, MPS and related tensor-network methods have emerged as useful nonperturbative tools for lattice field theory and real-time dynamics \cite{BanulsEtAl2013MPSLFT,BanulsEtAl2019LGTReview,MontangeroRicoSilvi2022LoopFree,MeuriceEtAl2022SnowmassTNHEP,Jha:2024jan}, with related developments for continuous quantum fields, continuum entanglement renormalization, and curved spacetime \cite{VerstraeteCirac2010cMPS,HaegemanEtAl2013cMERA,TilloyCirac2019ContinuousTN,ShacharZohar2022ContinuousTN,LewisVidal2020CurvedQFT,Bhat:2026epg}. There are also broader conceptual connections between tensor networks and de Sitter geometry \cite{Bao:2017kqt,Niermann:2021xht}. Very recently, there was an interesting study using MPS-based techniques to study two-point functions in a quenched set-up that mimics the expanding universe while bypassing the technical subtleties associated with the Bunch-Davies vacuum \cite{Budd:2026rmw}. Our focus in this paper is the $1+1$-dimensional $\phi^4$ scalar theory in de Sitter space---see \cite{Gorbenko:2019rza} for an analysis in 3+1d. In trying to apply MPS techniques to the de Sitter universe and to the patched universe described above, we encounter several obstacles. The primary one is the singular nature of the metric in the Poincar\'e patch in the infinite future. To facilitate numerics, we are forced to introduce a regulator, which in turn raises new questions that we discuss at length. \begingroup\color{black}
This also requires choosing a contour prescription for the real-time correlators. We discuss the possible choices in detail.
\endgroup

The MPS analysis also offers an interesting window into the entanglement structure of the states that arise in the computation of cosmological correlators. This gives us both a qualitative and a quantitative picture of where MPS techniques can be expected to struggle. For the operator $\phi^2(x)$, with $x$ in the center of the lattice, the picture that emerges is the following. For the in-in calculations, the entanglement in the MPS state remains comparable to what happens in flat space and in fact decreases as we approach late times. This decrease can be understood heuristically. In de Sitter spacetime, the effective mass and coupling become time-dependent. In the far past, or more precisely, when the magnitude of the conformal time is large, both the effective mass and the effective coupling vanish. In this sense, there is a natural turn-off, and the instantaneous interacting vacuum approaches the Bunch-Davies vacuum as the conformal time becomes large. At late times, both the effective mass and the effective coupling become large. From the lattice point of view, the onsite potential becomes stiff and the wavefunction is more locally trapped, leading to an approximate tensor-product structure. This underlies the decrease in entanglement. In the in-out MPS state, the situation changes once the universe is patched to a contracting one, and the entanglement begins to increase after the patching. This picture suggests, and our numerics support, the following:
\begin{itemize}
    \item While perturbative calculations benefit from the in-out formalism because there is only one kind of propagator, from a numerical point of view the in-in correlators are considerably easier to implement.
    \item In the in-out implementation in particular, one needs a fine-grained conformal-time grid near the patching slice.
    \item As we will explain, the implementation of the in-out formalism requires the preparation of two MPS states, thereby effectively doubling the complexity. This also raises the interesting possibility that, even in flat space, one may be able to use the in-in formalism to compute S-matrix elements using MPS techniques.
    \item \textcolor{black}{The light-field calculations remain finite at the stated regulators and cutoffs. The complex-contour tests below improve the in--in/in--out agreement, providing strong support that the perturbative infinities faced in the in-out calculations of \cite{Donath:2024} are resolved non-perturbatively.  }
    \item \textcolor{black}{We also study half-chain entanglement to understand how the cost of the MPS calculation grows. We find a strong dependence of entanglement on the contour used for real-time evolution. With the contour that gives agreement between in-in and in-out, the entanglement growth remains comparable in the two calculations. On the simpler real contour discussed below, entanglement grows substantially for the light field. }
    
\end{itemize}

While the main focus of our work is in $1+1$D, we will also briefly discuss how similar techniques can be useful in $3+1$D. There are interesting proposals to extend $1+1$d MPS methods to higher-dimensional systems, for instance using snaking-chain structures \cite{sukhireview} or more intrinsically higher-dimensional tensor-network ideas \cite{MagnificoEtAl2024Roadmap}. We will not pursue these directions here. Rather, we wish to point out that, using spherical symmetry, one can set up the analysis of correlators of the $s$-wave component of the fields using the same MPS techniques studied in the present work. There is an interesting difference between the effective Hamiltonian in 3+1d for the s-wave components and the $\phi^4$ Hamiltonian in 1+1d. The effective coupling in the lattice Hamiltonian in the former case is site dependent and decreases as one goes away from the origin. The s-wave dynamics is consequently less strongly interacting than the 1+1d dynamics, so its spectrum changes less as the bare coupling is increased.

Finally, we will briefly comment on the feasibility of studying cosmological correlators using near-term quantum computers. In the in-out calculation of $\phi^2(x)$, entanglement grows much faster for $m^2=0.1$ than for $m^2=1$, which is perturbatively safe for the in-in=in-out implementation. This strongly suggests that even in such simple examples, one may benefit from using quantum hardware.

\begingroup\color{black}
The paper is organized as follows. Section~\ref{sec:setup} specifies the background, observables and finite-time prescriptions. In Section~\ref{sec:BD_checks}, we check state preparation and finite-volume observables. In Section~\ref{sec:inin_inout_checks}, we present the original regulated real-contour calculations and the derivative-interaction calculation without a metric regulator. In Section~\ref{sec:asymptotic-projection}, we introduce the asymptotic contour prescription and distinguish its results from those of the original calculations. Section~\ref{sec:entanglement} discusses entanglement in the original implementation. Section~\ref{3plus1dS} discusses extensions, including quantum simulation, and Section~\ref{sec:conclusion} concludes. The appendices provide technical details, lattice perturbation theory and the effects of the regulator.
\endgroup

\section{Set-up and numerical framework}
\label{sec:setup}

\subsection{Background geometry and observables}
Let us start by considering the following conformally flat geometry in two dimensions
\begin{equation} \label{dsmetric}
ds^2 = \Omega(\eta)^2\left(-d\eta^2 + dx^2\right)\;.
\end{equation}
The case of flat space corresponds to
\begin{equation}
    \Omega(\eta)=1\;,\quad \quad -\infty<\eta<\infty\;.
\end{equation}
The Poincar\'e patch of dS$_2$ is given by 
\begin{equation}
    \Omega(\eta)=-\frac{1}{H\eta}\;,
\end{equation}
where $-\infty<\eta<0$ corresponds to the expanding universe and $0<\eta<\infty$ corresponds to the contracting universe. We will often set the Hubble constant $H$ to 1 for convenience. It should be noted that the flat space has two asymptotic time boundaries at $\eta=-\infty$ and $\eta=\infty$ where we can define in and out states. By contrast, dS$_2$ (in both contracting and expanding patches) has only one asymptotic boundary. 

\begin{figure}[!htb]
\centering
\includegraphics[width=\linewidth]{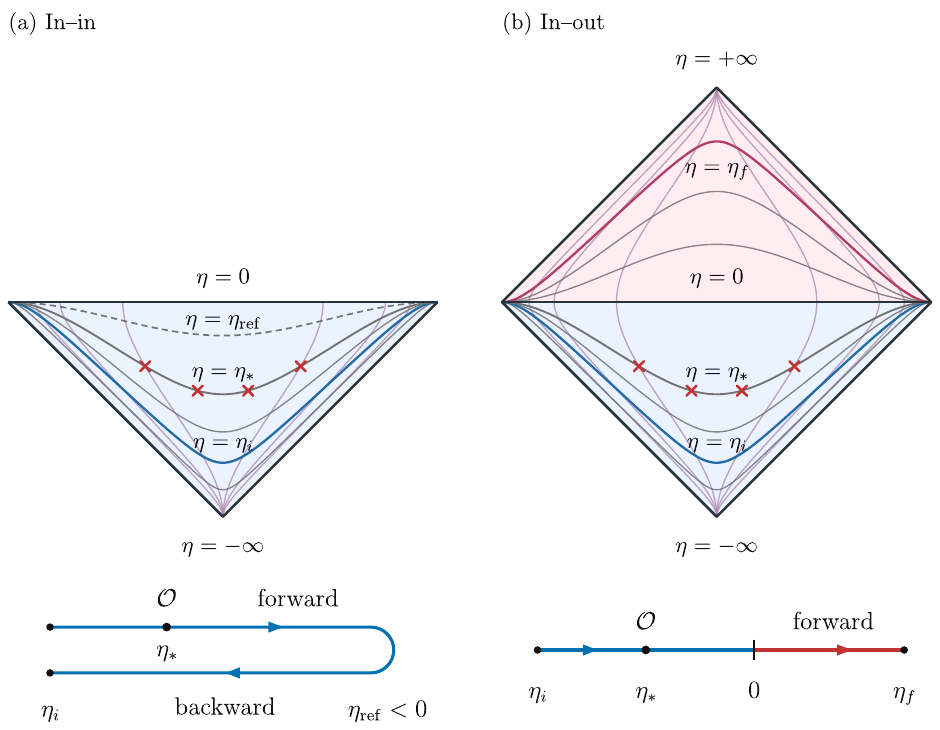}
\caption{Schematics of in-in and in-out formalisms. In the in-in formalism, only one copy of dS$_2$ is needed. The state is prepared at $\eta_i$ and operators are inserted at $\eta_*$ (marked by the red crosses). \textcolor{black}{The sketches below show the order of time evolution. Blue denotes evolution in the expanding patch and red in the contracting patch. The arrows indicate the direction of evolution. The two in-in legs are separated vertically for visibility. The symbol $\mathcal O$ represents the equal-time insertions marked by crosses above.} The in-in correlator is independent of the reference time $\eta_{\rm ref}$ as long as $\eta_{\rm ref}>\eta_*$. In the in-out formalism, we need to glue two copies of dS$_2$ along $\eta=0$. The in and out states are at $\eta_i$ and $\eta_f$ respectively.}
\label{fig:inin_inout_schematic}
\end{figure}

The basic claim of Donath and Pajer~\cite{Donath:2024} is that dS observables can be equivalently computed both in the in-in formalism and in the in-out formalism. To define the latter in a geometry with only one asymptotic boundary, they proposed that one should glue together the expanding and contracting dS patches along $\eta=0$, so that the total geometry has the same Penrose diagram as the flat space. For the dS$_2$ case which we will focus on mainly, this is illustrated in Figure \ref{fig:inin_inout_schematic}. To implement computer simulations, the evolution of the universe needs to start and end at finite times. More precisely, we define 
\begin{equation}
    \eta_i<\eta_*<0<\eta_f\;,
\end{equation}
where $\eta_i$, $\eta_f$ are the initial and final evolution times and $\eta_*$ is the moment when operators are inserted\footnote{For simplicity, we have inserted operators at equal time. However, we can also insert operators at different times.}. Moreover, to regularize the divergent metric at $\eta=0$---a vital step to keep the numerics stable---we introduce the following regulated conformal factor 
\begin{equation}
\Omega_{\rm reg}(\eta)=\frac{1}{H\sqrt{\eta^2+\eta_0^2}}\;,
\label{eq:Omega_reg}
\end{equation}
which approaches the dS$_2$ conformal factor in the $\eta_0\to0$ limit. There are important subtleties that show up in the in-out calculation with this regulator in place---these are discussed at length in the appendix.

Let us now introduce the in-in and in-out correlators. We start by recalling their definitions in the continuous limit \cite{Donath:2024}. The in-in correlator can be written as the contour-ordered expectation value
\begin{equation}
\label{eq:inindef}
    B_{\rm in\!-\!in} \equiv 
    \bra{0}\,
    \widetilde{\mathcal{T}}\!\left[\exp\!\left(i\int_{-\infty(1+i\epsilon)}^{\eta_{\rm ref}} H_{I,\rm int}\, d\eta\right)\right]\,
    \mathcal{T}\!\left[\mathcal{O}(\{\eta_*,\mathbf{x}\})\, \exp\!\left(-i\int_{-\infty(1-i\epsilon)}^{\eta_{\rm ref}} H_{I,\rm int}\, d\eta\right)\right]\,
    \ket{0}\,.
\end{equation}
Here $\mathcal{T}$ denotes time ordering (later times to the left) and $\widetilde{\mathcal{T}}$ denotes anti-time ordering (later times to the right). $|0\rangle$ is the Bunch-Davies vacuum. The operator $\mathcal{O}(\{\eta_*,\mathbf{x}\})$ is a generic string of interaction-picture operators, potentially inserted at different times, and $H_{I,\rm int}$ is the interaction Hamiltonian in the interaction picture. The quantity $B_{\rm in\!-\!in}$ is independent of the choice of the reference time $\eta_{\rm ref}>\eta_*$. The small contour tilt encoded by $\epsilon$ implements the usual adiabatic switching-on and switching-off of the interactions. For a Hermitian $\mathcal{O}$, $B_{\rm in\!-\!in}$ is manifestly real. On the other hand, the in-out correlator is defined as a vacuum-normalized
time-ordered expectation value \cite{Donath:2024}
\begin{equation}
\label{eq:Binout_DP_def}
B_{\rm in\!-\!out}\equiv\frac{\bra{0}\,\mathcal{T}\!\left[\mathcal{O}(\{\eta_*,\mathbf{x}\})\,\exp\!\left(-i\!\int_{-\infty(1-i\epsilon)}^{+\infty(1-i\epsilon)}\! d\eta\, H_{I,\rm int}(\eta)\right)\right]\ket{0}}{\bra{0}\,\mathcal{T}\!\left[\exp\!\left(-i\!\int_{-\infty(1-i\epsilon)}^{+\infty(1-i\epsilon)}\! d\eta\, H_{I,\rm int}(\eta)\right)\right]\ket{0}}\,,
\end{equation}
where the denominator (the vacuum persistence amplitude) removes vacuum-bubble contributions.

Since we will try to numerically simulate the dS physics, we cannot directly access the free vacuum $| 0 \rangle$ defined in the infinite past or infinite future. Instead, we must work with a finite time range of evolution bounded by the initial time $\eta_i<0$ and final time $\eta_f>0$. \textcolor{black}{The first set of calculations we discuss below use a strictly real contour and finite-time state preparation. In Section~\ref{sec:asymptotic-projection}, we instead use a complex contour with asymptotic vacuum projection. We distinguish these prescriptions throughout. Nonunitary evolution along the contour can be implemented classically with MPS, with appropriate checks of normalization and truncation.} For these reasons, we need to consider how the correlators can be correctly approximated in the finite time evolution. To this end, it is convenient to switch from the interaction picture and use the equivalent Schr\"odinger picture. For the in-in correlator, the free vacuum $| 0 \rangle$ can be approximated by the instantaneous vacuum $|0_{\eta_i}\rangle$ defined at a large negative time $\eta_i$. At $\eta=\eta_*<0$ where operators are inserted, the instantaneous vacuum is evolved to 
\begin{equation}\label{defPhi}
\ket{\Psi(\eta_*)}=U(\eta_*,\eta_i)\,\ket{0_{\eta_i}}\;.
\end{equation}
Here $U(\eta_2,\eta_1)$ is the Schr\"odinger-picture evolution operator generated by the full time-dependent Hamiltonian $H(\eta)$
\begin{equation}
i\,\partial_\eta U(\eta,\eta_1)=H(\eta)\,U(\eta,\eta_1),
\qquad
U(\eta_1,\eta_1)=\mathbbm{1}\,.
\end{equation}
Then the in-in correlator \eqref{eq:inindef} can be defined for finite time evolution as
\begin{equation}
B_{\rm in\!-\!in}(\eta_*) \equiv
\bra{\Psi(\eta_*)}\,\mathcal{O}\,\ket{\Psi(\eta_*)}\;.
\label{eq:Binin_def_setup}
\end{equation}
Given $H(\eta)$ and the initial state $\ket{0_{\eta_i}}$, the definition of $B_{\rm inin}(\eta_*)$ is unique and approaches \eqref{eq:inindef} in the limit $\eta_i\to-\infty$.

The situation is less direct for the in-out correlator. In that case, one must specify not only how the initial vacuum is approximated at finite time, but also how the outgoing vacuum sector is represented over the finite interval $[\eta_i,\eta_f]$. A natural finite-time analogue of the continuum expression \eqref{eq:Binout_DP_def} is
\begin{equation}
\label{eq:BinoutA_def_setup}
B_{\rm in\!-\!out}(\eta_*)
=
\frac{
\langle 0_{\eta_i}|\,
U_0^\dagger(\eta_f,\eta_i)\,
U(\eta_f,\eta_*)\,
\mathcal{O}\,
U(\eta_*,\eta_i)
\,|0_{\eta_i}\rangle
}{
\langle 0_{\eta_i}|\,
U_0^\dagger(\eta_f,\eta_i)\,
U(\eta_f,\eta_i)
\,|0_{\eta_i}\rangle
}\,,
\end{equation}
where $U(\eta_2,\eta_1)$ is the Schr\"odinger-picture evolution generated by the full interacting Hamiltonian $H(\eta)$, while $U_0(\eta_2,\eta_1)$ is generated by the free Hamiltonian
\[
H_0(\eta)\equiv H(\eta)\big|_{\lambda_{\rm eff}=0}\,.
\]
It is more transparent to rewrite \eqref{eq:BinoutA_def_setup} as
\begin{equation}
\label{eq:Binout_transition_form}
B_{\rm in\!-\!out}(\eta_*)
=
\frac{\bra{\chi_{\rm out}(\eta_*)}\mathcal{O}\ket{\psi_{\rm in}(\eta_*)}}
{\braket{\chi_{\rm out}(\eta_*) | \psi_{\rm in}(\eta_*)}}\;,
\end{equation}
where the in state $\ket{\psi_{\rm in}(\eta_*)}=|\Psi(\eta_*)\rangle$ was defined in \eqref{defPhi}, and the out state is
\begin{equation}\label{inoutbra}
\bra{\chi_{\rm out}(\eta_*)}=\bra{0_{\eta_i}}\,U_0^\dagger(\eta_f,\eta_i)\,U(\eta_f,\eta_*)\equiv \langle X(\eta_*)|\;.
\end{equation}
Physically, we evolve the instantaneous vacuum at $\eta_i$ to $\eta_f$ using the free Hamiltonian and then backwards to $\eta_*$ using the interacting Hamiltonian. The $U_0^\dagger$ cancels the contributions to the amplitude coming from the free evolution. Indeed, for the non-interacting case $U = U_0$ and therefore the denominator in~\eqref{eq:BinoutA_def_setup} is unity. However, the presence of this factor results in an additional evolution from $\eta_f$ to $\eta_i$. On the glued de Sitter contour, if one starts from the Bunch-Davies vacuum and evolves with the free Hamiltonian over the full contour, the resulting evolution takes the BD vacuum to itself, up to a phase
\begin{equation}
\label{eq:free_evol_phase}
    U_0(+\infty, -\infty)\ket{0}_{\rm BD} = e^{i \theta}\ket{0}_{\rm BD}\,.
\end{equation}
Motivated by this observation, one may consider a second definition, which resembles a transition amplitude,
\begin{equation}
\label{eq:BinoutB_def_setup}
B'_{\rm in\!-\!out}(\eta_*)
=
\frac{\bra{\chi'_{\rm out}(\eta_*)}\mathcal{O}\ket{\psi_{\rm in}(\eta_*)}}
{\braket{\chi'_{\rm out}(\eta_*) | \psi_{\rm in}(\eta_*)}}\;.
\end{equation}
In this case, the out state does not include the additional free evolution and is simply the instantaneous vacuum at $\eta_f$ evolved to $\eta_*$,
\begin{equation}
    \bra{\chi'_{\rm out}(\eta_*)}\equiv \bra{0_{\eta_f}}\,U(\eta_f,\eta_*)\;.
\end{equation}

It is instructive to compare the denominators appearing in these two definitions. For the first definition, one has
\begin{equation}
    \braket{\chi_{\rm out}(\eta_*) | \psi_{\rm in}(\eta_*)}
    =    \bra{0_{\eta_i}}\,U_0^\dagger(\eta_f,\eta_i)\,U(\eta_f,\eta_i)\ket{0_{\eta_i}}\;.
\end{equation}
When $\eta_i=-\eta_f$, this denominator measures the overlap between two states propagated over the same total time interval $2|\eta_i|$. One is obtained by evolving the initial vacuum with the full Hamiltonian, and the other by evolving it with the free Hamiltonian. By contrast, in the second definition the denominator is
\begin{equation}
    \braket{\chi'_{\rm out}(\eta_*) | \psi_{\rm in}(\eta_*)}   =\bra{0_{\eta_f}}U(\eta_f,\eta_i)\ket{0_{\eta_i}}\;,
\end{equation}
which instead measures the overlap between the initial vacuum evolved with the interacting Hamiltonian and the instantaneous vacuum at the final time.

In the infinite-time limit, with $|\eta_i|,|\eta_f|\to\infty$, we expect both definitions to reproduce the same in-out correlator defined in \eqref{eq:Binout_DP_def}. At finite time, however, the two definitions are markedly different. In particular, the second definition relies more directly on the assumption underlying \eqref{eq:free_evol_phase}, namely that the free evolution over the full glued contour returns the Bunch--Davies vacuum to itself up to an overall phase. Our numerical simulations indicate that this property is not well satisfied at finite evolution time. As a result, the equality between in--in and in--out deteriorates when \eqref{eq:BinoutB_def_setup} is used as the definition of the finite-time in-out correlator. \textcolor{black}{Equation~\eqref{eq:BinoutA_def_setup} is the finite-time definition used in the real-contour studies below. It differs at finite endpoints from the projected definition introduced in Section~\ref{sec:asymptotic-projection}.}

The finite-time in-in correlator \eqref{eq:Binin_def_setup}, together with the two definitions of finite-time in-out correlators \eqref{eq:BinoutA_def_setup} and \eqref{eq:BinoutB_def_setup}, can be written uniformly as
\begin{equation}
    \frac{
\langle 0_{\eta_i}|\,
V^\dagger(\eta_f,\eta_i)\,
U(\eta_f,\eta_*)\,
\mathcal{O}\,
U(\eta_*,\eta_i)
\,|0_{\eta_i}\rangle
}{
\langle 0_{\eta_i}|\,
V^\dagger(\eta_f,\eta_i)\,
U(\eta_f,\eta_i)
\,|0_{\eta_i}\rangle
}\;.
\end{equation}
The three cases then differ only in the choice of the reference evolution operator $V$:
\begin{itemize}
    \item $V=U$ gives the in-in correlator,
    \item $V=U_0$ gives the in-out correlator defined by \eqref{eq:BinoutA_def_setup},
    \item $V=1$ gives the in-out correlator defined by \eqref{eq:BinoutB_def_setup}, provided that $\langle 0_{\eta_f}|=\langle 0_{\eta_i}|$.
\end{itemize}
\begingroup\color{black}
The real-contour analysis concerns the finite-time regulated quantities in \eqref{eq:Binin_def_setup}, \eqref{eq:BinoutA_def_setup} and \eqref{eq:BinoutB_def_setup}. Their comparison tests agreement for these finite-time prescriptions. The massless derivative-interaction calculation below removes the metric regulator and treats switching explicitly. In Section~\ref{sec:asymptotic-projection}, we separately use tilted endpoints for the regulated $\phi^4$ theory. The analytic obstructions of the original contour are discussed in Appendix~\ref{app:cosmo}.
\endgroup

\subsection{Continuum theory}
\label{subsec:continuum_theory}
Although the construction presented is general enough to be applicable to any other theory, for this work, we restrict ourselves to the theory of a real scalar boson $\phi$ with a $\phi^4$ interaction. Before discretizing the theory, it is useful to recall the continuum scalar dynamics in the expanding Poincar\'e patch of dS$_2$ with the metric in~\ref{dsmetric}. The continuum action is
\begin{equation}
S=\int d\eta\,dx\,\sqrt{-g}\left[
-\frac12 g^{\mu\nu}\partial_\mu\phi\,\partial_\nu\phi
-\frac12 m^2\phi^2
-\frac{\lambda}{4!}\phi^4
\right].
\end{equation}
We can rewrite the above using \(\sqrt{-g}=\Omega^2\) and \(g^{\mu\nu}=\Omega^{-2}\eta^{\mu\nu}\)
\begin{equation}
S=\int d\eta\,dx\,
\left[
\frac12 (\partial_\eta\phi)^2
-\frac12 (\partial_x\phi)^2
-\frac{\Omega(\eta)^2}{2}m^2\phi^2
-\frac{\Omega(\eta)^2}{4!}\lambda\phi^4
\right].
\end{equation}
A special feature of \(1+1\) dimensions is that the conformal factors cancel out in the kinetic
term, so the time dependence appears only in the mass and interaction terms.

The canonical momentum is therefore
\begin{equation}
\pi(\eta,x)=\frac{\partial \mathcal{L}}{\partial(\partial_\eta\phi)}=\partial_\eta\phi(\eta,x)\;,
\end{equation}
and the Hamiltonian takes the form
\begin{equation}
\label{eq:hamiltonian_continuum}
H(\eta)=\int dx\,
\left[
\frac{\pi^2}{2}
+\frac{(\partial_x\phi)^2}{2}
+\frac{\Omega(\eta)^2}{2}m^2\phi^2
+\frac{\Omega(\eta)^2}{4!}\lambda\phi^4
\right].
\end{equation}
This is the continuum Hamiltonian that we will later discretize in space for the MPS
simulations.

\begingroup\color{black}
It is useful to distinguish the exact de Sitter free theory from the regulated free theory used in the numerical calculations. Their mode equations are
\begin{align}
(f_k^{\rm dS})''+\left[k^2+\frac{m^2}{H^2\eta^2}\right]f_k^{\rm dS}&=0,
& \Omega_{\rm dS}^2&=\frac1{H^2\eta^2},\label{eq:exact-mode-comparison}\\
(f_k^{\rm reg})''+\left[k^2+\frac{m^2}{H^2(\eta^2+\eta_0^2)}\right]f_k^{\rm reg}&=0,
& \Omega_{\rm reg}^2&=\frac1{H^2(\eta^2+\eta_0^2)}.\label{eq:regulated-mode-comparison}
\end{align}
At fixed $\eta_0>0$, $\eta=0$ is an ordinary point of the regulated equation: its solutions have a local Taylor expansion rather than the late-time power laws of exact de Sitter space. The regulated equation instead has complex singularities at $\eta=\pm i\eta_0$ when $m\ne0$. Both equations admit the same positive-frequency plane-wave condition in the asymptotic past for nonzero $k$, but their evolved solutions differ. Imposing an instantaneous vacuum at finite $\eta_i$ is an additional approximation to that asymptotic condition. The Hankel functions and scaling criterion derived immediately below refer to the \emph{exact}, unregulated equation. The regulated modes must be obtained from the regulated equation.
\endgroup

The free equation of motion is
\begin{equation}
\partial_\eta^2\phi-\partial_x^2\phi+\Omega(\eta)^2 m^2\phi=0\;.
\end{equation}
Expanding the field in Fourier modes,
\begin{equation}
\phi(\eta,x)=\int\frac{dk}{2\pi}\left[a_k\,f_k(\eta)e^{ikx}+a_k^\dagger\,f_k^*(\eta)e^{-ikx}\right]\;,
\end{equation}
with \([a_k,a_{k'}^\dagger]=2\pi\,\delta(k-k')\), the mode functions obey
\begin{equation}
\label{eq:modeeq_dS2}
f_k''(\eta)+\omega_k(\eta)^2 f_k(\eta)=0\;,\qquad\omega_k(\eta)^2=k^2+\Omega(\eta)^2 m^2= k^2+\frac{m^2}{H^2\eta^2}\;.
\end{equation}
They are normalized by the Wronskian condition:
\begin{equation}
\label{eq:modenormalization}
f_k f_k^{*\prime}-f_k' f_k^*=i.
\end{equation}

For each nonzero comoving momentum, as \(\eta\to-\infty\), the physical wavelength is much shorter than the curvature scale and the
modes approach flat-space plane waves. The Bunch--Davies choice is defined by requiring
positive-frequency behavior in this asymptotic past,
\begin{equation}
f_k(\eta)\xrightarrow[\eta\to-\infty]{}\frac{e^{-i|k|\eta}}{\sqrt{2|k|}}\;.
\end{equation}
The corresponding normalized mode functions are
\begin{equation}
\label{eq:BDmode_dS2}
f_k^{\rm BD}(\eta)=\frac{\sqrt{-\pi\eta}}{2}\,e^{\,i\pi(2\nu+1)/4}\,H_\nu^{(1)}(-|k|\eta)\;,\qquad\nu\equiv \sqrt{\frac14-\frac{m^2}{H^2}}\;.
\end{equation}
When \(m>H/2\), the index becomes imaginary and it is convenient to write
\begin{equation}
\nu=i\mu\;,\qquad \mu\equiv \sqrt{\frac{m^2}{H^2}-\frac14}\;.
\end{equation}
The Bunch--Davies vacuum is then the Fock vacuum annihilated by the operators \(a_k\),
\begin{equation}
a_k\,|{\rm BD}\rangle =0\qquad \forall\,k.
\end{equation}

The late-time behavior as \(\eta\to 0^{-}\) is important for the gluing at \(\eta=0\) used in the
Donath--Pajer in--out construction. For \(0\le m < H/2\), one has two real exponents
\begin{equation}
\Delta_\pm=\frac12\pm \nu\;,
\end{equation}
and the Bunch--Davies mode behaves as
\begin{equation}
f_k^{\rm BD}(\eta)\sim A_k(-\eta)^{\Delta_-}+B_k(-\eta)^{\Delta_+}\;,\qquad \eta\to 0^-\;.
\end{equation}
Thus the lighter mode decays as \((-\eta)^{\Delta_-}\), with the massless limit
\(\Delta_-=0\) giving the familiar freezing to a constant. For \(m>H/2\), the exponents are
complex,
\begin{equation}
\Delta_\pm=\frac12\pm i\mu\;,
\end{equation}
and the late-time behavior is oscillatory in \(\log(-\eta)\),
\begin{equation}
f_k^{\rm BD}(\eta)\sim(-\eta)^{1/2}\left[C_k\,(-|k|\eta)^{-i\mu}+D_k\,(-|k|\eta)^{+i\mu}
\right]\;,\qquad \eta\to 0^-\;,
\end{equation}
so that the envelope always decays as \((-\eta)^{1/2}\). The borderline case \(m=H/2\)
corresponds to \(\nu=0\) and involves the usual \(\sqrt{-\eta}\,\log(-\eta)\) behavior.

These asymptotics determine whether the interaction picture integrals remain finite near
\(\eta=0\). In the present dS$_2$ \(\phi^4\) theory, the quartic interaction contributes
\begin{equation}
\int d\eta\,dx\,\Omega(\eta)^2 \phi^4\;\sim\;\int d\eta\,(-\eta)^{-2+4\Delta}\;,
\end{equation}
where \(\Delta\) denotes the leading late-time decay exponent of the field. Finiteness of the
integral at \(\eta=0\) therefore requires
\begin{equation}
\label{eq:DPcondition_dS2}
4\Delta>1\;.
\end{equation}
For the light branch in the free theory, \(\Delta=\Delta_-=\frac12-\sqrt{\frac14-\frac{m^2}{H^2}}\), this gives
\begin{equation}
\frac{m^2}{H^2}>\frac{3}{16}\;.
\end{equation}
For the heavy branch, \(\Re\,\Delta=1/2\), so the condition is automatically satisfied.
\textcolor{black}{Accordingly, the time integral of the quartic vertex in exact dS$_2$ converges near the gluing surface for}
\begin{equation}
m>\frac{\sqrt{3}}{4}\,H\;,
\end{equation}
\begingroup\color{black}
while massless and sufficiently light fields fail this condition when described by free modes. This perturbative condition concerns the exact de Sitter vacuum. The regulated equation at fixed $\eta_0$ admits massless, light and heavy fields. At nonzero $\eta_0$, the gluing region on the real-time contour is regular for all these masses. The regulator-removal limit depends also on the contour and boundary states.
\endgroup

\subsection{Lattice discretizations and MPS}
\label{subsec:lattice_dirichlet}
Having defined the geometry and observables, let us set the stage for the numerical simulations. We will focus on $\phi^4$ theory on dS$_2$ and we first discuss how to discretize the theory on lattice. In the continuum, the theory is described by the Hamiltonian in eq.(\ref{eq:hamiltonian_continuum}).
To enable numerical simulations, the theory is placed on a finite lattice of size $L$ with $N$ dynamical sites separated by lattice spacing $a$. Dirichlet boundary conditions are imposed implicitly through the Hamiltonian. There are no dynamical boundary sites, and the field values at the endpoints are fixed to zero
\begin{equation}
\phi_0(\eta)=\phi_{N+1}(\eta)=0\;.
\end{equation}
The physical length of the lattice is therefore $L = (N + 1)a$, and the dynamical degrees of freedom reside at sites labelled by the discrete index $j$,
\begin{equation}
    x_j = ja - \frac{L}{2}\,, \qquad j = 1\,, \cdots \,, N\,.
\end{equation}
With these boundary conditions, the standard nearest-neighbor discretization of the spatial derivative term reads
\begin{equation}
\frac{1}{2}\int dx \, \left(\nabla \phi\right)^2 \rightarrow  \frac a2\sum_{j=0}^{N}\left(\frac{\phi_{j+1}-\phi_j}{a}\right)^2 = \frac{1}{a}\sum_{j=1}^{N}\phi_j^2 - \frac{1}{a}\sum_{j=1}^{N-1}\phi_j\phi_{j+1}\;,
\label{eq:grad_dirichlet}
\end{equation}
so that the endpoints act as springs attached to fixed walls at $\phi=0$.\footnote{%
An alternative ``free-end'' (Neumann-type) discretization modifies the endpoint coefficients and yields cosine normal modes with $k_n=\pi n/L$ (including a would-be zero mode), in contrast to the Dirichlet case. Since finite-volume correlators depend on this choice, all analytic and numerical comparisons must use the same endpoint convention.} \textcolor{black}{Unless specified otherwise, we set $a=1$ and report lattice units. We specify the different spacings used for refinement at fixed physical volume and for the calculation with separated fields on $N=9$ sites in Section~\ref{sec:asymptotic-projection}.}

Substituting \eqref{eq:grad_dirichlet}, the discretized version of the Hamiltonian~\eqref{eq:hamiltonian_continuum} becomes
\begin{align}
H(\eta) &= a\sum_{j=1}^{N}\left[\frac12\,\pi_j^2+ \frac12\,\Omega(\eta)^2 m^2\,\phi_j^2+ \frac{\Omega(\eta)^2\lambda_{\rm eff}(\eta)}{24}\,\phi_j^4\right]+ \frac{1}{a}\sum_{j=1}^{N-1}\left[-\phi_j\phi_{j+1}\right]+ \frac{1}{a}\sum_{j=1}^{N}\phi_j^2\;.
\label{eq:lattice_H_dirichlet}
\end{align}
Here $\lambda_{\rm eff}$ is defined in~\eqref{eq:lambda_eff_def}. The field $\phi$ and its canonical conjugate $\pi$ satisfy the equal-time commutation relation
\begin{equation}
    [\phi(x), \pi(x')] = i \delta(x - x')\,,
\end{equation}
where $\delta(x)$ is the Dirac delta function. To be consistent with the lattice prescription, we rescale the fields so that the discrete algebra
\begin{equation}
\label{eq:shoalgebra}
    [\Phi_k, \Pi_{\ell}] = i \delta_{k \ell}
\end{equation}
is satisfied. We use the simple choice
\begin{equation}
\label{eq:lattice_operators}
    \Phi_i = \phi(x_i)\,, \qquad \Pi_i = a \pi(x_i)\,.
\end{equation}
In terms of these lattice operators, the Hamiltonian \eqref{eq:lattice_H_dirichlet} takes the form
\begin{equation}
\label{eq:lattice_H_final}
    H(\eta) = \sum_{j= 1}^{N}\left[\frac{\Pi_j^2}{2a} + \frac{\Omega(\eta)^2 m^2 a}{2} \Phi_j^2 + \frac{\Omega(\eta)^2 \lambda_{\text{eff}}(\eta) a}{4!}\Phi_j^4\right] + \frac{1}{a}\sum_{j=1}^N\Phi_j^2 - \frac{1}{a}\sum_{j = 1}^{N-1}\Phi_j \Phi_{j + 1}\,.
\end{equation}
Finally, we define the lattice Hamiltonian
\begin{equation}
    H_{\text{lat}}(\eta) \equiv a H(\eta)\,,
\end{equation}
so that
\begin{equation}
    \label{eq:lattice_hamiltonian}
    H_{\text{lat}}(\eta) = \sum_{j= 1}^{N}\left[\frac{\Pi_j^2}{2} + \frac{\Omega(\eta)^2 \tilde{m}^2}{2} \Phi_j^2 + \frac{\Omega(\eta)^2 \tilde{\lambda}(\eta)}{4!}\Phi_j^4\right] + \sum_{j=1}^N\Phi_j^2 - \sum_{j = 1}^{N-1}\Phi_j \Phi_{j + 1}\,,
\end{equation}
where we have defined the dimensionless quantities $\tilde{m}^2 = m^2 a^2$ and $\tilde{\lambda}(\eta) = \lambda_{\rm eff}(\eta) a^2$. The lattice fields $\Phi_i$ and $\Pi_j$, at each site are then expanded in terms of creation and annihilation operators $a_i$ and $a_i^\dagger$ satisfying $[a_i, {a_j}^\dagger] = \delta_{ij}\,, [a_i, a_j] = [{a_i}^\dagger, {a_j}^\dagger] = 0$,
\begin{equation}
    \Phi_j = \frac{a_j + {a_j}^\dagger}{\sqrt{2}}\,, \qquad \Pi_j = i\frac{{a_j}^\dagger - a_j}{\sqrt{2}}\,, \qquad j = 1\,, \cdots \,, N\,.
\end{equation}
The full Hilbert space  at each site is infinite-dimensional. We render it finite by projecting onto a truncated subspace with the cutoff $N_{\text{max}}$ (which we take to be the same for all sites for simplicity). This truncation provides a controlled UV regulator at each site. 

The lattice discretization replaces the original continuum field theory by a chain of $N$ coupled anharmonic oscillators, with nearest-neighbour interactions arising from the discretized kinetic term and local self-interactions from the mass and quartic potential terms. We can then represent a many-body state (e.g. the ground state) with a matrix product ansatz
\begin{equation}
    \label{eq:MPSrepresentation}
    \ket{\psi} = \sum_{n_1,\dots, n_N} \sum_{\{\alpha_i\}}M^{[1]n_1}_{\alpha_1 \alpha_2} M^{[2]n_2}_{\alpha_2\alpha_3}\cdots M^{[N]n_N}_{\alpha_{N}\alpha_{N+1}}\ket{n_1\cdots n_N}\,,
\end{equation}
where $\alpha_1 = \alpha_{N+1} = 1$ for open boundaries and the basis states $\ket{n_1\cdots n_N}$ are the many-body occupation number eigenstates
\begin{equation}
    \ket{n_1 \cdots n_N} = \ket{n_1} \otimes \cdots \otimes \ket{n_N}\,,
\end{equation}
with $n_i$ taking values from $0$ to $N_{\text{max}}$. The MPS ground state can be obtained using the Density Matrix Renormalization Group (DMRG) algorithm and time evolution can be performed with the Time Dependent Variational Principle (TDVP) and Time Evolving Block Decimation (TEBD) algorithms. For a brief discussion of these algorithms we point the reader to Appendix~\ref{app:MPS}. Note that $N_{\max}$ must be chosen to be large enough so that the truncation remains inactive over the relevant time interval. This will be demonstrated in Section~\ref{sec:BD_checks}. Another important parameter is the bond dimension $\chi$, which limits the entanglement that an MPS can represent. Its dependence will also be explored in later sections.

\subsection{Adiabatic evolution}
An essential ingredient underlying the insensitivity of flat-space time-ordered correlators to the precise contour prescription is the \emph{Gell-Mann--Low construction}. Let \(H = H_0 + H_{\rm int}\), where \(H_0\) is the free Hamiltonian and \(H_{\rm int}\) the interaction term, and let \(\ket{\Phi_0}\) be an eigenstate of \(H_0\). Denote by \(U_{\epsilon}(t,t')\) the interaction-picture evolution operator generated by the adiabatically switched interaction
\begin{equation}
    H_\epsilon(t) = H_0 + e^{-\epsilon |t|} H_{\rm int}\,.
\end{equation}
If the limit
\begin{equation}
    \lim_{\epsilon\to 0^+}
    \frac{U_\epsilon(0,\pm\infty)\ket{\Phi_0}}
    {\bra{\Phi_0}U_\epsilon(0,\pm\infty)\ket{\Phi_0}}
\end{equation}
exists, then it defines an eigenstate of the full Hamiltonian \(H\), which we denote by \(\ket{\Psi_0}\). In the original work~\cite{PhysRev.84.350}, the existence of this limit was assumed order by order in perturbation theory. Later analyses clarified the relation of this construction to the adiabatic theorem and established its validity under suitable spectral assumptions~\cite{Molinari_2007}. Note, however, that the theorem by itself does not imply that \(\ket{\Psi_0}\) is the interacting ground state, even when \(\ket{\Phi_0}\) is the ground state of \(H_0\). To identify the adiabatically continued state with the interacting vacuum, one further requires the relevant adiabatic assumptions, such as isolation of the free ground state, continuous connection to a unique interacting vacuum, and the absence of obstructing level crossings.

In the discussion above, the \(\epsilon\)-dependence is only a switching regulator; the physical couplings are otherwise time independent. This is the setting in which the usual Gell--Mann--Low construction is formulated. By contrast, in a genuinely time-dependent background there is, in general, no single conserved Hamiltonian whose eigenstates define the interacting vacuum in the same way. In the de Sitter case, the fact that \(H(\eta)\) approaches a free Hamiltonian as \(\eta\to -\infty\) is enough to define an asymptotic in-state. However, by itself it does not imply that the state evolved to a finite time \(\eta\) is the instantaneous ground state of \(H(\eta)\). For the relation between the in--in and in--out correlators, however, it is sufficient that the vacuum returns to itself up to a phase under the full interacting evolution,
\begin{equation}
    U_I(+\infty,-\infty)\ket{0}
    =
    \ket{0}\,\braket{0|U_I(+\infty,-\infty)|0}\;.
\end{equation}

If this relation holds, then unitarity implies that the proportionality factor \(\braket{0|U_I(+\infty,-\infty)|0}\) is a pure phase. Explicit time dependence generically produces excitations, so one expects the evolved state to develop components outside the vacuum sector rather than return to \(\ket{0}\) up to a phase.

\subsection{Two numerical implementations}
\label{subsec:two_approaches}
Let us now describe in more detail how we compute the in-out correlators on the lattice. We use two complementary approaches which differ in their time-evolution algorithms and treatment of the interaction.

\paragraph{Approach A using TDVP with adiabatic coupling switching.} In Approach~A we fix the evolution endpoints $(\eta_i,\eta_f)$ and introduce an explicit adiabatic switching profile for the coupling
\begin{equation}
\lambda_{\rm eff}(\eta)=\lambda\,s(\eta)\;,
\label{eq:lambda_eff_def}
\end{equation}
chosen so that interactions are suppressed outside a finite time window. Following~\cite{PhysRev.84.350}, in the simulations we use a plateau with exponential tails
\begin{equation}
s(\eta)=
\begin{cases}
1, & |\eta|\le T_A,\\[2pt]
\exp\!\big[-\mu\,(|\eta|-T_A)\big]\;, & |\eta|>T_A\;,
\end{cases}
\label{eq:switching_profile}
\end{equation}
where $T_A>0$ is chosen to be sufficiently smaller than $\min\{-\eta_i,\eta_f\}$. The parameter $T_A$ sets the width of the interacting region, and $\mu>0$ sets the switching rate. Here $\mu$ is independent of the heavy-field mode index introduced in the continuum discussion.\footnote{%
Other smooth choices (e.g.\ $C^\infty$ compact-support profiles or logistic/tanh ramps) can be used in place of~\eqref{eq:switching_profile}; our convergence tests treat the profile as part of the regulator data.}
This switching plays a practical role similar to the standard $i\epsilon$ prescription in analytic in-in proofs. It makes the far past effectively free and suppresses contributions from the tails of oscillatory time integrals. It also avoids introducing an artificial interaction quench at the endpoints of the evolution interval. A sudden jump from $\lambda_{\rm eff}=0$ to $\lambda$ at $\eta_i$, or from $\lambda$ back to $0$ at $\eta_f$, would itself produce nonadiabatic excitations and additional entanglement, contaminating the comparison between $B_{\rm in\!-\!in}$ and its in--out counterpart. By turning the coupling on and off over a timescale set by $\mu^{-1}$, the initial and final states remain closer to the adiabatic continuation of the free vacuum. In this setting we use the normalized in--out ratio $B_{\rm in\!-\!out}$ in~\eqref{eq:BinoutA_def_setup}, which matches the structure of the Donath--Pajer proposal and is the quantity directly compared to $B_{\rm in-in}$ in our main tests. Because the interaction is effectively switched off at $\eta_i$ and $\eta_f$, the instantaneous vacua are essentially free and better approximate the Bunch-Davies vacuum. For this approach, we use the TDVP algorithm to evolve the states. The TDVP algorithm is implemented using both the \texttt{Julia} package \texttt{MPSKit.jl}~\cite{devos_2026_18792879} and \texttt{TenPy}~\cite{tenpy2024} in \texttt{Python}.

\paragraph{Approach B using TEBD without adiabatic coupling switching.} In Approach B, we do not use an adiabatic ramp for the interaction coupling and the interaction is always turned on. For in-out correlators, we choose for convenience the evolution contours to have symmetric end points
$\eta_i=-T_B$ and $\eta_f=T_B$,
and
\begin{equation}
\lambda_{\rm eff}(\eta)=\lambda\;, \qquad \text{for }\eta\in[-T_B,+T_B]\;.
\label{eq:always_on_def}
\end{equation}
Note that $T_B$ should not be confused with $T_A$ of Approach A, as in the latter $[-T_A,T_A]$ is a much smaller interval contained within $[\eta_i,\eta_f]$. Because the coupling is kept on, interactions have a greater effect on the instantaneous vacua than in Approach A at the same $\eta_i$ and $\eta_f$. These vacua are therefore further away from the Bunch-Davies vacuum. In Approach B we evolve the states using the TEBD algorithm implemented in \texttt{TenPy}. Both approaches use the normalized in--out ratio in~\eqref{eq:BinoutA_def_setup}. They differ in their coupling profiles, initial states, and time-evolution algorithms, and we compare their results where feasible. \textcolor{black}{We describe the two evolution algorithms in Appendix~\ref{app:MPS}.}

In both approaches, we compute the in-out correlators as ratios of overlaps between evolved states. We work with the states \( |\Psi(\eta_*)\rangle = U(\eta_\ast,\eta_i)\,|0_{\eta_i}\rangle \) and \( |X(\eta_*)\rangle = U(\eta_\ast,\eta_f)\,U_0(\eta_f,\eta_i)\,|0_{\eta_i}\rangle \). The ket $|\Psi\rangle$ evolves forward from $\eta_i$ to $\eta_*$. We obtain $|X\rangle$ by evolving the initial state freely from $\eta_i$ to $\eta_f$, then backward with the interacting Hamiltonian to $\eta_*$. We evaluate their overlaps at the operator insertion time $\eta_*$, and the in-out correlator is given by the ratio of \( \langle X(\eta_*) |\,O\,|\Psi(\eta_*)\rangle \) and \( \langle X(\eta_*)|\Psi(\eta_*)\rangle \). This avoids evolving the state after operator insertion across $\eta=0$, where increased entanglement can reduce the numerical accuracy. In both approaches, we use a time grid with smaller steps near $\eta\approx 0$, where $\Omega(\eta)$ varies rapidly. More precisely, we use $d\eta=0.1 |\eta|$ near this region, subject to a minimum step $d\eta_{\min}$ so that the evolution can cross $\eta=0$ and a maximum step $d\eta_{\max}$ to prevent the grid from becoming too sparse. We check stability under grid refinement (increasing the number of time points and/or reducing $d\eta_{\max}$), and in the free theory we compare the MPS results with numerical solutions of the mode equations in Section~\ref{sec:BD_checks}.

We present a convenient summary of the differences in the two approaches in Table \ref{tab:approachA_vs_B}. For very small \(\eta_0\), the time-dependent Hamiltonian becomes highly stiff near \(\eta=0\), since the effective scale factor behaves as \(\Omega^2(\eta) \sim (\eta^2+\eta_0^2)^{-1}\), so that the interaction terms become sharply enhanced in the central region. In our TDVP implementation, the local integrator ceases to converge reliably once the Hamiltonian varies too rapidly relative to the timestep and the available MPS manifold at fixed bond dimension. TEBD applies a sequence of short-time local gates obtained from a Trotter decomposition, so it can remain operational in these regimes. This does not mean that TEBD is automatically more accurate at small \(\eta_0\); it shows stronger weight near the bond-dimension cutoff, as we will illustrate in Section \ref{sec:entanglement}.

\begin{table}[t]
\centering
\begin{tabular}{|p{0.27\textwidth}|p{0.31\textwidth}|p{0.31\textwidth}|}
\hline
 & \textbf{Approach A} & \textbf{Approach B} \\
\hline
Time-evolution algorithm
& TDVP
& TEBD \\[4pt]

Interaction profile
& Adiabatically switched coupling,
$\lambda_{\rm eff}(\eta)=\lambda\, s(\eta)$
& Constant coupling,
$\lambda_{\rm eff}(\eta)=\lambda$ \\[4pt]

Nature of endpoint vacua
& Endpoints are effectively free, so the instantaneous vacua are close to free vacua
& Interactions affect the instantaneous endpoint vacua more strongly \\[4pt]

Practical advantage
& Suppresses interactions at the time endpoints
& Useful for probing small-$\eta_0$ regimes \\
\hline
\end{tabular}
\caption{Summary of the two numerical implementations used in this work. Both approaches evaluate the normalized in--out ratio in~\eqref{eq:BinoutA_def_setup}, but differ in their coupling profiles, initial states, and time-evolution algorithms.}
\label{tab:approachA_vs_B}
\end{table}

\section{Checks at finite volume}
\label{sec:BD_checks}
IR observables in de~Sitter are sensitive to the choice of initial state. At finite volume and with a local Hilbert-space truncation, it is therefore necessary to check that the prepared MPS state is close to the Bunch--Davies vacuum in the range of modes used later in the analysis. In practice, this requires three ingredients: choosing $\eta_i$ such that the relevant modes begin well inside the horizon; preparing the ground state of the instantaneous Hamiltonian accurately within the MPS ansatz; and comparing the subsequent free evolution with the decoupled mode-function equations.

\subsection{Ground state preparation}
The first step in the real-time MPS simulation is the preparation of the initial ground state. For fixed bond dimension $\chi$ and local Hilbert-space cutoff $N_{\rm max}$, we begin from a randomly initialized MPS and use the DMRG algorithm to optimize it to the ground state of the instantaneous Hamiltonian at $\eta=\eta_i$.

To test how well the instantaneous ground state prepared at $\eta_i$ approximates the Bunch--Davies vacuum, it is useful to consider the free theory, $\lambda = 0$, where analytic results are available. In this limit, the lattice modes behave as a set of coupled harmonic oscillators. Because the lattice Hamiltonian in~\eqref{eq:lattice_hamiltonian} is defined with Dirichlet boundary conditions, translational invariance is broken and the eigenstates of the discrete Laplacian are no longer plane waves. Instead, the normal modes are standing waves, with the appropriate orthonormal basis given by the Fourier sine modes

\begin{equation}
\label{eq:sine_modes}
u_j(n)=\sqrt{\frac{2}{N+1}}\sin\!\Big(\frac{\pi j n}{N+1}\Big),\qquad j=1,\dots,N\,.
\end{equation}

We use this basis both to compare with analytic results in the free theory and to extract momentum-space observables from the MPS data. In particular, the mode occupation number is
\begin{equation}
N_j(\eta)=\frac{1}{2\omega_j(\eta)}
\Big(\langle P_j^2\rangle_\eta+\omega_j(\eta)^2\langle Q_j^2\rangle_\eta\Big)-\frac12\,,
\label{eq:Nj_def}
\end{equation}
where
\begin{equation}
Q_j(\eta)=\sqrt{a}\sum_{n=1}^N u_j(n)\,\phi_n(\eta)\;,\qquad
P_j(\eta)=\sqrt{a}\sum_{n=1}^N u_j(n)\,\pi_n(\eta)\;
\end{equation}
are canonical mode coordinates and momenta. Here $\pi_n=\Pi_n/a$ and
$[\phi_n,\pi_m]=i\delta_{nm}/a$, so the factors of $\sqrt a$ give
$[Q_j,P_\ell]=i\delta_{j\ell}$. They are unity in our simulations, where
$a=1$. The corresponding instantaneous mode frequencies are
\begin{equation}
\label{eq:lattice_dispersion}
\omega_j(\eta)=\sqrt{\hat k_j^2+m^2\Omega(\eta)^2}\;,
\end{equation}
with
\begin{equation}
\hat k_j=\frac{2}{a}\sin\!\Big(\frac{k_j a}{2}\Big)\;,\qquad
k_j=\frac{\pi j}{a(N+1)}\;.
\end{equation}

The free part of $H(\eta)$ in~\eqref{eq:lattice_H_final} is
$H_0=\frac12\sum_j(P_j^2+\omega_j^2Q_j^2)$ and generates evolution in
conformal time $\eta$. Each mode $Q_j$ therefore satisfies
\begin{equation}
Q_j''(\eta) + \omega_j^2(\eta)Q_j(\eta) = 0\,.
\label{eq:mode_fn_eq}
\end{equation}
For either the regulated or unregulated conformal factor, we write
\begin{equation}
\label{eq:gensol}
Q_j(\eta)=b_j v_j(\eta)+b_j^\dagger v_j^*(\eta)\,,
\qquad
P_j(\eta)=b_j v_j'(\eta)+b_j^\dagger {v_j'}^*(\eta)\,,
\end{equation}
where $v_j(\eta)$ and $v_j^*(\eta)$ are independent solutions of~\eqref{eq:mode_fn_eq}.
The operators $b_j$ annihilate normal-mode quanta, in contrast to the
site oscillators $a_n$ used to construct the local basis. The canonical
commutators require
\begin{equation}
\label{eq:lattice_mode_wronskian}
[b_j,b_\ell^\dagger]=\delta_{j\ell}\,,\qquad
v_j v_j^{*\prime}-v_j'v_j^*=i\quad\text{for each }j\,.
\end{equation}

For $\eta_0=0$ and $\eta<0$, the mode equation has Hankel-function
solutions; for $\eta_0\neq0$, we solve it numerically. In either case, to
compare with the free instantaneous vacuum at $\eta=\eta_i$, we impose
\begin{equation}
\label{eq:mode_initial_data}
v_j(\eta_i) = \frac{1}{\sqrt{2 \omega_j(\eta_i)}}\,, \qquad
v'_j(\eta_i) = -i\, \omega_j(\eta_i)v_j(\eta_i)\,.
\end{equation}
With these initial conditions, $b_j|0_{\eta_i}\rangle=0$. Using~\eqref{eq:gensol}, one obtains
\begin{equation}
\braket{0_{\eta_i}|\,Q_j^2(\eta)\,|0_{\eta_i}}= |v_j(\eta)|^2\,,
\qquad
\braket{0_{\eta_i}|\,P_j^2(\eta)\,|0_{\eta_i}} = |v_j'(\eta)|^2\,,
\end{equation}
and hence
\begin{equation}
N_j(\eta)=
\frac{|v_j'(\eta)|^2+\omega_j^2(\eta)\,|v_j(\eta)|^2}{2\omega_j(\eta)}
-\frac12\,.
\end{equation}

In the MPS calculation, the same observables for each mode are obtained by projecting the equal-time correlator matrices onto the sine basis:
\begin{equation}
P_\phi^{\rm MPS}(k_n;\eta)\equiv a\,\vec{u}_n^{\mathsf T}C^\phi(\eta)\vec{u}_n\;,\qquad
P_\pi^{\rm MPS}(k_n;\eta)\equiv a\,\vec{u}_n^{\mathsf T}C^\pi(\eta)\vec{u}_n\;,
\end{equation}
where $\vec{u}_n$ is the vector whose entries are the sine modes in~\eqref{eq:sine_modes},
\begin{equation}
C^\phi_{ij}= \braket{\phi_i \phi_j}\,,
\end{equation}
and the momentum correlator $C^\pi$ is defined analogously using $\pi_n=\Pi_n/a$.
Thus $P_\phi^{\rm MPS}=\langle Q_n^2\rangle$ and
$P_\pi^{\rm MPS}=\langle P_n^2\rangle$. At $\eta=\eta_i$, these projected
correlators should reproduce the free-theory vacuum values determined above.
Equivalently, they can be combined into an effective occupation number,
\begin{equation}
N_n^{\rm MPS}(\eta)\equiv \frac12\left(\omega_n(\eta)\,P_\phi^{\rm MPS}(k_n;\eta)+\frac{P_\pi^{\rm MPS}(k_n;\eta)}{\omega_n(\eta)}\right)-\frac12\,,
\end{equation}
which should vanish at the initial time for a perfectly prepared vacuum.

A separate issue is the onsite Hilbert-space truncation. Even if $\chi$ is large, taking the local cutoff $N_{\max}$ too small can distort the vacuum when the effective mass $m_{\rm eff}^2(\eta_i)=m^2\Omega(\eta_i)^2$ is small. We therefore monitor the mean onsite occupation $\langle N\rangle$ and the probability $P_{\rm top}$ of the highest local level in the DMRG vacuum at $\eta_i$. In practice, we regard the cutoff as harmless when $P_{\rm top}\ll 1$ throughout the bulk; in the runs reported below we typically require $P_{\rm top}\lesssim 10^{-4}$.

A necessary, though not sufficient, condition for Bunch--Davies-like initial conditions is that the modes later used in the infrared analysis begin well inside the horizon. For Dirichlet modes,
\begin{equation}
k_n=\frac{\pi n}{a(N+1)}\,,
\end{equation}
this requires
\begin{equation}
k_{n_{\min}}|\eta_i|\gg 1,
\label{eq:subhorizon_band}
\end{equation}
where $n_{\min}$ is the smallest mode included in the range used for the analysis. For example, if the analysis uses the range $n=4,\dots,12$, it is sufficient to require~\eqref{eq:subhorizon_band} for the lowest mode in that interval rather than for the full spectrum.

The free theory therefore provides a check of both the preparation of the initial state and the subsequent real-time evolution. From the mode functions $v_n(\eta)$ obtained above by solving~\eqref{eq:mode_fn_eq}, the reference equal-time power spectrum is
\begin{equation}
P_{\rm ODE}(k_n,\eta)=|v_n(\eta)|^2.
\end{equation}
We compare this with the MPS result extracted from the projected correlator matrix,
\begin{equation}
P_{\rm MPS}(k_n,\eta)\equiv P_\phi^{\rm MPS}(k_n;\eta)\,.
\end{equation}
As a representative example, Table~\ref{tab:free_dS2_mode_by_mode_clean} shows the mode-by-mode agreement at $\eta_*=-0.2$. We quantify the mismatch by
\begin{equation}
\textcolor{black}{r_n\equiv \ln\!\left[\frac{P_{\rm MPS}(k_n,\eta_*)}{P_{\rm ODE}(k_n,\eta_*)}\right].}
\end{equation}
For the range $n=4\,,\dots\,,12$, the deviation remains below $3\times 10^{-3}$.
\begin{table}[htb]
\centering
\caption{Comparison of the MPS power spectrum with the numerical solution of the mode equations in free ($\lambda=0$) dS$_2$ theory, after enforcing Dirichlet boundary conventions and choosing
an onsite cutoff for which truncation is inactive. The parameters are
$L=40$, $m^2=0.1$, $\eta_i=-20$, $\eta_*=-0.2$, $\eta_f=5$, $\eta_0=0.05$,
with MPS parameters $N_{\max}=6$, $\chi=80$.
\textcolor{black}{The final column reports $r_n=\ln(P_{\rm MPS}/P_{\rm ODE})$, in which the common nonzero momentum factor cancels.}}
\label{tab:free_dS2_mode_by_mode_clean}
\vspace{1mm}
\begin{tabular}{|c|c|c|c|c|}
\hline
$n$ & $k_n$ & $P_{\rm MPS}(k_n,\eta_*)$ & $P_{\rm ODE}(k_n,\eta_*)$ & $r_n$ \\
\hline
 4 & 0.3065 & 1.169830\,$\times 10^{0}$ & 1.166717\,$\times 10^{0}$ & $+2.66\times 10^{-3}$ \\
 5 & 0.3831 & 9.733294\,$\times 10^{-1}$ & 9.718581\,$\times 10^{-1}$ & $+1.51\times 10^{-3}$ \\
 6 & 0.4597 & 8.371679\,$\times 10^{-1}$ & 8.364116\,$\times 10^{-1}$ & $+9.04\times 10^{-4}$ \\
 7 & 0.5364 & 7.370432\,$\times 10^{-1}$ & 7.366903\,$\times 10^{-1}$ & $+4.79\times 10^{-4}$ \\
 8 & 0.6130 & 6.600636\,$\times 10^{-1}$ & 6.598959\,$\times 10^{-1}$ & $+2.54\times 10^{-4}$ \\
 9 & 0.6896 & 5.990168\,$\times 10^{-1}$ & 5.989956\,$\times 10^{-1}$ & $+3.54\times 10^{-5}$ \\
10 & 0.7662 & 5.493729\,$\times 10^{-1}$ & 5.494225\,$\times 10^{-1}$ & $-9.04\times 10^{-5}$ \\
11 & 0.8429 & 5.082418\,$\times 10^{-1}$ & 5.083551\,$\times 10^{-1}$ & $-2.23\times 10^{-4}$ \\
12 & 0.9195 & 4.736155\,$\times 10^{-1}$ & 4.737600\,$\times 10^{-1}$ & $-3.05\times 10^{-4}$ \\
\hline
\end{tabular}
\end{table}

In the later interacting analysis, the overall consistency between the two numerical approaches (A and B) and the expected parameter dependence provides an additional indirect check that the state preparation and evolution are under adequate control.

\subsection{Interacting dynamics}
\label{sec:finite_volume_interacting}
Having established the quality of the initial state and the free-theory evolution, we now turn to the interacting theory. For each coupling $\lambda$, the full interacting ground state is prepared at the initial time $\eta_i$ by DMRG for the instantaneous Hamiltonian $H(\eta_i;\lambda)$. Starting from this state, we evolve in real time from $\eta_i=-20$ to $\eta_*=-1$ with the same coupling $\lambda$, $\eta_0=0$, and with identical numerical parameters across $\lambda$. Note that these initial states are the instantaneous vacua used in Approach B. Using these states removes the interaction quench that would be present if one started from the free vacuum at $\eta_i$, and isolates the redistribution of occupation numbers generated during the subsequent evolution. For Approach A, where free vacua are used, we find the results are nevertheless qualitatively the same. The adiabatic window for the coupling also suppresses the quench.

We use the mode occupations defined in~\eqref{eq:Nj_def} to identify departures from the instantaneous Gaussian vacuum in the chosen oscillator basis. We quantify the net production during the evolution by
\begin{equation}
\Delta N_j^{\rm prod}(\lambda)\equiv N_j(\eta_*;\lambda)-N_j(\eta_i;\lambda)\;.
\end{equation}
To isolate the interaction-induced modification relative to the free theory, we define
\begin{equation}
\Delta N_j^{\rm int\text{-}only}(\lambda)\equiv
\Delta N_j^{\rm prod}(\lambda)-\Delta N_j^{\rm prod}(0),
\end{equation}
and the cumulative sum
\begin{equation}
S_{\rm int}(J;\lambda)\equiv \sum_{j\le J}\Delta N_j^{\rm int\text{-}only}(\lambda).
\label{eq:Sint_def}
\end{equation}

\begin{figure}[!htbp]
    \centering
    \captionsetup[subfigure]{justification=raggedright}
    \begin{subfigure}[t]{0.48\textwidth}
        \centering
        \includegraphics[width=\linewidth]{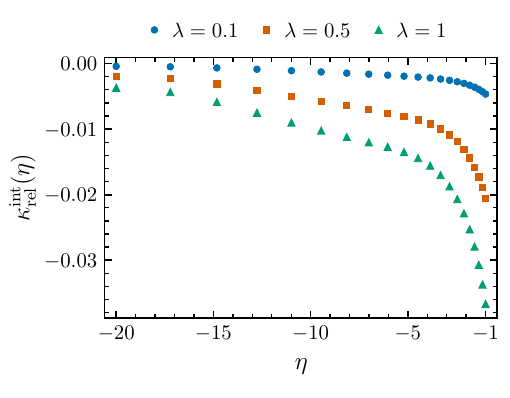}
        \caption{Bulk non-Gaussianity $\kappa_{\rm rel}^{\rm int}(\eta)$ for $\lambda=0.1,0.5,1$.}
        \label{fig:kappa_rel_vs_eta}
    \end{subfigure}
    \hfill
    \begin{subfigure}[t]{0.48\textwidth}
        \centering
        \includegraphics[width=\linewidth]{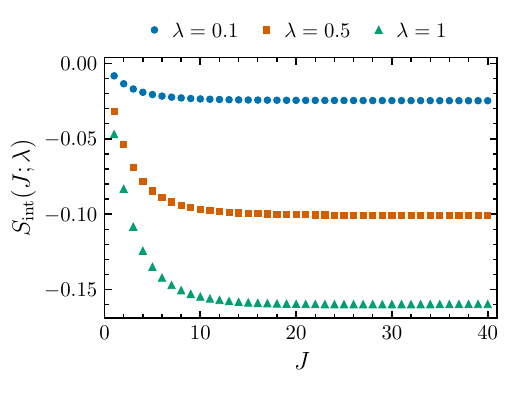}
        \caption{Cumulative shift due to the interaction $S_{\rm int}(J;\lambda)$ showing suppression mainly in the infrared modes.}
        \label{fig:Sint_plateau}
    \end{subfigure}

    \caption{Interaction effects on non-Gaussianity and mode production for $N=40$, $N_{\max}=15$, $\chi=40$, $m^2=1$, $a=1$, $\eta_0=0$, $\eta_i=-20$ and $\eta_*=-1$. The left panel shows the dimensionless excess kurtosis $\kappa_{\rm rel}^{\rm int}(\eta)$ defined in \eqref{eq:kappa_def} for $\lambda=0.1,0.5,1$. The right panel shows the cumulative interaction-induced shift $S_{\rm int}(J;\lambda)$ defined in \eqref{eq:Sint_def}, obtained from the net occupation change $\Delta N_j^{\rm prod}$ built from the instantaneous mode occupation \eqref{eq:Nj_def}.}
    \label{fig:non_gaussianity_and_Sint}
\end{figure}

As an independent check of interaction effects beyond Gaussian squeezing, we monitor the bulk-averaged connected fourth moment,
\begin{equation}
\kappa(\eta)\equiv \overline{\langle \phi_j^4\rangle-3\langle \phi_j^2\rangle^2}\;,
\qquad
\kappa_{\rm rel}(\lambda,\eta)\equiv \frac{\kappa(\eta)}{\overline{\langle \phi_j^2\rangle}^{\,2}}\;,
\end{equation}
where the overbar denotes an average over central sites in order to minimize boundary effects. In Figure~\ref{fig:non_gaussianity_and_Sint}, we average over sites $j=11,\ldots,30$ of the $N=40$ chain. The connected fourth moment is formed at each site before this average is taken. For the field-reflection-symmetric states considered here, a Gaussian state has $\kappa=\kappa_{\rm rel}=0$. To isolate the interaction-induced contribution, we subtract the free result and define
\begin{equation}
\kappa_{\rm rel}^{\rm int}(\eta)=\kappa_{\rm rel}(\lambda,\eta)-\kappa_{\rm rel}(0,\eta)\;.
\label{eq:kappa_def}
\end{equation}

The results at finite coupling are presented in Figure \ref{fig:non_gaussianity_and_Sint}. Figure~\ref{fig:kappa_rel_vs_eta} shows $\kappa_{\rm rel}^{\rm int}(\eta)$ as a function of conformal time for $\lambda=0.1,0.5,1$. As the interaction strength is increased, the state develops a progressively stronger non-Gaussian component and $\kappa_{\rm rel}^{\rm int}(\eta)$ becomes negative at late times, corresponding to a platykurtic local distribution. Figure~\ref{fig:Sint_plateau} shows the cumulative shift due to the interaction $S_{\rm int}(J;\lambda)$ defined in~\eqref{eq:Sint_def}. The negative plateau shows that turning on $\lambda$ suppresses the net production relative to the free squeezing result, while the rapid saturation with $J$ shows that this suppression is dominated by the lowest modes. In the cases shown here, the plateau is essentially reached by $J\simeq 10$, which justifies restricting the sum over modes to $j\le 10$.

\begingroup\color{black}
\subsubsection{Connected fourth moment at weak coupling}
\label{sec:kurtosis_weak_coupling}

To check the magnitude and time dependence of the kurtosis in
Figure~\ref{fig:kappa_rel_vs_eta}, we repeat the calculation at smaller
couplings and compare with first-order perturbation theory. The connected
fourth moment receives contributions from a single quartic vertex during
evolution and from the interacting initial vacuum. Both are needed for
the initial state used in Figure~\ref{fig:non_gaussianity_and_Sint}.

Let $\kappa_j$ denote the local connected moment whose spatial average
gives $\kappa$ as defined above. Writing
$\kappa_j(\lambda,\eta)=\lambda\kappa_j^{(1)}(\eta)+O(\lambda^2)$
and $\kappa^{(1)}=\overline{\kappa_j^{(1)}}$, the relative moment has the
expansion
\begin{equation}
\kappa_{\rm rel}^{\rm int}(\eta)
=\lambda\,\frac{\kappa^{(1)}(\eta)}
{\overline{G_j(\eta)}^{\,2}}+O(\lambda^2).
\label{eq:kurtosis_relative_first_order}
\end{equation}
Here $G_j(\eta)=\langle0|\phi_{I,j}^2(\eta)|0\rangle$ is the local
variance under free evolution from the free instantaneous vacuum at $\eta_i$.
We derive the two contributions to $\kappa^{(1)}$ in
Appendix~\ref{app:kurtosis-first-order}.

To compare with the MPS result, we subtract the moment measured in the free calculation
and divide by the coupling. The resulting quantity,
$q(\lambda,\eta)=[\kappa(\lambda,\eta)-\kappa(0,\eta)]/\lambda$,
should approach $\kappa^{(1)}(\eta)$ as the coupling decreases. This
compares the size of the interaction effect with its predicted value.
On three sites, the MPS values at the final time and smallest coupling
agree with the first-order predictions within $0.2\%$ for both free and
interacting initial vacua. The comparisons with direct Schr\"odinger
evolution are given in Appendix~\ref{app:kurtosis_small_chain}.

Figure~\ref{fig:kurtosis_weak_coupling} shows the comparison on the lattice
and over the time interval used in Figure~\ref{fig:non_gaussianity_and_Sint},
with the interacting initial vacuum. The MPS values of $q$ follow the
first-order time dependence, becoming more negative during the early
evolution and then increasing towards the final time.
The remainder varies nonmonotonically with the coupling at several times.
At the numerical settings specified in the caption, the connected fourth
moment after subtracting the free result agrees with its first-order
prediction within $2\%$ at all 21 observation times and all three weak
couplings.

\begin{figure}[!htbp]
    \centering
    \color{black}
    \captionsetup{font+={color=black},labelfont+={color=black}}
    \includegraphics[width=\textwidth]{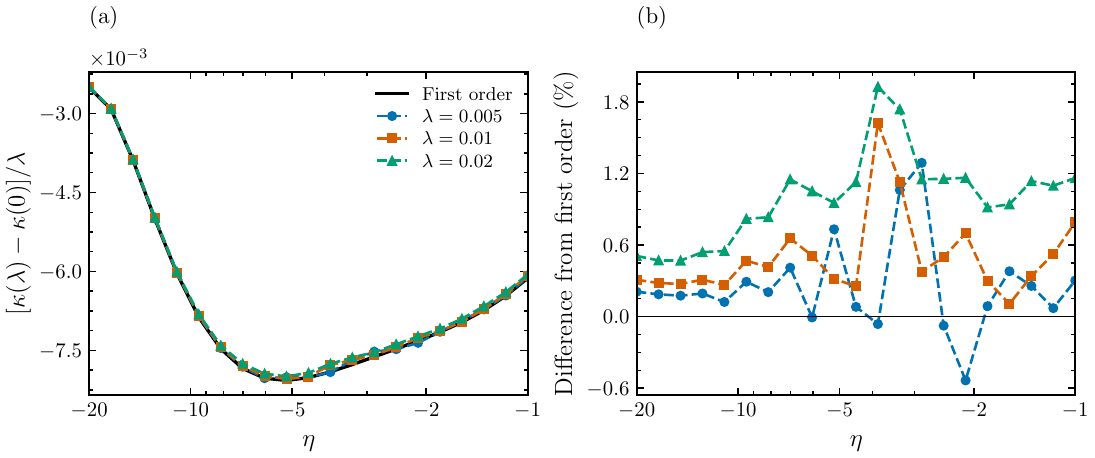}
    \caption{The connected fourth moment at weak coupling on forty sites.
    (a) The points show $q(\lambda,\eta)=[\kappa(\lambda,\eta)-\kappa(0,\eta)]/\lambda$.
    The black curve gives $\kappa^{(1)}(\eta)$, including the contribution from the interacting
    initial vacuum. (b) The signed percentage difference
    $100[q(\lambda,\eta)-\kappa^{(1)}(\eta)]/|\kappa^{(1)}(\eta)|$.
    Lines join the 21 logarithmically spaced observations from
    $\eta_i=-20$ to $\eta_*=-1$. The couplings are
    $\lambda=0.005,0.01,0.02$. We use $a=H=m^2=1$,
    $\eta_0=0$, $N_{\max}=15$, $\chi=40$, a maximum time step of $0.1$,
    and average the local connected moments over sites $11,\ldots,30$.
    The initial vacuum is prepared with the interaction at $\eta_i=-20$.}
    \label{fig:kurtosis_weak_coupling}
\end{figure}
\endgroup

\section{\texorpdfstring{{Finite-time comparisons and a test without a metric regulator}}{Finite-time comparisons and a test without a metric regulator}}
\label{sec:inin_inout_checks}
For our main nonperturbative test, we compare the in--in expectation value $B_{\rm in\!-\!in}(\eta_*)$ given in \eqref{eq:Binin_def_setup} to the normalized in--out ratio $B_{\rm in\!-\!out}(\eta_*)$ as in \eqref{eq:BinoutA_def_setup}. The perturbative argument in~\cite{Donath:2024} includes conditions ensuring that time integrals remain finite
near the gluing surface and both at early and late times.  In our dS$_2$ setup these conditions translate into constraints on the falloff of the interaction-picture operator insertions and on the switching profile
$s(\eta)$ used in $\lambda_{\rm eff}(\eta)$ as discussed in Section \ref{subsec:two_approaches}.  We will use the free-theory ODE solution and the vacuum checks
of Section~\ref{sec:BD_checks} to ensure that any observed deviation between $B_{\rm in\!-\!in}$ and $B_{\rm in\!-\!out}$ is not an artifact of IR contamination from the initial state.

\begingroup\color{black}
The replacement $\Omega\to\Omega_{\rm reg}$ changes the analytic structure. The coefficient $\Omega_{\rm reg}^2$ has poles at $\eta=\pm i\eta_0$, and massive mode functions generally have logarithmic monodromy. These changes affect the contour argument. The difference between the correlators on the real contour can depend on the regulator, continuation and boundary states. Testing the original relation therefore requires controlling these choices and the other limits as $\eta_0$ tends to zero. We examine these limits in Section~\ref{sec:asymptotic-projection}. The real-contour comparisons are given in Appendix~\ref{app:cosmo}.
\endgroup

\subsection{Set up}
\paragraph{Implementation of $B_{\rm in\!-\!in}$.}
To test our implementation, we consider the equal-time two-point function obtained by choosing the operator insertion $\mathcal{O}$ in~\eqref{eq:Binin_def_setup} to be two spatially separated field operators evaluated at $\eta=\eta_*$:
\begin{equation}
    B^{\rm in\text{-}in}_{mn}(\eta_*)
    \equiv
    \braket{\Psi(\eta_*) | \phi_m \phi_n | \Psi(\eta_*)}\;.
\end{equation}
Note that the above time evolution from the ground state at $\eta_i$ to the state at $\eta_*$ can be implemented using both approach A and approach B. Here $\phi_m$ denotes the Schrödinger-picture field operator at lattice site $m$, and $\ket{0_{\eta_i}}$ is the instantaneous vacuum prepared at the initial time $\eta_i$.

In order to check our numerical results, we also consider perturbation theory at weak coupling. Expanding the evolution operator to first order in the interaction gives
\begin{equation}
\label{eq:inin-first-order-commutator}
    B^{\rm in\text{-}in}_{mn}(\eta_*)
    =
    W_{mn}(\eta_*,\eta_*)
    + i \int_{\eta_i}^{\eta_*} d\eta\,
    \braket{
        0_{\eta_i}
        \left|
        \left[
            H_I(\eta),
            \phi_{I,m}(\eta_*) \phi_{I,n}(\eta_*)
        \right]
        \right|
        0_{\eta_i}
    }
    + \mathcal{O}(\lambda^2)\;,
\end{equation}
where $\phi_{I,m}(\eta)$ is the free field in the interaction picture. For the quartic interaction considered here, Wick contraction gives
\begin{equation}
\label{eq:inin-first-order-wightman}
    B^{\rm in\text{-}in}_{mn}(\eta_*)
    =
    W_{mn}(\eta_*,\eta_*)
    -
    \int_{\eta_i}^{\eta_*} d\eta\,
    \lambda(\eta)\,\Omega^2(\eta)\,
    \sum_j a\,
    \Im\left[
        W_{jm}(\eta,\eta_*)\,
        W_{jn}(\eta,\eta_*)\,
        W_{jj}(\eta,\eta)
    \right]
    + \mathcal{O}(\lambda^2)\;.
\end{equation}
Here
\begin{equation}
    W_{mn}(\eta_1,\eta_2)
    \equiv
    G^>_{mn}(\eta_1,\eta_2)
    \equiv
    \braket{
        0_{\eta_i}
        \left|
        \phi_{I,m}(\eta_1)\phi_{I,n}(\eta_2)
        \right|
        0_{\eta_i}
    }\;,
\end{equation}
is the free Wightman function in the instantaneous vacuum at $\eta_i$. For Dirichlet boundary conditions, with normalized sine modes~\eqref{eq:sine_modes} the above takes the form
\begin{equation}
\label{eq:wightmanfree}
    W_{mn}(\eta_1,\eta_2)
    =
    \frac{2}{a(N+1)}
    \sum_{r=1}^{N}
    v_r(\eta_1)\,v_r^*(\eta_2)\,
    \sin\frac{m r \pi}{N+1}\,
    \sin\frac{n r \pi}{N+1}\;.
\end{equation}
Since the in--in contour lies entirely in the region $\eta<0$, the regulator $\eta_0$ in the conformal factor~\eqref{eq:Omega_reg} may be taken to the $\eta_0\to 0$ limit in this calculation, in which case the mode functions appearing above are simply linear combinations of Hankel functions.

A further comparison for our numerical results can be obtained by considering the Hartree--Fock approximation to the equal-time correlator
\begin{equation}
    B^{\rm HF}_{mn}(\eta_*)    \equiv    \braket{\Psi_{\rm HF}(\eta_*) | \phi_m \phi_n | \Psi_{\rm HF}(\eta_*)}\;.
\end{equation}
For states with vanishing one-point function, $\braket{\phi_j}_{\rm HF}=0$, we decompose the quartic operator into its Gaussian contractions and a normal-ordered remainder,
\begin{equation}
    \phi_j^4
    \to    6\,\braket{\phi_j^2}_{\rm HF}\,\phi_j^2-3\,\braket{\phi_j^2}_{\rm HF}^2+ :\phi_j^4:\;.
\end{equation}
\textcolor{black}{The Hartree--Fock approximation discards the normal-ordered remainder $:\phi_j^4:$, with normal ordering defined relative to the self-consistent Gaussian state.} Denoting
\begin{equation}
    F_j(\eta)
    \equiv
    B^{\rm HF}_{jj}(\eta)\;,
\end{equation}
this gives the effective quadratic interaction
\begin{equation}
    H_I^{\rm HF}(\eta)=  \sum_j a\,    \frac{\lambda(\eta)\Omega^2(\eta)}{4}\,
    F_j(\eta)\,\phi_j^2    -    \sum_j a\,    \frac{\lambda(\eta)\Omega^2(\eta)}{8}\,
    F_j^2(\eta)\;.
\end{equation}
The second term is a $c$-number vacuum-energy shift and does not affect the two-point function. The nontrivial effect of the Hartree--Fock approximation is therefore a self-consistent local correction to the quadratic kernel. If the free Hamiltonian is written as
\begin{equation}
    H_0(\eta)=  \frac{a}{2}\sum_j \pi_j^2+    \frac{a}{2}\sum_{j,k}\phi_j\,K^{(0)}_{jk}(\eta)\,\phi_k\;,
\end{equation}
then the Hartree--Fock Hamiltonian takes the form
\begin{equation}
    H_{\rm HF}(\eta)  =
    \frac{a}{2}\sum_j \pi_j^2 +  \frac{a}{2}\sum_{j,k}\phi_j\,K^{\rm HF}_{jk}(\eta)\,\phi_k  +    E_0^{\rm HF}(\eta),
\end{equation}
with
\begin{equation}
    K^{\rm HF}_{jk}(\eta)
    =
    K^{(0)}_{jk}(\eta)
    +
    \delta_{jk}\,
    \frac{\lambda(\eta)\Omega^2(\eta)}{2}\,
    F_j(\eta)\;,
\end{equation}
and
\begin{equation}
    E_0^{\rm HF}(\eta)   =    -  \sum_j a\,\frac{\lambda(\eta)\Omega^2(\eta)}{8}\,
    \bigl(F_j(\eta)\bigr)^2\;.
\end{equation}
The local mass shift is
\begin{equation}\label{hf1}
    \delta M_j^2(\eta)
    =
    \frac{\lambda(\eta)\Omega^2(\eta)}{2}\,
    F_j(\eta)\;,
\end{equation}
where $F_j(\eta)=\braket{\phi_j^2(\eta)}_{\rm HF}$.

Since $H_{\rm HF}(\eta)$ is quadratic, the corresponding state remains Gaussian, and the equal-time two-point function is completely determined by the Hartree--Fock mode functions. Let $f_j^{(\alpha)}(\eta)$ denote a complete set of solutions of
\begin{equation}
    \partial_\eta^2 f_j^{(\alpha)}(\eta)
    +
    \sum_k K^{\rm HF}_{jk}(\eta)\,f_k^{(\alpha)}(\eta)
    =
    0,
\end{equation}
with initial conditions fixed by the instantaneous vacuum at $\eta_i$. The Hartree--Fock Wightman function is then
\begin{equation}
    W^{\rm HF}_{mn}(\eta_1,\eta_2)
    =
    \frac{1}{a}
    \sum_{\alpha=1}^N
    f_m^{(\alpha)}(\eta_1)\,
    f_n^{(\alpha)\,*}(\eta_2),
\end{equation}
and the equal-time correlator is
\begin{equation}
    B^{\rm HF}_{mn}(\eta)
    =
    W^{\rm HF}_{mn}(\eta,\eta)
    =
    \frac{1}{a}
    \sum_{\alpha=1}^N
    f_m^{(\alpha)}(\eta)\,
    f_n^{(\alpha)\,*}(\eta).
\end{equation}
The self-consistency condition is therefore
\begin{equation}\label{hf2}
    B^{\rm HF}_{jj}(\eta)
    =
    \frac{1}{a}
    \sum_{\alpha=1}^N
    \bigl|f_j^{(\alpha)}(\eta)\bigr|^2.
\end{equation}

This shows that the Hartree--Fock approximation resums the local tadpole contribution into a time-dependent, site-dependent effective mass. For Dirichlet boundary conditions, $B^{\rm HF}_{jj}(\eta)$ is in general not independent of $j$, so the Hartree--Fock correction is local but not translationally invariant. Consequently, the free sine modes do not in general diagonalize the full Hartree--Fock kernel, even though they do diagonalize the free theory. At leading order in $\lambda$, one may replace $B^{\rm HF}_{jj}(\eta)$ on the right-hand side by the free coincidence correlator $W_{jj}(\eta,\eta)$, showing that the Hartree--Fock approximation reproduces the same local tadpole structure that appears in the perturbative result~\eqref{eq:inin-first-order-wightman}.

\paragraph{Implementation of $B_{\rm in-out}$.} We implement $B_{\rm in\!-\!out}$ using Eq.~\eqref{eq:Binout_transition_form}, by constructing the numerator and denominator as overlaps between MPS states evolved over the appropriate time segments. To compare with the in--in results, we consider the same operator string $\mathcal{O} = \phi_n \phi_m$, i.e., we examine the in--out correlator
\begin{equation}
    B^{\text{in-out}}_{nm} = \frac{\braket{0_{\eta_i} | U_0^\dagger(\eta_f, \eta_i) U(\eta_f, \eta_*) \phi_n \phi_m U(\eta_*, \eta_i) | 0_{\eta_i}}}{\braket{0_{\eta_i} | U_0^\dagger(\eta_f, \eta_i) U(\eta_f, \eta_i)|0_{\eta_i}}}\;.
\end{equation}

For comparison, we also present the first-order perturbative result for the in--out correlator. Defining the in--out Wightman functions
\begin{equation}
    G^>_{mn}(\eta_1,\eta_2)
    \equiv
    \frac{
        \braket{
            0_{\eta_f}
            \left|
            \phi_{I,m}(\eta_1)\phi_{I,n}(\eta_2)
            \right|
            0_{\eta_i}
        }
    }{
        \braket{0_{\eta_f}|0_{\eta_i}}
    },
    \qquad
    G^<_{mn}(\eta_1,\eta_2)
    \equiv
    \frac{
        \braket{
            0_{\eta_f}
            \left|
            \phi_{I,n}(\eta_2)\phi_{I,m}(\eta_1)
            \right|
            0_{\eta_i}
        }
    }{
        \braket{0_{\eta_f}|0_{\eta_i}}
    },
\end{equation}
and performing the Wick contractions, 
the disconnected vacuum-bubble contributions cancel, leaving
\begin{align}
\label{eq:inout-one-loop}
    B^{\rm in\text{-}out}_{nm}(\eta_*)
    &=
    G^>_{nm}(\eta_*,\eta_*)
    \nonumber\\
    &\quad
    -
    \frac{i}{2}
    \int_{\eta_i}^{\eta_*} d\eta\,
    \lambda(\eta)\,\Omega^2(\eta)\,
    \sum_j a\,
    G^<_{jn}(\eta,\eta_*)\,
    G^<_{jm}(\eta,\eta_*)\,
    G^>_{jj}(\eta,\eta)
    \nonumber\\
    &\quad
    -
    \frac{i}{2}
    \int_{\eta_*}^{\eta_f} d\eta\,
    \lambda(\eta)\,\Omega^2(\eta)\,
    \sum_j a\,
    G^>_{jn}(\eta,\eta_*)\,
    G^>_{jm}(\eta,\eta_*)\,
    G^>_{jj}(\eta,\eta)
    +
    \mathcal{O}(\lambda^2)\;.
\end{align}
Note that when the in and out vacua coincide, $G^{\gtrless} = W$ is the Wightman function defined in~\eqref{eq:wightmanfree}.

\subsection{Checks in the heavy- and light-mass regimes}
\label{subsec:perturbative_checks}
In this section, we summarize the checks for the case when the operator string is the one-point insertion $\mathcal{O} = \phi^2$ at the center of the lattice, focusing separately on the heavy case \(m^2=1\) and the light case \(m^2=0.1\). We demonstrate that the heavy case behaves as expected perturbatively, whereas the light case is much more subtle. A na\"ive perturbative in--out estimate can be badly enhanced, but the non-perturbative data show a strong improvement with bond dimension and leave open the possibility that interaction effects shift the effective late-time scaling so that the in--out construction still gives the correct correlator.

Dropping the site labels, we write equation~\eqref{eq:inin-first-order-commutator} to the first order in $\lambda$ as
\begin{equation}
B_{\rm in\text{-}in}(\eta_*)=B^{(0)}+\lambda B^{(1)}_{\rm in\text{-}in}+\mathcal{O}(\lambda^2)\;,
\end{equation}
Analogously, for the finite-time in--out quantity we write~\eqref{eq:inout-one-loop} as
\begin{equation}
    B_{\text{in-out}}(\eta_*) = B^{(0)} + \lambda B^{(1)}_{\text{in-out}} + \mathcal{O}(\lambda^2)\,.
\end{equation}
In the above, the free theory piece $B^{(0)}$ is the same for both cases as the in and out vacuum are taken to be the same.

\paragraph{Heavy field ($m^2 = 1$).} We begin with the set of parameters used in the main comparison
\[
N=11,\qquad \eta_i=-10,\qquad \eta_*=-1,\qquad \eta_f=10,\qquad \eta_0=0.3,
\]
where we will first focus on Approach A. For the adiabatically switched profile we use the same adiabatic turn-on as in the main text, with
\(T_A=2\) and \(\mu=0.3\). We also compare against the case when the coupling is kept constant on the interval
\([\eta_i,\eta_*]\).

\begin{table}[H]
\centering
\begin{tabular}{|l|c|c|c|c|}
\hline
Profile & $B^{(0)}$ & $B^{(1)}_{\text{in-in}}$ & $B^{\rm pert}_{\text{in-in}}(0.1)$ & $B^{\rm pert}_{\text{in-in}}(0.5)$ \\
\hline
Adiabatic & $0.3672291$ & $-0.0297393$ & $0.3642552$ & $0.3523595$ \\
Constant  & $0.3672291$ & $-0.0268671$ & $0.3645424$ & $0.3537956$ \\
\hline
\end{tabular}
\caption{Perturbative in--in results for $m^2=1$, $N=11$, $\eta_0=0.3$, $\eta_*=-1$.}
\label{tab:pert_m1_inin}
\end{table}

Table~\ref{tab:pert_m1_inin} shows that the in--in quantity is only weakly sensitive to the
switching prescription.
At $\lambda=0.1$ the difference between the adiabatic and constant profiles is $2.9\times10^{-4}$,
and even at $\lambda=0.5$ it remains only $1.4\times10^{-3}$.
The same conclusion holds for the finite-time in--out quantity, as can be seen in
Table~\ref{tab:pert_m1_inout}.
The real parts of $B_{\rm in\text{-}out}$ under the two profiles agree to better than $3\times10^{-4}$
at $\lambda=0.1$, and remain close to the corresponding in--in values, consistent with a sub-percent
mismatch in $B_{\rm in\text{-}in}=B_{\rm in\text{-}out}$
at this regulator value ($\eta_0=0.3$).
The imaginary part of $B_{\rm in\text{-}out}$ is nonzero but small
($\mathrm{Im}[B_{\rm in\text{-}out}(0.1)]\approx2\times10^{-3}$).
\textcolor{black}{This imaginary part is obtained with the regulated real contour and the specified preparation at finite times. Section~\ref{sec:asymptotic-projection} instead studies the asymptotically projected observable on a complex contour.}
\begin{table}[H]
\centering
\begin{tabular}{|l|c|c|c|}
\hline
Profile & $B_{\text{in-out}}^{(1)}$ & $B^{\rm pert}_{\rm in\text{-}out}(0.1)$ & $B^{\rm pert}_{\rm in\text{-}out}(0.5)$ \\
\hline
Adiabatic & $-0.05523+0.01887\,i$ & $0.36171+0.00189\,i$ & $0.33961+0.00943\,i$ \\
Constant  & $-0.05244+0.02184\,i$ & $0.36198+0.00218\,i$ & $0.34101+0.01092\,i$ \\
\hline
\end{tabular}
\caption{Finite-time in--out perturbative results for $m^2=1$, $N=11$, $\eta_0=0.3$, $\eta_*=-1$.}
\label{tab:pert_m1_inout}
\end{table}

\begin{figure}[!htb]
\centering
\includegraphics[width=\linewidth]{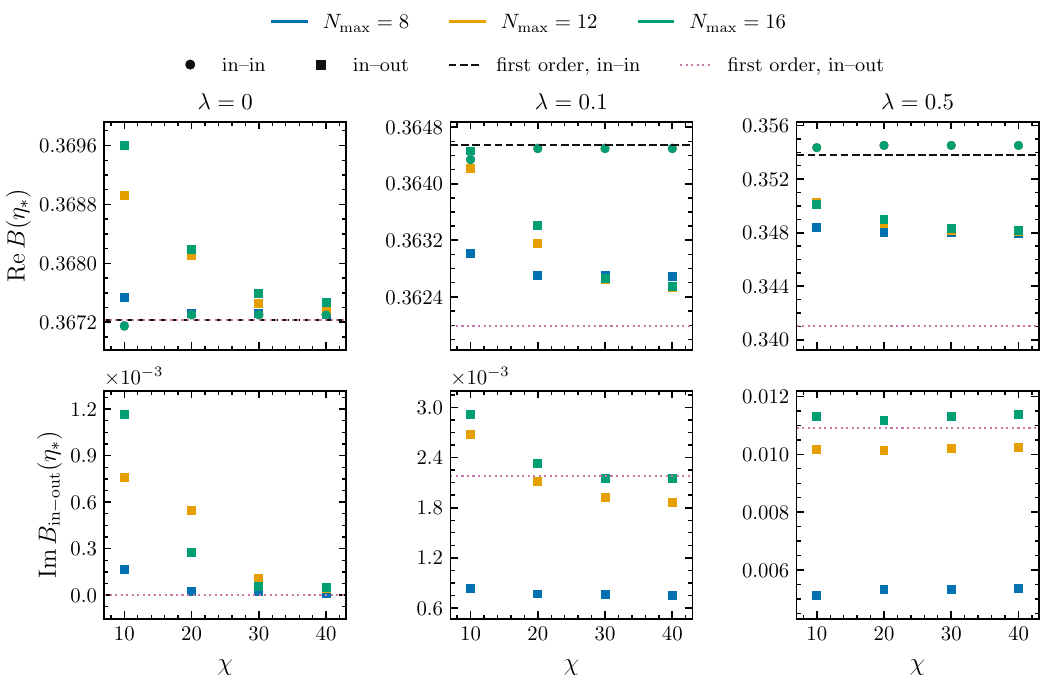}
\caption{Dependence of the correlators on the bond dimension $\chi$ and oscillator cutoff $N_{\rm max}$, for $\lambda=0,0.1,0.5$ (left to right). The top row shows the real parts of the in--in (circles) and in--out (squares) correlators. The bottom row shows the signed imaginary part of the in--out correlator. Horizontal lines show the first-order perturbative results. Approach B is used with $N=11$, $m^2=1$, $\eta_i=-10$, $\eta_*=-1$, $\eta_f=10$, and $\eta_0=0.3$. \textcolor{black}{The scans show the dependence on each cutoff. 
}}
\label{fig:binin_binout_vs_chi_lambda0}
\end{figure}

\begingroup\color{black} 
The cutoff dependence in Figure~\ref{fig:binin_binout_vs_chi_lambda0} is nonmonotone, especially for the imaginary part. At zero and small coupling the results are close to the perturbative curves, while the disagreement is larger at \(\lambda=0.5\). We therefore do not extrapolate these data.
\endgroup

\begin{figure}[t]
    \centering
    \includegraphics[width=0.9\linewidth]{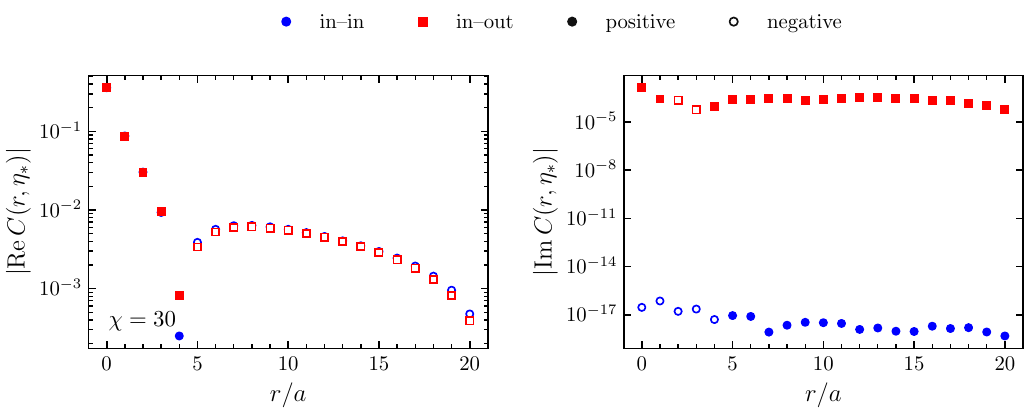}

    \vspace{0.5cm}

    \includegraphics[width=0.9\linewidth]{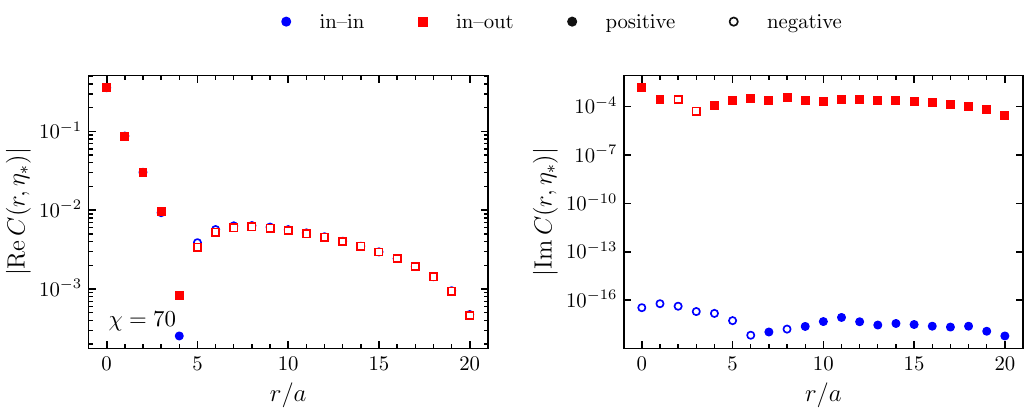}

    \caption{Magnitudes of the real and imaginary parts of $C(r, \eta_*)$ on logarithmic axes, with $\chi = 30$ (top) and $\chi = 70$ (bottom), $N = 40, N_{\rm max} = 15, \eta_* = -1, \eta_i = -20, \eta_f = 20, m^2 = 1, \lambda = 0.1, \eta_0 = 0.3, \mu = 0.3$. Filled markers denote positive values whereas hollow ones denote negative. Exact zeros are omitted from the logarithmic axes. Increasing the bond dimension improves the agreement at larger values of $r$. However, close to zero crossing there is a large disagreement in the values of the in--in and in--out correlators. Approach A was used.}
    \label{fig:correlator}
\end{figure}

While the checks here have focused on coincident operator insertion, we can complement them with tests for the separated correlator \(C(r,\eta_\ast)\), as shown in Figure~\ref{fig:correlator}. Here one field is inserted at site 20 of the $N=40$ chain and the other at site $j\geq20$, with $r=(j-20)a$. Unlike the midpoint quantity \(O=\phi^2\), this correlator retains the dependence on spatial separation and is correspondingly more sensitive to residual numerical and regulator effects, especially near zero-crossings. We therefore regard it as a supplementary qualitative check rather than the primary test of the Donath--Pajer comparison in the heavy-mass regime.

\paragraph{Nonperturbative results for a light field (\(m^2=0.1\)).}

The light-field regime is more revealing. To isolate the effect, we performed an additional constant-profile perturbative check\footnote{Since we are working on the lattice with various truncations in place, it makes sense to ask how big the perturbative corrections are in order to avoid numerical coincidences.} with 
\[
N=11,\quad N_{\max}=10,\quad \lambda=0.1,\quad
\eta_i=-10,\quad \eta_*=-2.5,\quad \eta_f=10,\quad \eta_0=0.01,
\]
and no adiabatic damping.

\begingroup\color{black}
\begin{table}[H]
\centering
\small
\begin{tabular}{|c|c|c|c|}
\hline
$m^2$ & $B^{(0)}$ & $B_{\rm in\text{-}out}^{(1)}$ & $B_{\rm in\text{-}out}^{\rm pert}(0.1)$\\
\hline
$0.1$ & $0.619$ & $-4.87\times10^4-2.81\times10^3\,i$ & $\simeq-4.87\times10^3-2.81\times10^2\,i$\\
$1$ & $0.502$ & $-1.73\times10^{-2}+3.67\times10^{-2}\,i$ & $\simeq0.500+3.67\times10^{-3}\,i$\\
\hline
\end{tabular}
\caption{\textcolor{black}{First-order \emph{perturbative} in--out quantities with constant coupling at $\eta_*=-2.5$, $\eta_0=0.01$ and $N=11$. The final column is the truncated series $B_{\rm in\text{-}out}^{\rm pert}(0.1)=B^{(0)}+0.1B_{\rm in\text{-}out}^{(1)}$, rounded to the precision of the reported coefficients. The table contains only the free term and the first-order in--out results. The MPS correlators are reported separately below.}}
\label{tab:pert_light_vs_heavy}
\end{table}
Table~\ref{tab:pert_light_vs_heavy} contains the free term and the first-order in--out series. We report the independently computed MPS correlators at $\lambda=0.1$ in Table~\ref{tab:smallm2highchi}, where the $N=11$ in--in value is $0.613422$. Extracting $B_{\rm in\text{-}in}^{(1)}$ from that value by subtracting $B^{(0)}$ and dividing by $0.1$ would require control of the higher orders. We therefore retain the MPS value as a finite-coupling result.

For $m^2=0.1$, the reported first-order in--out correction is much larger than the free term, whereas for $m^2=1$ it is small. The first-order truncation of the finite-time, regulated series is therefore unreliable for the light field. The exact de Sitter bound $m^2>3/16$ concerns uniform regulator removal. 
\endgroup

\begingroup\color{black}
The nonperturbative results at finite cutoffs behave differently from this truncated series. Table~\ref{tab:smallm2highchi} compares $N=9,10,11$. For even $N=10$ the insertion is slightly off centre. For $N=9$ the displayed relative difference between the real parts reaches $0.39\%$. For $N=11$ the scan is nonmonotone, and the additional reported point $N_{\max}=14$, $\chi=150$ gives $2.20\%$. The real-part differences are not converged, while the real-contour imaginary parts are sizable. We examine the numerical uncertainty below.
\endgroup

\begin{table}[!htb]
    \centering
    \caption{Correlators for $N=9,10,11$, with $m^2=0.1$, $\eta_0=0.01$, $\eta_*=-2.5$, $\eta_f=-\eta_i=10.0$, $N_{\max}=10$. The simulations use Approach B. \textcolor{black}{The last column gives the relative difference of the real parts. Numerical uncertainty and the imaginary part enter separately when testing equality of the full correlators.}}
    \vspace{0.2cm}
    \begin{tabular}{|c|c|c|c|c|c|}
        \hline
        $N$ & $\chi$ & $\text{Re(in-in)}$ & $\text{Re(in-out)}$ & $\text{Im(in-out)}$ & \textcolor{black}{Real difference (\%)} \\
        \hline
        
        \multirow{4}{*}{9} 
        & 30  & 0.588990 & 0.581400 & 0.235206 & 1.29\% \\
        & 50  & 0.588990 & 0.580242 & 0.234314 & 1.49\% \\
        & 80  & 0.588990 & 0.584478 & 0.240846 & 0.77\% \\
        & 100 & 0.588990 & 0.586668 & 0.242954 & 0.39\% \\
        \hline
        
        \multirow{4}{*}{10} 
        & 30  & 0.600400 & 0.581154 & 0.231005 & 3.21\% \\
        & 50  & 0.600401 & 0.579556 & 0.237189 & 3.47\% \\
        & 80  & 0.600401 & 0.579365 & 0.240389 & 3.50\% \\
        & 100 & 0.600401 & 0.583061 & 0.242595 & 2.89\% \\
        \hline
        
        \multirow{4}{*}{11} 
        & 30  & 0.613422 & 0.579702 & 0.233253 & 5.50\% \\
        & 50  & 0.613422 & 0.576153 & 0.232869 & 6.08\% \\
        & 80  & 0.613422 & 0.574734 & 0.239489 & 6.31\% \\
        & 100 & 0.613422 & 0.581209 & 0.241648 & 5.25\% \\
        \hline
    \end{tabular}\label{tab:smallm2highchi}
\end{table}

\begingroup\color{black}
\paragraph{Numerical uncertainty in the real-contour calculations.}
For a direct check using the displayed data, define
$\delta_\chi X=X_{\chi=100}-X_{\chi=80}$ at fixed $N_{\max}=10$.
The last available increments and final real-part differences are
\[
\begin{array}{|c|c|c|c|}
\hline
N & |\delta_\chi\operatorname{Re}B_{\rm in\text{-}out}|
  & |\delta_\chi\operatorname{Im}B_{\rm in\text{-}out}|
  & |B_{\rm in\text{-}in}-\operatorname{Re}B_{\rm in\text{-}out}|_{\chi=100}\\ \hline
9 & 0.002190 & 0.002108 & 0.002322\\
10 & 0.003696 & 0.002206 & 0.017340\\
11 & 0.006475 & 0.002159 & 0.032213\\
\hline
\end{array}
\]
For $N=9$ the last real-part increment is $94\%$ of the final real-part difference. The increments measure the change with bond dimension. Because the sequence is nonmonotone, they provide no bound on the remaining error, and the small difference between the real parts remains indistinguishable from numerical error. The imaginary part is much larger than the last bond-dimension increment. Establishing its convergence requires increasing the cutoffs and reducing the time step.

A quantitative error estimate for the larger lattices would require independent variations of \(\chi\), \(N_{\max}\), and the time step at fixed physical parameters. We therefore do not assign a combined error to the smallest differences in Figure~\ref{fig:binin_binout_vs_chi_lambda0} and Table~\ref{tab:smallm2highchi}. Variations of the endpoints and geometric regulator are treated separately because they change the finite-time prescription. The ED--MPS comparisons and time-step tests below provide checks on the smaller systems.

\endgroup

{ \color{black}
To test whether the sizable imaginary part observed in the light-mass in--out quantity is merely an MPS truncation artifact, we compared TEBD with exact diagonalization on a smaller system. For a five-site system with \(N_{\max}=4\), and with a shortened contour chosen for numerical feasibility, exact evolution reproduces the same qualitative behavior and agrees very well with TEBD for both the real and imaginary parts of \(B_{\rm in\text{-}out}\). For example, at \(\eta_0=0.01\) we find \(\Im B_{\rm in\text{-}out}=0.20939\) in exact evolution and \(0.20925\) in TEBD, while at \(\eta_0=0.03\) the corresponding values are \(0.03392\) and \(0.03397\). The imaginary part also decreases sharply as \(\eta_0\) is increased, becoming very small by \(\eta_0=0.3\). These checks strongly disfavor interpreting the light-mass imaginary part as a simple MPS truncation artifact.

The effect is also clearly due to interactions and stronger for the lighter field. In both TEBD and Hartree calculations, the imaginary part is strongly reduced when \(\lambda\) is lowered from \(0.1\) to \(0.01\), and it vanishes within numerical accuracy at \(\lambda=0\). Likewise, for fixed \(\lambda\) and \(\eta_0\), the heavy scalar case \(m^2=1\) shows only a very small imaginary part, whereas the light scalar case \(m^2=0.1\) shows a much larger one. This is consistent with the expectation that light fields are especially sensitive to the regulated gluing region, since their superhorizon fluctuations decay slowly and are more easily squeezed by the time-dependent effective mass.

These observations support interpreting \(B_{\rm in\text{-}out}\) as a genuine transition quantity,
\[
B_{\rm in\text{-}out}
=
\frac{\langle \chi|O|\psi\rangle}{\langle \chi|\psi\rangle},
\]
rather than as an ordinary expectation value. For Hermitian \(O\), such a ratio is generically complex unless the out-branch state \(|\chi\rangle\) is proportional to the in-branch state \(|\psi\rangle\). In the present regulated problem, propagation through the \(\eta=0\) region produces Bogoliubov mixing and squeezing, so the ket and bra states arrive at \(\eta_\ast\) with different phase-space structures. The real part can therefore remain close to the in--in value while an additional imaginary part is generated by a mismatch in the conjugate quadrature, equivalently by a difference in the squeeze angles of the two branches.

A Hartree calculation supports this interpretation at the qualitative level, but not as a complete quantitative explanation. In the Hartree approximation, the quartic interaction generates the self-consistent time-dependent local mass shift~\eqref{hf1}, so the dynamics remains Gaussian while still incorporating interaction-induced squeezing. For the same light-mass parameters as in Table~\ref{tab:smallm2highchi}, namely \(N=9\), \(m^2=0.1\), \(\lambda=0.1\), \(\eta_i=-10\), \(\eta_\ast=-2.5\), \(\eta_f=10\), and \(\eta_0=0.01\), we find
\[
\Re B_{\rm in\text{-}in}^{\rm HF}=0.588995,\qquad
\Re B_{\rm in\text{-}out}^{\rm HF}=0.552683,\qquad
\Im B_{\rm in\text{-}out}^{\rm HF}=0.107402.
\]
Thus Hartree already produces a substantial nonzero imaginary part, showing that a difference in the squeezing of the two Gaussian states is sufficient to generate the effect, but it reproduces only part of the full MPS value, for which \(\Im B_{\rm in\text{-}out}\approx 0.23\). The most natural interpretation is therefore that Gaussian mass-renormalization and squeezing account for an important part of the mechanism, while the full MPS result also contains additional corrections absent from the Hartree approximation.}

\begingroup\color{black}
The Hartree--Fock calculation illustrates how mass shifts and squeezing affect the correlators. On the real contour, the full complex correlators still differ because the imaginary part is nonzero. Agreement of the real parts therefore leaves the proposed equivalence and any change in the allowed late-time scaling dimensions unresolved. To isolate an improvement over perturbation theory due to nonperturbative effects, we must compare results obtained with the same contour and state prescription. For related mass-shift analyses see \cite{massper1,massper2,massper3,massper4,Gorbenko:2019rza}.
\endgroup

\begin{figure}[ht]
    \centering
    \includegraphics[width=\textwidth]{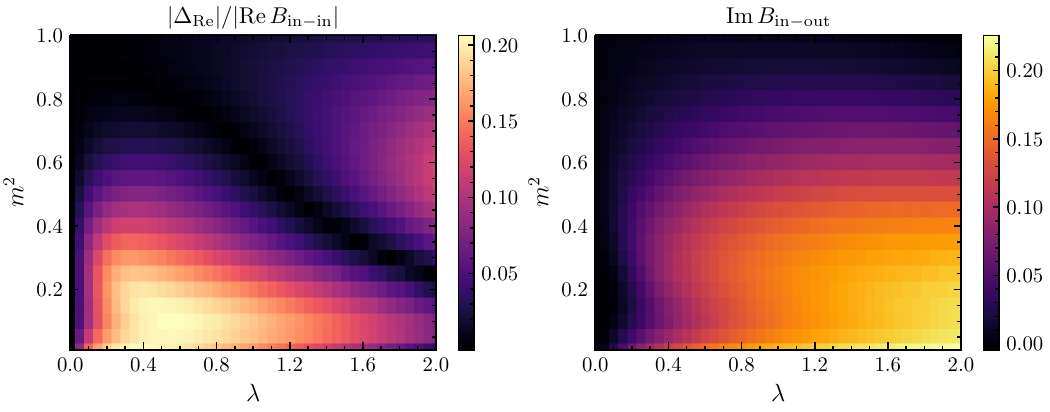}
    \caption{Midpoint correlator $\langle\phi^2\rangle$ over the $m^2$--$\lambda$ parameter space with $N=11$, $\chi=30$, $N_{\rm max}=12$, $\eta_*=-2.5$, $\eta_0=0.3$. The left panel shows the relative difference between the real parts of the in-in and in-out correlators, $|\Re\langle\phi^2\rangle_{\mathrm{in\text{-}in}} - \Re\langle\phi^2\rangle_{\mathrm{in\text{-}out}}| / |\Re\langle\phi^2\rangle_{\mathrm{in\text{-}in}}|$. The right panel shows the signed imaginary part of the in-out correlator, $\Im\langle\phi^2\rangle_{\mathrm{in\text{-}out}}$, with values increasing from dark purple through orange to yellow.}
    \label{fig:heatmap_phisq}
\end{figure}

\paragraph{General mass and coupling.} The comparison of in-in and in-out correlators across general ranges of $m^2$ and $\lambda$ is presented in Figure \ref{fig:heatmap_phisq}. In the left panel, darker regions indicate closer agreement of the real parts, especially at very small couplings and larger masses. In the plot for the relative error in the real parts, there is also a noticeable dark ridge which anti-correlates $m^2$ and $\lambda$. However, the imaginary part of the in-out correlator remains substantial for this parameter subspace.

\subsection*{Comparison between approaches A and B}

We now compare approaches A and B, starting with the real contour prescription for in-out computations. Their different switching profiles and MPS implementations mean that they are not expected to give identical results. For the heavy-field parameters tested, we find comparable values, with representative differences shown in Table~\ref{tab:inin_inout_loggrid_fixed}. For small $\eta_0$, Approach B, which uses TEBD, is more suitable.

\begin{table}[H]
\centering
\caption{In--in vs in--out for $O=\phi^2$ at the center site (site 5). 
Parameters are
$N=9$, $N_{\max}=10$, $\chi=30$, $m^2=1$, $\eta_i=-10$, $\eta_*=-1$, $\eta_f=10$,
$\eta_0=0.3$, $\mu=0.3$, $T_A=2$, $\mathrm{dt}_{\max}=0.1$, $N_{\mathrm{steps}}=12$.
The relative difference of the real parts is
$\Delta_{\rm DP}=|\Re B_{\rm in\text{-}out}-B_{\rm in\text{-}in}|/|B_{\rm in\text{-}in}|$.}
\begin{tabular}{|c|c|c|c|c|c|c|c|}
\hline
Approach & $\lambda$ & ${\rm den}$ & $\Re B_{\rm in\text{-}in}$ & $\Im B_{\rm in\text{-}in}$ & $\Re B_{\rm in\text{-}out}$ & $\Im B_{\rm in\text{-}out}$ & $\Delta_{\rm DP}$ \\
\hline
A & 0.1 & 0.99904 & 0.36540 & $\sim 0$ & 0.36094 & $3.20\times10^{-3}$ & 1.22\% \\
B & 0.1 & 0.99955 & 0.36553 & $\sim 0$ & 0.36288 & $1.51\times10^{-3}$ & 0.73\% \\
A & 0.5 & 0.98389 & 0.35483 & $\sim 0$ & 0.34263 & $1.86\times10^{-2}$ & 3.44\% \\
B & 0.5 & 0.99462 & 0.35567 & $\sim 0$ & 0.34795 & $8.52\times10^{-3}$ & 2.17\% \\
\hline
\end{tabular}
\label{tab:inin_inout_loggrid_fixed}
\end{table}

\begingroup
\color{black}
\subsection*{Derivative interaction without a regulator}
\label{subsec:derivative_interaction_without_a_regulator}

We found that for light scalar fields the agreement between the real part of in-out is excellent with the in-in, suggestive of a resolution of the light field singularities encountered in perturbation theory. However, the real contour gave a residual imaginary part for the in-out. There are two possibilities that we need to examine now to see if this residual is an artifact of the regulator or the simplistic real time contour--- a) The metric regulator b) The real time contour. We start with the metric regulator.

We consider a massless scalar with a derivative interaction that is regular
at the gluing surface \(\eta=0\). We compare the correlators calculated with
the exact de Sitter metric and with \(\Omega_{\rm reg}\) to determine the
effect of the metric regulator. The covariant interaction is
\begin{equation}
 S_{\partial}=-\frac{\lambda_{\partial}}{4!}
 \int d^2x\,\sqrt{-g}\,\big[(\nabla\phi)^2\big]^2 .
\end{equation}
In conformal coordinates, its contribution to the Lagrangian density is
\begin{equation}
 {\cal L}_{\partial}=-\frac{\lambda_{\partial}}{4!\,\Omega^2(\eta)}
 \left[\phi_{\eta}^2-(\partial_x\phi)^2\right]^2 .
\end{equation}
Here \(\phi_\eta=\partial_\eta\phi\). We work to first order in \(\lambda_{\partial}\). Since the interaction
contains a time derivative, it changes the canonical momentum to
\begin{equation}
 \pi=\phi_{\eta}-\frac{\lambda_{\partial}}{6\,\Omega^2}
 \phi_{\eta}\left[\phi_{\eta}^2-(\partial_x\phi)^2\right],
\end{equation}
so inverting this relation perturbatively gives
\begin{equation}
 \phi_{\eta}=\pi+\frac{\lambda_{\partial}}{6\,\Omega^2}
 \pi\left[\pi^2-(\partial_x\phi)^2\right]
 +O(\lambda_{\partial}^2).
\end{equation}
The Legendre transform therefore gives
\begin{equation}
 H=H_0+\frac{\lambda_{\partial}}{4!\,\Omega^2(\eta)}
 \int dx\,\left[\pi^2-(\partial_x\phi)^2\right]^2
 +O(\lambda_{\partial}^2).
\end{equation}
For the exact de Sitter metric, \(\Omega^{-2}(\eta)=H^2\eta^2\), so the
interaction vanishes at \(\eta=0\). On the lattice we impose Dirichlet
boundary conditions and discretize the spatial derivatives symmetrically:
\begin{equation}
 K_j=\Pi_j^2-\frac12\left[(\Phi_{j+1}-\Phi_j)^2
 +(\Phi_j-\Phi_{j-1})^2\right],\qquad
 H_{\partial,\mathrm{lat}}=\frac{1}{4!}\sum_j K_j^2,
\end{equation}
with \(\Phi_0=\Phi_{N+1}=0\) and dimensionless coupling
\(g=\lambda_{\partial}/a^2\). In dimensionless lattice units, we evolve with
\begin{equation}
 H_{\rm lat}(\eta)=H_{\rm lat,0}
 +g\,s(\eta)\,\Omega^{-2}(\eta)\,H_{\partial,\mathrm{lat}}
 +O(g^2).
\end{equation}

The switching function \(s(\eta)=\exp(-\mu\eta^2)\) turns the interaction
on and off smoothly, suppressing sudden changes at the time endpoints.
We keep this function fixed when comparing the two metrics and calculate
the correlators by exact diagonalization (ED) in the finite lattice Hilbert
space.

We extract the coefficients \(dB/dg\) at \(g=0\) from correlators evaluated
at small couplings symmetric about zero. For \(N=4\), \(\eta_i=-4\),
\(\eta_*=-0.5\), \(\eta_f=4\), and \(\mu=0.5\), the results with the exact
metric at two oscillator cutoffs are

\begin{center}
\begin{tabular}{|c|c|c|}
\hline
\(N_{\max}\) & \(dB_{\mathrm{in-in}}/dg\)
 & \(dB_{\mathrm{in-out}}/dg\) \\
\hline
7 & 0.03644 & \(0.02744-0.01016\,i\) \\
8 & 0.03620 & \(0.02729-0.01010\,i\) \\
\hline
\end{tabular}
\end{center}

We repeat the calculation with the regulated conformal factor used in the
main simulations,
\begin{equation}
 \Omega_{\mathrm{reg}}(\eta)=\frac{1}{H\sqrt{\eta^2+\eta_0^2}},
 \qquad \eta_0=0.3 .
\end{equation}
Since the field is massless, the free Hamiltonian is the same for both
metrics. The regulator changes the interaction coefficient from
\(H^2\eta^2\) to \(H^2(\eta^2+\eta_0^2)\). At \(N_{\max}=8\) we find
\begin{equation}
 \left.\frac{dB_{\mathrm{in-in}}}{dg}\right|_{\mathrm{unreg.}}=0.03620,
 \quad
 \left.\frac{dB_{\mathrm{in-out}}}{dg}\right|_{\mathrm{unreg.}}
 =0.02729-0.01010\,i,
\end{equation}
and
\begin{equation}
 \left.\frac{dB_{\mathrm{in-in}}}{dg}\right|_{\mathrm{reg.}}=0.03847,
 \quad
 \left.\frac{dB_{\mathrm{in-out}}}{dg}\right|_{\mathrm{reg.}}
 =0.03077-0.00844\,i.
\end{equation}
Introducing the metric regulator increases \(dB_{\mathrm{in-in}}/dg\) by
approximately 6\%. The in--out coefficient has a nonzero imaginary part for
both metrics. Further ED calculations show that this imaginary part changes
with the switching function and remains nonzero when the initial state is
the instantaneous interacting state at \(\eta_i\). These observations
suggest that the finite-time switching and the choice of states at the
endpoints contribute to the difference between the correlators.

With the switching function held fixed, the correlators depend on the metric regulator and remain distinct for both metrics. To test their asymptotic equality, we next remove the switching function, computing the first-order perturbative coefficients directly and checking them independently with MPS.

\subsubsection*{Removal of switching and comparison of ED and MPS at first order}
\label{subsec:derivative_verified}
We now set $m^2=0$, $\eta_0=0$, $a=H=1$ and $\eta_*=-0.5$. Let
$H_0=H_{\rm lat,0}$ and $V=H_{\partial,\rm lat}$ with the same symmetric
$K_j$ as above. The finite oscillator matrices are multiplied before
forming $K_j^2$. ED and MPS use identical matrices. The free Dirichlet
Hamiltonian is time independent. We use its normalized, nondegenerate
ground state $|0\rangle$ at both endpoints. Its positive finite-volume excitation energies remove
the zero-mode ambiguity for this test.

Write $B=B^{(0)}+gB^{(1)}+O(g^2)$ and
$f_\mu(\eta)=\eta^2e^{-\mu\eta^2}$. The coefficients below are computed
directly from the first-order Dyson expansion. Define
\begin{align}
C(\eta)&=\langle0|V_I(\eta)O_I(\eta_*)|0\rangle_c,\qquad
I_-=\int_{-T}^{\eta_*} f_\mu(\eta)C(\eta)\,d\eta,\qquad
I_+=\int_{\eta_*}^{T} f_\mu(\eta)C(\eta)\,d\eta,\\
B_{\rm in\text{-}in}^{(1)}&=-2\,\operatorname{Im}I_-,
\qquad
B_{\rm in\text{-}out}^{(1)}=-i(I_-^*+I_+).
\label{eq:derivative_dyson_v6}
\end{align}
Here the connected subtraction removes the vacuum term in the normalized
in--out denominator. We evaluate the two integrals independently to
compare the in--in and in--out coefficients directly.
Consequently,
\begin{equation}
B_{\rm in\text{-}out}^{(1)}-B_{\rm in\text{-}in}^{(1)}
=-i\int_{-T}^{T}f_\mu(\eta)C(\eta)\,d\eta .
\label{eq:derivative_difference_v6}
\end{equation}
We test the full complex difference.

\paragraph{Analytic control of the switching limit.}
For $H_0|n\rangle=E_n|n\rangle$, let
$\omega_n=E_n-E_0>0$ and $c_n=\langle0|V|n\rangle\langle n|O|0\rangle$.
At a fixed finite Hilbert-space cutoff,
$C(\eta)=\sum_{n>0}c_n e^{-i\omega_n(\eta-\eta_*)}$. Taking $T\to\infty$
first gives the explicitly calculable difference
\begin{align}
B_{\rm in\text{-}out}^{(1)}-B_{\rm in\text{-}in}^{(1)}
&=-i\sum_{n>0}c_n e^{i\omega_n\eta_*}F_\mu(\omega_n),\\
F_\mu(\omega)
&=\sqrt{\frac{\pi}{\mu}}e^{-\omega^2/(4\mu)}
\left(\frac{1}{2\mu}-\frac{\omega^2}{4\mu^2}\right)
\ \longrightarrow\ 0\quad(\mu\to0^+,\ \omega>0).
\label{eq:derivative_fourier_v6}
\end{align}
Thus the regulator-free equality holds for these first-order finite-lattice
coefficients after the switching function is removed. At finite $\mu$ their difference need
not decrease monotonically. We use the Gaussian to ensure convergence on
the real axis and remove it after the endpoint limit. The geometry remains
unregulated throughout, and the argument is independent of any assumed
decay on complex semicircles. Exchanging the switching limit with
finite-coupling evolution of the Hamiltonian truncated at $O(g)$ requires
a separate justification.

\paragraph{Numerical implementation and switching check.}
ED uses spectral matrix elements and adaptive integration of each half
of \eqref{eq:derivative_dyson_v6}. As an independent implementation,
we apply the sum of products of local operators in $V$ to the vacuum MPS,
subtract its vacuum component, and evolve the resulting
states with second-order free TEBD. MPS overlaps give $C(\eta)$,
which we integrate by the composite Simpson rule to obtain both coefficients. The
ground state is imported from ED to isolate the tests of the interaction
insertion, evolution and contractions. Vacuum preparation by DMRG lies
outside this comparison. The MPS insertion vectors agree with
the independently assembled sparse $K_j^2$ operators to below
$2.4\times10^{-14}$ in the $N=3,d=12$ check.

For $N=3$, $d=N_{\max}+1=8$ and $O=\Phi_2^2$ (the central site), the
ED results are
\begin{center}
\begin{tabular}{|c|c|c|c|}
\hline
$\mu$ & $B_{\rm in\text{-}in}^{(1)}$ &
$B_{\rm in\text{-}out}^{(1)}$ &
$|B_{\rm in\text{-}out}^{(1)}-B_{\rm in\text{-}in}^{(1)}|$\\
\hline
$0.5$ & $0.04171898$ & $0.02379319-0.02042524i$ & $2.72\times10^{-2}$\\
$0.05$ & $-0.03051613$ & $-0.03080509-0.000300882i$ & $4.17\times10^{-4}$\\
$0.02$ & $-0.02409235$ & $-0.02409235-7.18\times10^{-11}i$ & $9.96\times10^{-11}$\\
$0.01$ & $-0.02264350$ & $-0.02264350+O(10^{-14})i$ & $<10^{-13}$\\
\hline
\end{tabular}
\end{center}
The $\mu=0.5$ row uses $T=24$ and the remaining rows use $T=60$.
The numerical full-line differences agree with
\eqref{eq:derivative_fourier_v6} to below $10^{-13}$ for these cases.
The adaptive integrator's absolute error estimates reach
$3.2\times10^{-12}$. These estimates leave absolute accuracy at the
$10^{-14}$ level unresolved. At $\mu=0.01$,
reducing the endpoint magnitude to $T=40$ instead gives a difference of
$4.79\times10^{-6}$, demonstrating why the endpoint and switching limits
must be controlled together.

\paragraph{MPS convergence and a second lattice.}
At $\mu=0.01$ and $T=60.5$, define
$E_{\rm MPS}=\max_{X={\rm in\text{-}in,in\text{-}out}}
|B_X^{(1),{\rm MPS}}-B_X^{(1),{\rm ED}}|$.
The following table separates implementation error from agreement
between the two correlators:
\begin{center}
\begin{tabular}{|c|c|c|c|c|}
\hline
$(N,d)$ & $\chi$ & $\Delta\eta$ & $E_{\rm MPS}$ &
$|B_{\rm in\text{-}out}^{(1)}-B_{\rm in\text{-}in}^{(1)}|_{\rm MPS}$\\
\hline
$(3,8)$ & $8$ & $0.04$ & $2.27\times10^{-5}$ & $1.58\times10^{-7}$\\
$(3,8)$ & $8$ & $0.02$ & $5.69\times10^{-6}$ & $9.89\times10^{-9}$\\
$(3,8)$ & $8$ & $0.01$ & $1.42\times10^{-6}$ & $6.18\times10^{-10}$\\
$(4,6)$ & $36$ & $0.04$ & $7.17\times10^{-5}$ & $6.35\times10^{-7}$\\
$(4,6)$ & $36$ & $0.02$ & $1.79\times10^{-5}$ & $3.97\times10^{-8}$\\
$(4,6)$ & $36$ & $0.01$ & $4.48\times10^{-6}$ & $2.48\times10^{-9}$\\
\hline
\end{tabular}
\end{center}
Here $O=\Phi_2^2$ for both lattice sizes. The MPS--ED errors decrease by
approximately four on each step halving, as expected. The full bond
dimensions in the table permit exact representation at these small
sizes, so no Schmidt weight is discarded. Reducing $\chi$ to $4$ for
$(N,d)=(3,8)$ gives a difference of $1.77\times10^{-2}$ between the complex correlators, while
$\chi=12$ for $(4,6)$ gives $5.46\times10^{-4}$ at $\Delta\eta=0.02$.
At these reduced bond dimensions, the calculations do not meet the required accuracy.
Extrapolating the $0.02$ and $0.01$ steps with the independently checked
second-order law gives MPS--ED in--out differences
$2.51\times10^{-9}$ and $9.96\times10^{-9}$ for the three- and four-site
cases, respectively.

A separated insertion $O=\Phi_1\Phi_3$ at $N=3,d=8$, with the same
$\mu,T,\chi=8$ and $\Delta\eta=0.01$, gives ED coefficient
$-0.01150773137$ and MPS coefficient
$-0.01150695402+3.08\times10^{-10}i$. Its complex in--in/in--out
difference is $3.08\times10^{-10}$. Together, these finite-volume
comparisons test both coincident and separated insertions at the stated
lattice sizes.

\paragraph{Convergence of the first-order coefficient.}
We next examine the convergence of the value on which the two correlators
agree. For the real matrix elements of the observables used here, the
unswitched first-order coefficient has the spectral expression
\begin{equation}
B^{(1)}_{\mu=0}
=2\sum_{n>0}c_n
\left(\frac{2}{\omega_n^3}-\frac{\eta_*^2}{\omega_n}\right).
\label{eq:derivative_unswitched_v6}
\end{equation}
This expression follows by evaluating the damped half-line integral
before removing its damping. At $N=3$, the oscillator-cutoff sequence is
\begin{center}
\begin{tabular}{|c|c|c|}
\hline
$d$ & $N_{\max}$ & $B^{(1)}_{\mu=0}$\\
\hline
$8$ & $7$ & $-0.02139219444$\\
$10$ & $9$ & $-0.02114252844$\\
$12$ & $11$ & $-0.02112797463$\\
$14$ & $13$ & $-0.02112730234$\\
\hline
\end{tabular}
\end{center}
The last increment is $6.72\times10^{-7}$, or $3.18\times10^{-5}$ relatively. This measures the change between the last two cutoffs. At $d=12,\mu=0.01,T=60.5,\chi=12,\Delta\eta=0.01$, MPS gives $B_{\rm in\text{-}in}^{(1)}=-0.02236326087$ and
$B_{\rm in\text{-}out}^{(1)}=-0.02236326087+6.05\times10^{-10}i$,
with MPS--ED error $1.44\times10^{-6}$. The remaining shift of this common value from
\eqref{eq:derivative_unswitched_v6} is switching dependence.
For example, at fixed $d=8$ the ED values at
$\mu=0.002,0.001,0.0005$ are
$-0.02162912,-0.02150989,-0.02145085$ ($T=300$), approaching
$-0.02139219$. The unswitched coefficient is defined by the analytic
limit as $\mu$ tends to zero.

The Legendre transform used here retains terms through $O(g)$, and the
calculations keep a finite spatial cutoff. In this setting, the MPS
calculation reproduces the first-order result and tests equality of the
correlators without a metric regulator. 
\endgroup

\section{Asymptotic projection and the imaginary part}
\label{sec:asymptotic-projection}

\begingroup\color{black}
The light-field calculation on the real contour gives a sizeable imaginary part in the finite-time in--out ratio, which persists over the range of cutoffs tested. To compare this result with the asymptotic identity, we must specify the observable and how the boundary states are prepared.

The geometric regulator $\eta_0$ modifies the Hamiltonian near the gluing surface and changes its complex analytic structure. The endpoint tilt $\epsilon$ implements asymptotic state preparation. The tests below support suppression of the imaginary part and of the full in--in/in--out difference as the endpoints move farther from the origin with nonzero tilt. We then examine removal of the regulator at fixed lattice cutoffs. To assess the proposed ordered limits, we follow the full correlator through this sequence.

\paragraph{Choice of regulator.}
At finite $\eta_0$ the real-contour Hamiltonian is Hermitian and local, while $\Omega_{\rm reg}^2(z)$ has conjugate poles. The regulator therefore changes the analytic structure of exact de Sitter space. A one-sided complex regulator can avoid poles in a chosen half-plane but generally sacrifices real-time Hermiticity. Introducing the regulator in the interaction also requires checking both properties. Here we retain the known regulator, choose a contour below its lower pole, and test removal of $\eta_0$ for the projected observable. We implement the resulting nonunitary evolution classically with MPS. A direct unitary-circuit implementation is unavailable for this evolution, which was the reason behind choosing the simplistic real time contour. This calculation uses a different prescription from the derivative-interaction test without a metric regulator above.
\endgroup

\subsection{Analytic continuation and vacuum projection}
\label{subsec:continuation-projection}

After complexifying conformal time, $\eta\to z$, the regulated scalar Hamiltonian contains
\begin{equation}
  \Omega_{\mathrm{reg}}^2(z)
  =\frac{1}{H^2\bigl(z^2+\eta_0^2\bigr)}.
  \label{eq:omega-complex}
\end{equation}
This coefficient is single-valued and has simple poles at $z=\pm\ii\eta_0$.
For nonzero mass, the scalar mode functions generally have logarithmic monodromy about
these points, as described in Appendix~\ref{app:cosmo:pert}. We choose cuts extending
upward from $+\ii\eta_0$ and downward from $-\ii\eta_0$, and continue to the outgoing patch
along a path below the lower singularity. The labels Sheet~I and Sheet~II distinguish
the branches encountered before and after crossing the lower cut once. The branch of
the solution, together with its position in the complex $z$-plane, specifies its value.
Further windings around a logarithmic
branch point generally give further branches.

\begin{figure}[H]
  \centering
  \includegraphics[width=\linewidth]{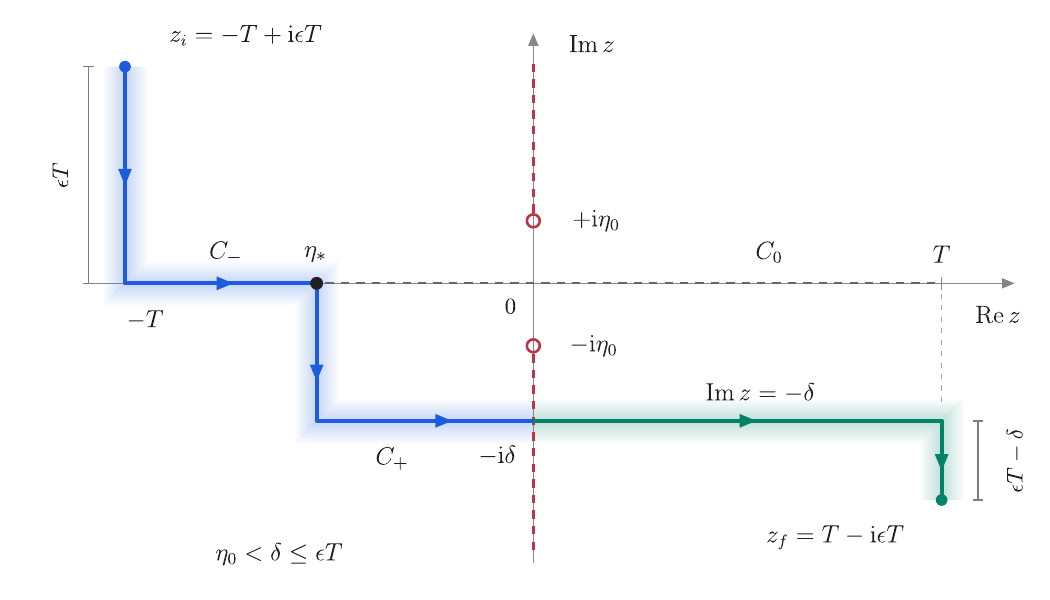}
  \caption{Contour used for asymptotic vacuum projection. The path $C_-$ reaches the
  insertion at $\eta_*$. The path $C_+$ descends to $\operatorname{Im}z=-\delta$, crosses the
  lower cut, and reaches $z_f$. The brackets mark the imaginary-time lengths of the
  endpoint projection pieces. The red dashed rays show the chosen cuts. Blue and green
  follow the solution along the path before and after crossing the lower cut. The fading
  bands emphasize these path segments. The dashed real-time
  contour $C_0$ shares its segment from $-T$ to $\eta_*$ with $C_-$.}
  \label{fig:projected-contour}
\end{figure}

For endpoints of magnitude $T$, the vacuum prescription is implemented by
\begin{equation}
  z_i=-T+\ii\epsilon T,
  \qquad
  z_f=+T-\ii\epsilon T,
  \qquad \epsilon>0.
  \label{eq:tilted-endpoints-section}
\end{equation}
For the numerical calculation we use the contour
\begin{align}
  C_- &: z_i\longrightarrow -T\longrightarrow\eta_*,
  \nonumber\\
  C_+ &: \eta_*\longrightarrow\eta_*-\ii\delta
  \longrightarrow T-\ii\delta\longrightarrow z_f,
  \qquad \eta_0<\delta\le\epsilon T,
  \label{eq:implemented-contour}
\end{align}
shown in Figure \ref{fig:projected-contour}.  The implementation adds integration breakpoints to
the long horizontal piece at $\operatorname{Im}z=-\delta$.  Thus every segment of $C_-$ and
$C_+$ is either vertical or horizontal.  The future path descends at $\eta_*$ and crosses the
lower cut along the fixed-depth segment.  The two endpoint pieces have imaginary-time lengths
$\epsilon T$ and $\epsilon T-\delta$.  These lengths grow without bound as $T\to\infty$ and
project the incoming and outgoing states.  The horizontal segment gives the required
continuation to Sheet~II while remaining at fixed imaginary depth $\delta$.

To define the states entering the transition matrix element, let
\begin{equation}
  \mathcal U[C]
  =\mathcal P\exp\!\left[-\ii\int_C\dd z\,H(z)\right]
  \label{eq:contour-evolution-operator}
\end{equation}
denote evolution along an oriented contour $C$.  The incoming ket and outgoing bra evaluated
at the insertion are
\begin{equation}
  |R_T\rangle=\mathcal U[C_-]|\Omega_{\rm in}\rangle,
  \qquad
  \langle L_T|=\langle\Omega_{\rm out}|\mathcal U[C_+].
  \label{eq:left-right-definition}
\end{equation}
In the scalar theory the mass and interaction terms vanish at large $|z|$.  We may therefore
take both boundary states to be the ground state of the asymptotic lattice Hamiltonian
$H_{\rm as}$.  Their evolutions to $\eta_*$ follow $C_-$ and $C_+$, respectively.

For a Hermitian insertion $O$, the finite-time normalized ratio is
\begin{equation}
  B_{\mathrm{in\text{-}out}}(T,\epsilon,\eta_0)
  =\frac{\langle L_T|O|R_T\rangle}{\langle L_T|R_T\rangle}.
  \label{eq:finite-ratio-section}
\end{equation}
At finite $T$, $\langle L_T|$ and $|R_T\rangle$ are two different states.  Consequently,
(\ref{eq:finite-ratio-section}) can be complex even when $O$ is Hermitian.  Below we test
whether the two states become conjugate after the endpoint projection.

\begingroup\color{black}
\paragraph{Relation to the original finite-time definition.}
In \eqref{eq:BinoutA_def_setup}, the reference out-state is obtained by free evolution of the finite-time incoming state. Here the boundary ket and bra are independently specified by the asymptotic Hamiltonian and evolved on $C_-$ and $C_+$. Both the finite-time boundary preparation and the contour have changed. The appropriate comparison is therefore with $B_R$ constructed from the same incoming branch. Whether the states become Hermitian conjugates at the insertion depends on their evolution from the common boundary ground state. We test this relation through $F_{LR}$ and the full complex difference between the correlators.
\endgroup

\subsection{Finite-volume spectral argument and the order of limits}
\label{subsec:spectral-projection}

At fixed lattice size and occupation cutoff, the asymptotic Hamiltonian has a discrete
spectrum,
\begin{equation}
  H_{\mathrm{as}}|n\rangle=E_n|n\rangle,
  \qquad E_0<E_1\le E_2\le\cdots .
\end{equation}
\textcolor{black}{Approximating the Hamiltonian on the vertical endpoint segments of (\ref{eq:implemented-contour}) by its asymptotic form gives factors of the form}
\begin{equation}
  \e^{-\epsilon T(H_{\mathrm{as}}-E_0)}|\psi\rangle
  =c_0|0\rangle+
   \sum_{n>0}c_n\e^{-\epsilon T(E_n-E_0)}|n\rangle.
  \label{eq:spectral-projector-section}
\end{equation}
Thus every excited component is exponentially suppressed when $T\to\infty$ at fixed
$\epsilon>0$.  The outgoing leg gives the same suppression with
$\epsilon T\to\epsilon T-\delta$, with the same asymptotic rate because the shift is fixed.

Let $\Delta E_{\mathrm{rel}}$ be the smallest gap that can contribute in the symmetry sector
of the evolution and the observable.  We normalize the two states according to
\begin{equation}
  |\widehat R_T\rangle=\frac{|R_T\rangle}{\|R_T\|},
  \qquad
  \langle\widehat L_T|=\frac{\langle L_T|}{\|L_T\|}.
\end{equation}
Here $\|R_T\|^2=\sum_n|\langle n|R_T\rangle|^2$ and
$\|L_T\|^2=\sum_n|\langle L_T|n\rangle|^2$ in any orthonormal basis.  The fidelity used
below is
\begin{equation}
  F_{LR}(T)
  =|\langle\widehat L_T|\widehat R_T\rangle|^2
  =\frac{|\langle L_T|R_T\rangle|^2}
  {\|L_T\|^2\|R_T\|^2},
  \qquad 0\le F_{LR}\le1.
  \label{eq:fidelity-definition}
\end{equation}
\textcolor{black}{In the regime where endpoint projection dominates the residual errors, we test the following large-$T$ form after choosing phases continuously:}
\begin{align}
  |\widehat R_T\rangle
  &=|R_*\rangle
    +\order\!\left(\e^{-\epsilon T\Delta E_{\mathrm{rel}}}\right),\\
  \langle\widehat L_T|
  &=\langle L_*|
    +\order\!\left(\e^{-\epsilon T\Delta E_{\mathrm{rel}}}\right).
  \label{eq:left-right-expansion}
\end{align}
\textcolor{black}{The data below support $F_{LR}\to1$ at the tested cutoffs. In this limit, Cauchy--Schwarz gives $\langle L_*|=\e^{\ii\alpha}\langle R_*|$, so the two projected states agree up to a phase.}

To obtain the leading dependence on the imaginary-time interval, set
$q=\e^{-\epsilon T\Delta E_{\mathrm{rel}}}$.  The fixed shift $\delta$ on the outgoing
branch only changes the coefficient of $q$.  With phases chosen as above, we may write the
leading corrections as
\begin{align}
  |\widehat R_T\rangle
  &=|R_*\rangle+q|r_\perp\rangle+o(q),\\
  \langle\widehat L_T|
  &=\e^{\ii\alpha}\bigl(\langle R_*|+q\langle l_\perp|+o(q)\bigr),
  \label{eq:leading-state-corrections}
\end{align}
where the first-order corrections are orthogonal to $|R_*\rangle$.  Defining
the ket $|l_\perp\rangle$ as the Hermitian conjugate of $\langle l_\perp|$ and
$B_*=\langle R_*|O|R_*\rangle$, the normalized transition matrix element becomes
\begin{equation}
  \Binout
  =B_*+q\left[
  \langle l_\perp|(O-B_*)|R_*\rangle
  +\langle R_*|(O-B_*)|r_\perp\rangle
  \right]+o(q).
  \label{eq:transition-expansion}
\end{equation}
Since $B_*$ is real for $O=O^\dagger$, (\ref{eq:transition-expansion}) gives
\begin{align}
  \operatorname{Im}\Binout
  &=\order\!\left(\e^{-\epsilon T\Delta E_{\mathrm{rel}}}\right),
  \label{eq:imag-gap-rate}\\
  1-F_{LR}
  &=\order\!\left(\e^{-2\epsilon T\Delta E_{\mathrm{rel}}}\right),
  \label{eq:fidelity-gap-rate}
\end{align}
where the second line follows more explicitly from
\begin{equation}
  1-F_{LR}
  =q^2\bigl\|\,|r_\perp\rangle-|l_\perp\rangle\,\bigr\|^2+o(q^2).
  \label{eq:fidelity-expansion}
\end{equation}
The phase $\alpha$ cancels between the numerator and denominator of
(\ref{eq:finite-ratio-section}).  Hence
\begin{equation}
  \lim_{T\to\infty}\Binout
  =\frac{\langle R_*|O|R_*\rangle}
  {\langle R_*|R_*\rangle}
  =\Binin\in\mathbb R,
  \qquad O=O^\dagger.
  \label{eq:projected-reality-section}
\end{equation}

\textcolor{black}{At fixed spatial and oscillator cutoffs, the proposed asymptotic prescription is the following ordered limit}
\begin{equation}
  B_{\mathrm{phys}}
  =\lim_{\epsilon\to0^+}
   \lim_{\eta_0\to0^+}
   \lim_{T\to\infty}
   B^{\mathrm{II}}_{\mathrm{in\text{-}out}}(T,\epsilon,\eta_0).
  \label{eq:ordered-limits-section}
\end{equation}
The innermost limit sends $T\to\infty$ at fixed $\eta_0>0$ and fixed $\epsilon>0$.  The
contour remains on Sheet~II and the endpoint states are projected onto the asymptotic vacuum.
We next take $\eta_0\to0^+$, keeping the contour below the lower singularity. A fixed
$\delta>0$ satisfies this condition once $\eta_0<\delta$.  This removes the geometric
regulator from the already projected matrix element.  Finally we take $\epsilon\to0^+$.
Taking $\epsilon\to0$ before the large-$T$ limit removes the projection.  Taking
$\eta_0\to0$ at finite $T$ changes the continuation near the gluing surface while leaving the
finite-time excited components in place.

In practice the first limit requires
\begin{equation}
  \epsilon T\Delta E_{\mathrm{rel}}\gg1.
  \label{eq:projection-condition-section}
\end{equation}
The spatial continuum and infinite-volume limits may be interleaved with the $\eta_0$ limit
when the convergence is uniform and (\ref{eq:projection-condition-section}) continues to hold.
If the finite-volume gap decreases as the volume is enlarged, $T$ must be increased at the
same time.

\subsection{Finite-volume numerical results}
\label{subsec:finite-volume-results}

We first test (\ref{eq:imag-gap-rate}) and (\ref{eq:fidelity-gap-rate}) by exact evolution on an
interacting scalar lattice with five sites.  The parameters are
\begin{equation}
  N=5,\qquad d=N_{\max}+1=6,\qquad m^2=0.1,\qquad
  \lambda=0.1,\qquad \eta_0=0.01,\qquad \eta_*=-2.5,
  \label{eq:small-system-parameters}
\end{equation}
with a fixed Sheet-II depth $\delta=0.5$.  The time-dependent mass and quartic terms are even
under $\phi\to-\phi$.  The lowest one-particle excitation is odd and therefore does not
contribute.  The smallest excitation gap permitted by the symmetries is
\begin{equation}
  \Delta E_{\mathrm{rel}}=1.027608.
  \label{eq:scalar-relevant-gap}
\end{equation}

In the left panel of Figure \ref{fig:projection-tests} we vary the imaginary-time interval $\epsilon T$.
A fit to the imaginary part gives
\begin{equation}
  |\operatorname{Im}\Binout|\propto\e^{-1.059230\epsilon T},
  \qquad R^2=0.999932,
  \label{eq:imag-fit-section}
\end{equation}
within $3.1\%$ of (\ref{eq:scalar-relevant-gap}).  The left--right infidelity gives
\begin{equation}
  1-F_{LR}\propto\e^{-2.115619\epsilon T},
  \qquad R^2=0.999927,
  \label{eq:fidelity-fit-section}
\end{equation}
within $2.9\%$ of $2\Delta E_{\mathrm{rel}}$.  Both exponents agree with the
finite-volume spectrum.  The finite imaginary part is therefore controlled by the excited
components left at the temporal endpoints.

For comparison, we also evaluate the expectation value using the incoming branch alone,
\begin{equation}
  B_R(T)=\frac{\langle R_T|O|R_T\rangle}{\langle R_T|R_T\rangle}.
  \label{eq:projected-ket-expectation}
\end{equation}
At $T=40$ and $\epsilon=0.2$, exact evolution gives
$\operatorname{Im}\Binout=1.69\times10^{-9}$ and
$|\Binout-B_R|=2.13\times10^{-9}$.  The future branch has therefore reached the Hermitian
conjugate of the incoming state at this scale.  The same $N=5$ calculation also permits a
direct comparison with MPS evolution.  At $T=20$, extrapolating the MPS results at
$|\Delta z|=0.02,0.01,0.005$ to zero step gives
$0.5169425173+2.13123\times10^{-6}\ii$, while exact evolution gives
$0.5169425048+2.08391\times10^{-6}\ii$.  Their absolute difference is
$4.89\times10^{-8}$.

\begin{figure}[t]
  \centering
  \color{black}
  \begin{tikzpicture}
    \begin{groupplot}[
      group style={group size=2 by 1,horizontal sep=1.35cm},
      width=0.44\textwidth,
      height=0.34\textwidth,
      ymode=log,
      grid=both,
      major grid style={gray!25},
      minor grid style={gray!12},
      tick label style={font=\footnotesize},
      label style={font=\small},
      legend style={font=\footnotesize,draw=none,fill=white,fill opacity=0.9,
        text opacity=1},
      every axis plot/.append style={line width=1.0pt}
    ]
    \nextgroupplot[
      xlabel={$\epsilon T$},
      ylabel={$|\operatorname{Im}\Binout|$ and $1-F_{LR}$},
      xmin=0.35,xmax=4.15,
      ymin=2e-11,ymax=2e-4,
      legend style={at={(0.5,1.03)},anchor=south,legend columns=2,
        /tikz/every even column/.append style={column sep=0.8em}}
    ]
      \addplot[paperblue,mark=*,only marks,mark size=2.2pt]
        coordinates {(0.5,8.4725675e-5) (1.0,5.0627495e-5)
        (1.5,3.0064656e-5) (2.0,1.7769452e-5)
        (3.0,6.1309609e-6) (4.0,2.0839059e-6)};
      \addlegendentry{$|\operatorname{Im}\Binout|$}
      \addplot[paperblue,dashed,domain=0.45:4.05,samples=80,forget plot]
        {exp(-8.83165973-1.05923041*x)};
      \addplot[paperred,mark=square*,only marks,mark size=2.0pt]
        coordinates {(0.5,5.7332739e-8) (1.0,2.0505544e-8)
        (1.5,7.2499149e-9) (2.0,2.5376676e-9)
        (3.0,3.0278202e-10) (4.0,3.5022651e-11)};
      \addlegendentry{$1-F_{LR}$}
      \addplot[paperred,dashed,domain=0.45:4.05,samples=80,forget plot]
        {exp(-15.58614476-2.11561888*x)};

    \nextgroupplot[
      xmode=log,
      xlabel={$\eta_0$},
      ylabel={$|\operatorname{Im}\Binout|$},
      yticklabel pos=right,
      ylabel near ticks,
      xmin=7e-4,xmax=0.43,
      ymin=8e-14,ymax=2.0e-12,
      legend style={at={(0.5,1.03)},anchor=south,legend columns=1}
    ]
      \addplot[papergreen,mark=*,mark size=2.2pt]
        coordinates {(0.001,1.9255726e-13) (0.003,2.2715494e-13)
        (0.01,2.0854330e-13) (0.03,1.2071580e-12)
        (0.1,1.4092057e-13) (0.3,7.3118701e-13)};
      \addlegendentry{$T=80$, $\epsilon=0.2$}
      \addplot[gray,densely dashed,domain=7e-4:0.43,samples=2] {1e-12};
      \addlegendentry{numerical floor}
    \end{groupplot}
  \end{tikzpicture}
  \caption{Asymptotic projection in interacting scalar $\phi^4$ theory.  Left: in the $N=5$
  calculation the imaginary part decays at a rate set by the smallest excitation gap permitted by the symmetries, and the
  infidelity decays with twice that gap.  The dashed lines are the fits in
  (\ref{eq:imag-fit-section}) and (\ref{eq:fidelity-fit-section}).
  Right: dependence of the projected result on $\eta_0$ in the $N=3$, $d=8$ calculation.
  The residual values are set by the finite endpoint and numerical precision.}
  \label{fig:projection-tests}
\end{figure}

The role of the endpoint tilt can also be seen in a separate $N=3$, $d=12$ calculation with
the same values of $m^2$, $\lambda$, $\eta_0$, and $\eta_*$.  We find
\begin{equation}
\begin{array}{|c|r|r|r|r|}
  \hline
  &T=10&T=20&T=40&T=80\\ \hline
  \epsilon=0
  &+1.115\!\times\!10^{-4}&-5.950\!\times\!10^{-5}
  &-1.661\!\times\!10^{-5}&+3.691\!\times\!10^{-7}\\
  \epsilon=0.1
  &+3.714\!\times\!10^{-5}&-2.742\!\times\!10^{-6}
  &-3.537\!\times\!10^{-8}&+7.479\!\times\!10^{-12}\\
  \epsilon=0.2
  &+9.361\!\times\!10^{-6}&-1.341\!\times\!10^{-7}
  &-6.101\!\times\!10^{-11}&+1.268\!\times\!10^{-13}\\
  \hline
\end{array}.
\label{eq:scalar-endpoint-scan-section}
\end{equation}
The first row sets $\epsilon$ exactly to zero, so its temporal endpoints lie on the real axis
and no endpoint projection is present.  We include this row only to display the finite-time
behavior.  The ordered prescription keeps $\epsilon>0$ while $T$ is increased.  The
\begingroup\color{black}
imaginary part then decreases rapidly. At $T=80$, $\epsilon=0.2$, $N=3$ and $d=8$, the regulator scan gives $|\operatorname{Im}\Binout|\lesssim1.3\times10^{-12}$. To examine convergence of the full observable, we also compute its real part and the complex difference from the projected expectation value. The same calculations give
\begin{center}
\begin{tabular}{|c|c|c|}
\hline
$\eta_0$ & $\operatorname{Re}\Binout$ & $|\Binout-B_R|$\\
\hline
$0.03$  & $0.457343573430$ & $2.10\times10^{-12}$\\
$0.01$  & $0.457343171252$ & $6.07\times10^{-13}$\\
$0.003$ & $0.457343125500$ & $5.52\times10^{-13}$\\
$0.001$ & $0.457343121477$ & $2.96\times10^{-13}$\\
\hline
\end{tabular}
\end{center}
The real part and the full complex difference thus support regulator convergence at the fixed lattice cutoffs. The last two real parts differ by $4.02\times10^{-9}$.  This shows the residual $\eta_0$-dependence is much larger than the in--in/in--out difference. 
\endgroup

The parameter $\delta$ specifies the deformation below the lower singularity.  At $T=40$,
$\epsilon=0.2$, $d=12$, and $\eta_0=0.01$, changing
$\delta=0.25,0.5,1.0$ changes the real part by less than $1.1\times10^{-10}$ and leaves
$|\operatorname{Im}\Binout|<7.0\times10^{-11}$.  These three contours give the same result
within the numerical accuracy.

\FloatBarrier

\subsection{Numerical convergence and lattice size}
\label{subsec:numerical-convergence}

\begingroup\color{black}
We use $P_{\rm top}$ to denote the largest occupation probability of the highest retained oscillator level $d-1$, taking the maximum over sites and the two normalized states at the insertion. We evaluate this quantity only at the insertion. $P_{\rm top}$ monitors the local occupation cutoff. For the $N=3$ scan at $T=40$ and $\epsilon=0.2$ we find
\endgroup
\begin{center}
\begin{tabular}{|c|c|c|}
  \hline
  $d$ & $P_{\rm top}$ & $\operatorname{Im}\Binout$ \\
  \hline
  4  & $4.60\times10^{-3}$  & $-2.94\times10^{-10}$ \\
  6  & $6.37\times10^{-5}$  & $-5.47\times10^{-11}$ \\
  8  & $1.25\times10^{-6}$  & $-6.95\times10^{-11}$ \\
  10 & $2.52\times10^{-8}$  & $-6.23\times10^{-11}$ \\
  12 & $5.13\times10^{-10}$ & $-6.10\times10^{-11}$ \\
  16 & $2.17\times10^{-13}$ & $-6.97\times10^{-11}$ \\
  \hline
\end{tabular}
\end{center}
The $d=4$ row exceeds our chosen threshold for the occupation of the highest retained level.  From $d=6$ onward,
$P_{\rm top}$ decreases rapidly, while the imaginary part varies only at the
$10^{-11}$ level.

Exact evolution and MPS evolution also show the expected second-order time-step convergence.
We denote the step along each straight contour segment by $\Delta z$.  For an $N=4$, $d=6$
calculation, halving $|\Delta z|$ from $0.02$ to $0.01$ and
then to $0.005$ reduces the absolute exact--MPS error from $4.06\times10^{-5}$ to
$1.02\times10^{-5}$ and $2.54\times10^{-6}$.  The ratios, $3.994$ and $4.006$, agree with
second-order convergence, and no Schmidt weight is discarded at this size.

\begingroup\color{black}
The calculation with spatially separated fields on $N=9$ sites uses $a=0.4$, $L=(N+1)a=4$, $d=6$ ($N_{\max}=5$), $\lambda=0.1$, $m_R^2=0.1$, $\eta_0=0.01$, $\eta_*=-2.5$, $T=20$, $\epsilon=0.2$ and $\delta=0.5$. Normal ordering with respect to the massless vacuum at finite volume gives the bare mass squared $m^2=0.06963963147$. The correlator is the mean of the connected correlators for the site pairs $(1,6)$ and $(2,7)$, with site numbering starting at zero. Both pairs have physical separation $2$. This provides a separate check of the calculation. In addition to the contour, the observable, spacing, local cutoff and mass convention differ from those in Table~\ref{tab:smallm2highchi}.
\endgroup

Separated equal-time insertions give the same behavior.  They were studied by exact
evolution at $N=3,5$ and by MPS evolution at $N=7,9$.  To remove the leading time-step
error of the second-order evolution, we use
\begin{equation}
  C(0)\simeq \frac{4C(0.01)-C(0.02)}{3},
  \label{eq:zero-step-extrapolation}
\end{equation}
which follows from $C(h)=C(0)+c h^2+\order(h^4)$ with $h=|\Delta z|$.  For the $N=9$
calculation this gives
$C_{\phi\phi}=0.0527477783+4.38857\times10^{-6}\ii$. Its relative imaginary part is
$8.32\times10^{-5}$.  The expectation value formed from the projected ket is
$0.0527513882$ and is real to numerical precision.  The two branches reach maximum bond
dimensions $60$ and $61$, below the imposed cap $64$, with accumulated discarded weights
$7.27\times10^{-9}$ and $9.07\times10^{-9}$.  The highest-level occupation is
$6.72\times10^{-5}$.  \begingroup\color{black}
The mean correlators before extrapolation are $0.0525895224-7.20994\times10^{-4}i$ and $0.0527082143-1.76957\times10^{-4}i$ at steps $0.02$ and $0.01$, respectively. Their imaginary parts decrease by a factor $4.07$, consistent with leading second-order error. The extrapolated full difference from the projected expectation is approximately $5.68\times10^{-6}$, or $1.08\times10^{-4}$ relatively. The two time steps support convergence with the expected leading power. 
\endgroup

We have also refined the lattice at fixed $L=4$ while choosing $T$ so that
$\epsilon T\Delta E_{\mathrm{rel}}\simeq10$.  For $N=1,3,5$ the relative imaginary parts of
the local projected correlator are $2.29\times10^{-9}$, $1.08\times10^{-9}$, and
$3.37\times10^{-10}$, respectively.  The relevant even-sector gap approaches
$2\pi/L=1.570796$ as $a\to0$, which motivates maintaining the projection condition along further fixed-volume refinement. The real coincident correlator also requires the usual mass and composite-operator renormalization in the continuum limit.

At fixed temporal endpoints, the states $\langle L_T|$ and $|R_T\rangle$ contain different
excited-state components.  A transition matrix element between these states can be complex
for a Hermitian insertion.  The MPS calculation on the original contour had converged to
this finite-time quantity.

The asymptotic contour removes the excited components.  The decay rate of
$\operatorname{Im}\Binout$ agrees with the smallest excitation gap permitted by the symmetries, and the decay rate of
$1-F_{LR}$ agrees with twice that gap.  The in--out ratio then approaches the expectation
value formed from $|R_T\rangle$.  The same behavior is found as $\eta_0$ is reduced, as the
occupation cutoff and time step are varied, for separated fields, and along the fixed-volume
lattice-refinement sequence.

\begingroup\color{black}
{\bf \underline{Summary:}} 
After these lengthy discussions, we summarize the main findings. The metric regulator is necessary in performing numerics for in-out calculations and with the modified complex contour accounts correctly for the extra pole it introduces. 
For the equal-time Hermitian observables studied here, the finite-lattice results support recovery of the full complex relation
\begin{equation}
\Binin=\Binout,\qquad \operatorname{Im}\Binout=0,
\label{eq:final-equality-section}
\end{equation}
with the asymptotic prescription, conditional on alignment of the limiting states and convergence as the regulator is removed. The evidence concerns the original relation between the full complex correlators. To extend it to the continuum, we must follow a renormalized line of constant physics, control the oscillator cutoff, and establish uniform convergence in the endpoint and regulator limits. Throughout that extrapolation we must maintain $\epsilon T\Delta E_{\rm rel}\gg1$. We take $\epsilon\to0^+$ after endpoint projection.
\endgroup

\section{Entanglement during evolution}
\label{sec:entanglement}

\begingroup\color{black}
We discuss here how entanglement in the MPS behaves in the regulated real-contour implementation and the complex contour implementation. The key finding is that the naive real time contour sees a larger entanglement during time evolution past $\eta\approx 0$ in the in-out calculations while the complex contour sees entanglement comparable to the in-in calculations. 
\endgroup

During our simulations, we also monitor the growth of the entanglement entropy, since it is directly related to the bond dimension required for the MPS.  For a normalized state, the entanglement entropy across the bond between sites $i$ and $i+1$ is obtained from the Schmidt spectrum,
\begin{equation}
  S_i(\eta)=-\sum_k p_{i,k}\log p_{i,k}\;,
  \qquad
  p_{i,k}=s_{i,k}^2\;,
  \qquad
  \sum_k s_{i,k}^2=1\;,
  \label{eq:entanglement-entropy}
\end{equation}
where $\{s_{i,k}\}_{k=1}^{\chi}$ are the Schmidt coefficients for this bipartition.  For the states considered here, the bipartite entanglement entropy attains its maximum near the center of the chain.  We therefore use the half-chain entropy
\begin{equation}
  S_{\rm hc}(\eta)\equiv S_{i_{\rm hc}}(\eta)\;,
  \qquad
  i_{\rm hc}=\left\lfloor\frac{N+1}{2}\right\rfloor\;,
  \label{eq:half-chain-cut}
\end{equation}
corresponding to the cut after site $(N+1)/2$ when $N$ is odd.  Since $S_{\rm hc}$ contains short-distance contributions that depend on the cutoff, we examine its dependence on $\lambda$ and on the lattice size.

\begingroup\color{black}
Two related quantities will be useful below.  The entropy at the real insertion time measures the cost of representing the states that enter the correlator.  The largest entropy encountered during an evolution measures the cost of following the chosen evolution path.  For evolution along a complex contour, this maximum can depend on the shape of the contour.  At every point on such a contour, we normalize the state before evaluating \eqref{eq:entanglement-entropy}.
\endgroup

\begin{figure}[H]
  \centering
  \begin{subfigure}[T]{0.48\textwidth}
    \includegraphics[width=\textwidth]{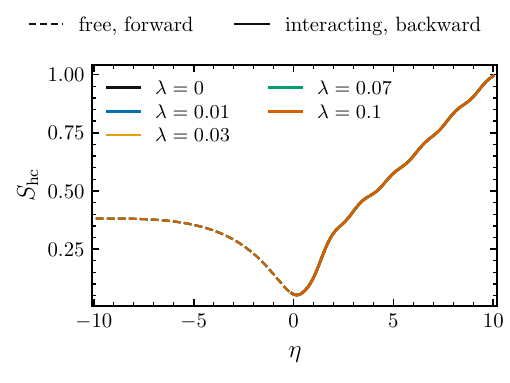}
    \caption{Half-chain entanglement entropy $S_{\rm hc}(\eta)$ during preparation of $|X_t\rangle$ at different couplings.}
    \label{fig:ee_inout}
  \end{subfigure}
  \hfill
  \begin{subfigure}[T]{0.48\textwidth}
    \includegraphics[width=\textwidth]{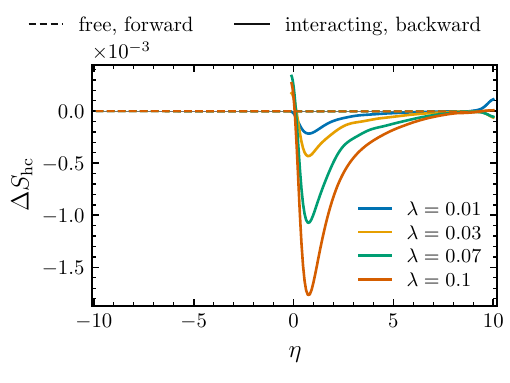}
    \caption{Difference $\Delta S_{\rm hc}=S_{\rm hc}(\lambda)-S_{\rm hc}(0)$.}
    \label{fig:ee_diff}
  \end{subfigure}
  \caption{Entanglement during bra preparation in the finite-time construction, computed via TDVP with $N=41$, $N_{\max}=15$, $\chi=30$, $m^2=1$, $\eta_i=-10$, $\eta_*=-0.1$, $\eta_f=10$, and $\eta_0=0.6$.  Dashed curves show free forward evolution; solid curves show interacting backward evolution.  The difference in the right panel is taken relative to the $\lambda=0$ calculation at matching times and stages of preparation.}
  \label{fig:ee}
\end{figure}

\begin{figure}[!htb]
  \centering
  \begin{subfigure}[t]{0.48\textwidth}
    \centering
    \includegraphics[width=\textwidth]{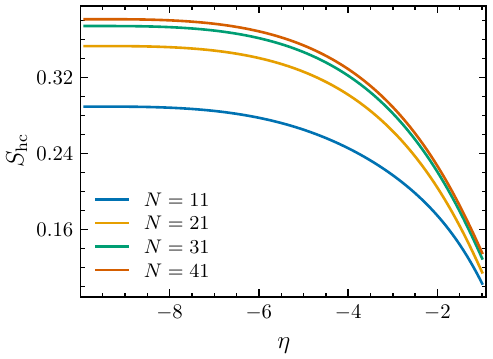}
    \caption{Entanglement entropy for the ket evolution in the in-in formalism.}
    \label{fig:ee-halfchain-inin}
  \end{subfigure}
  \hfill
  \begin{subfigure}[t]{0.48\textwidth}
    \centering
    \includegraphics[width=\textwidth]{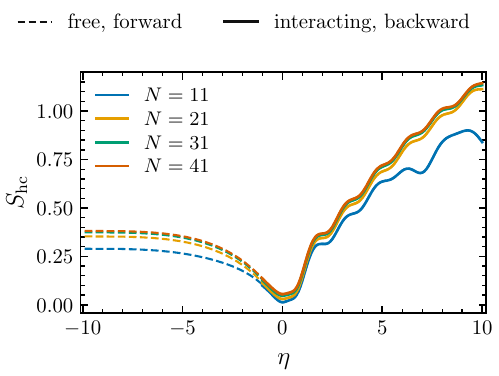}
    \caption{Entanglement entropy for the bra side of the finite-time in-out construction.  Dashed curves show free evolution from $\eta_i$ to $\eta_f$, and solid curves show interacting backward evolution from $\eta_f$ to $\eta_*$.}
    \label{fig:ee-halfchain-inout}
  \end{subfigure}
  \caption{Half-chain entanglement entropy $S_{\rm hc}(\eta)$, computed using Approach A with $T_A=0$, as a function of conformal time for $\lambda=0.1$, $\chi=30$, $m^2=1$, $\eta_*=-1$, and $\eta_0=0.3$, with $N\in\{11,21,31,41\}$.  For each size, the entropy is evaluated across the cut after site $(N+1)/2$.}
  \label{fig:ee-halfchain}
\end{figure}

\begin{figure}[!htb]
  \centering
  \begin{subfigure}[t]{0.48\textwidth}
    \centering
    \includegraphics[width=\textwidth]{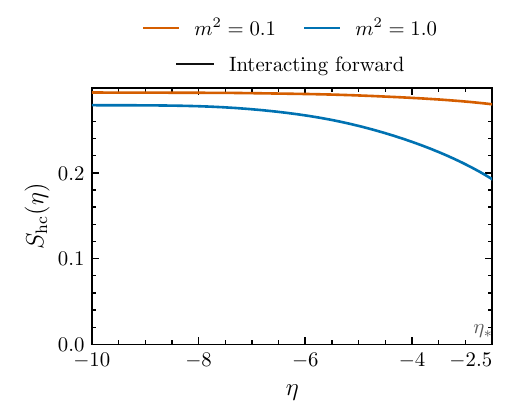}
    \caption{Half-chain entanglement entropy of the forward ket branch $|\Psi_s\rangle$.}
    \label{fig:ee-halfchain-psi-star}
  \end{subfigure}
  \hfill
  \begin{subfigure}[t]{0.48\textwidth}
    \centering
    \includegraphics[width=\textwidth]{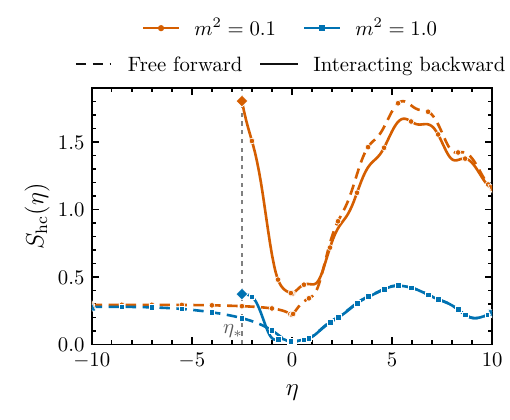}
    \caption{Half-chain entanglement entropy during preparation of $|X_t\rangle$.}
    \label{fig:ee-halfchain-chi-star}
  \end{subfigure}
  \caption{Half-chain entanglement entropy $S_{\rm hc}(\eta)$ in the finite-time construction, using Approach B for $N=11$, $N_{\max}=10$, $\chi=30$, $\lambda=0.1$, $\eta_i=-10$, $\eta_*=-2.5$, $\eta_f=10$, and $\eta_0=0.01$, for $m^2=0.1$ and $m^2=1.0$.  In the right panel, dashed lines denote free forward evolution from $\eta_i$ to $\eta_f$, while solid lines denote interacting backward evolution to $\eta_*$.}
  \label{fig:ee-halfchain-adapted-approachB}
\end{figure}

We first describe the finite-time construction used in the original numerical comparison.  We track the ket and bra histories
\begin{equation}
\begin{split}
  |\Psi_s\rangle={}&U(\eta_i+2Ts,\eta_i)\,|0_{\eta_i}\rangle\;,
  \qquad s\in[0,s_*]\;,\\
  |X_t\rangle={}&
  \begin{cases}
    U_0(\eta_f+2Tt,\eta_i)|0_{\eta_i}\rangle\;, & -1\leq t\leq0\;,\\
    U(\eta_f-2Tt,\eta_f)U_0(\eta_f,\eta_i)|0_{\eta_i}\rangle\;,
      & 0<t\leq t_*\;,
  \end{cases}
\end{split}
\label{eq:finite-time-entanglement-histories}
\end{equation}
where $s$ and $t$ parameterize the evolution paths, with
\begin{equation}
  s_*=\frac{\eta_*-\eta_i}{2T}\;,
  \qquad
  t_*=\frac{\eta_f-\eta_*}{2T}=1-s_*\;,
  \label{eq:finite-time-entanglement-parameters}
\end{equation}
and $T=-\eta_i=\eta_f$.  Since $\eta_*\in(\eta_i,0)=(-T,0)$, one has $s_*\in(0,\tfrac12)$ and $t_*\in(\tfrac12,1)$.  The endpoints are $|\Psi_{s_*}\rangle=|\Psi(\eta_*)\rangle$, defined in \eqref{defPhi}, and $|X_{t_*}\rangle=|X(\eta_*)\rangle$, whose adjoint is the bra defined in \eqref{inoutbra}.  Figures~\ref{fig:ee}--\ref{fig:ee-halfchain-adapted-approachB} and Table~\ref{tab:schmidt_tail_adapted_code} refer to this finite-time construction.

Figure~\ref{fig:ee} shows the half-chain entanglement entropy during preparation of $|X_t\rangle$, for $N=41$ and several couplings.  The dashed curves follow the free evolution from $\eta_i$ to $\eta_f$; the solid curves follow the interacting evolution backward from $\eta_f$ to $\eta_*=-0.1$.  The entropy is evaluated across the cut after site 21.  To examine its dependence on the interaction, we plot
\begin{equation}
  \Delta S_{\rm hc}(\eta;\lambda)
  =S_{\rm hc}(\eta;\lambda)-S_{\rm hc}(\eta;0)\;,
  \label{eq:entropy-difference}
\end{equation}
subtracting the free result at the same conformal time and at the same stage of preparation.  For these couplings, the interacting entropy stays close to the free result.  The small differences between the dashed free-preparation curves arise from the separate numerical preparations.

Figure~\ref{fig:ee-halfchain} compares the ket and bra entropies as the lattice size is increased, at $\lambda=0.1$.  Over the displayed range, the entanglement approaches a size-independent profile.  Thus the observed cost is mainly associated with the dynamics near the bipartition once the volume is sufficiently large.

The numerator can also be evaluated at $\eta_f$.  We take the overlap of $U(\eta_f,\eta_*)O(\eta_*)|\Psi(\eta_*)\rangle$ with $\langle0_{\eta_i}|U_0^\dagger(\eta_f,\eta_i)$, so that the ket evolves through the operator insertion.  We examined this form for the parameters of Figure~\ref{fig:ee-halfchain} and found that, apart from a small jump at the insertion, the entropy curves are essentially unchanged.

We now turn to the light-field case with the small regulator $\eta_0=0.01$.  We use Approach B because TDVP gives large integration errors for small $\eta_0$.  The results are shown in Figure~\ref{fig:ee-halfchain-adapted-approachB}.  The forward ket entropy remains small and is comparable for $m^2=0.1$ and $m^2=1$ up to the insertion time $\eta_*$, as shown in Figure~\ref{fig:ee-halfchain-psi-star}.  In the finite-time bra preparation, the entropy is much larger for the lighter mass, as shown in Figure~\ref{fig:ee-halfchain-chi-star}.

We supplement the half-chain entropy by monitoring the Schmidt spectrum across the same bipartition.  At each sampled conformal time $\eta$, we compute the ordered Schmidt coefficients $\{s_r(\eta)\}_{r=1}^{\chi}$, with $s_1\geq s_2\geq\cdots$, and the cumulative tail weights
\begin{equation}
  \tau_r(\eta)\equiv\sum_{k\geq r}s_k(\eta)^2\;.
  \label{eq:schmidt-tail}
\end{equation}
The entropy can remain nearly unchanged while Schmidt coefficients near the retained rank $\chi$ are still appreciable.  The behavior of $s_{20}$, $s_{30}$, and $\tau_{20}$ therefore indicates whether the bond dimension is limiting the evolution.  For $|X_t\rangle$, Table~\ref{tab:schmidt_tail_adapted_code} shows that coefficients near the cutoff remain large over a longer interval for $m^2=0.1$ than for $m^2=1$.  The light-mass finite-time calculation consequently requires a larger bond dimension, and its detailed entropy profile is not fully converged at $\chi=30$.

\begin{table}[H]
\centering
\begin{tabular}{|c|c|c|c|c|c|}
\hline
$m^2$ & branch & $\max_\eta S_{\rm hc}$ & $\max_\eta s_{30}$ & $\max_\eta\tau_{20}$ & $s_{30}>10^{-2}$ \\
\hline
$0.1$ & $|\Psi_s\rangle$ & $0.294$ & $3.61\times10^{-6}$ & $1.83\times10^{-9}$ & none \\
$1.0$ & $|\Psi_s\rangle$ & $0.279$ & $2.25\times10^{-6}$ & $1.15\times10^{-9}$ & none \\
$0.1$ & $|X_t\rangle$ & $1.762$ & $2.47\times10^{-2}$ & $1.33\times10^{-2}$ & $[-2.5,10.0]$ \\
$1.0$ & $|X_t\rangle$ & $0.434$ & $1.02\times10^{-3}$ & $3.42\times10^{-5}$ & none \\
\hline
\end{tabular}
\caption{Schmidt coefficients near the bond-dimension cutoff and their cumulative weights at the half-chain cut in the finite-time construction.  The parameters are $N=9$, $\chi=30$, $N_{\max}=10$, $\lambda=0.1$, $\eta_i=-10$, $\eta_*=-2.5$, $\eta_f=10$, and $\eta_0=0.01$.  The last column gives the sampled conformal-time interval over which $s_{30}>10^{-2}$.}
\label{tab:schmidt_tail_adapted_code}
\end{table}

\begingroup\color{black}
\captionsetup{font+={color=black},labelfont+={color=black}}
The preceding results describe the finite-time definition in \eqref{eq:BinoutA_def_setup}.  The asymptotically projected contour of Section~\ref{sec:asymptotic-projection} gives a different entropy history.  We evaluated the entropies of $|R_T\rangle$ and $\langle L_T|$ from \eqref{eq:left-right-definition} for
\begin{equation}
\begin{gathered}
  N=7\;,
  \qquad d=N_{\max}+1=6\;,
  \qquad a=0.5\;,
  \qquad m^2=0.1\;,
  \qquad \lambda=0.1\;,\\
  \eta_0=0.01\;,
  \qquad T=20\;,
  \qquad \eta_*=-2.5\;,
  \qquad \epsilon=0.2\;.
\end{gathered}
  \label{eq:projected-entanglement-parameters}
\end{equation}
The contour depth is $\delta=0.5$, the time step is $0.02$, and the maximum bond dimension is $\chi=64$.  Table~\ref{tab:projected-entanglement} compares these results with the real-axis in-in ket and with the finite-time bra history at the same lattice and field parameters.

\begin{table}[H]
\centering
\color{black}
\begin{tabular}{|l|c|c|}
\hline
Evolution history & $\max S_{\rm hc}$ & $S_{\rm hc}(\eta_*)$ \\
\hline
Real-axis in-in ket & $0.2418$ & $0.2371$ \\
Finite-time bra history & $2.7178$ & $2.5284$ \\
Projected ket $|R_T\rangle$ & $0.2420$ & $0.23843$ \\
Projected bra $\langle L_T|$ & $0.27553$ & $0.23855$ \\
\hline
\end{tabular}
\caption{Half-chain entropy in the light-field calculation of \eqref{eq:projected-entanglement-parameters}.  The finite-time bra evolution is limited by the retained bond dimension, and its entropy values are included only to show the size of the entropy growth.  The two projected histories are converged at the quoted precision.}
\label{tab:projected-entanglement}
\end{table}

The projected ket and bra have nearly equal entropies at the insertion, and both values remain close to the in-in result.  The accumulated discarded weights are $3.74\times10^{-9}$ and $4.50\times10^{-9}$ for the projected ket and bra, respectively.  Their left--right fidelity, defined in \eqref{eq:fidelity-definition}, is $F_{LR}=0.99999894$.  The same behavior is found on the larger $N=9$ lattice: at time step $0.01$, the insertion entropies are $0.278124$ for the projected ket and $0.277886$ for the projected bra.

The entropy reached at intermediate points of the complex contour is path dependent.  We verified this directly by exact evolution for $N=3$, $d=10$, $m^2=1$, and $\eta_0=0.3$.  Increasing the depth $\delta$ of the horizontal Sheet-II segment from $0.7$ to $2.0$ changes the largest entropy from $1.08866$ to $0.15561$, while the entropy at the real insertion remains $0.07058$.  Thus the transient maximum measures the cost of evaluating the in-out matrix element along the chosen contour.

The large light-field entropy in Figures~\ref{fig:ee-halfchain-adapted-approachB} and Table~\ref{tab:schmidt_tail_adapted_code} is therefore a feature of the finite-time construction, in which the reference state is evolved across the regulated region before the interacting backward evolution.  In the projected calculations examined here, the bra and ket entropies at the insertion are close to one another and to the in-in entropy.  The in-out calculation still requires two independently evolved MPS states, and its numerical cost also depends on the complex contour.
\endgroup

\section{Extensions toward \texorpdfstring{$3+1$}{3+1} dimensions and quantum hardware}

We now consider how these techniques can be applied to a class of problems in $3+1$ dimensions and outline a quantum-circuit implementation of the correlators.

\subsection{dS in 3+1 dimensions}\label{3plus1dS}

While the present work focuses on $\phi^4$ theory in $1+1$ dimensional de Sitter spacetime, it is useful to note that tensor-network methods can also be applied to certain sectors of higher-dimensional theories. In particular, a scalar field in $3+1$ dimensional de Sitter space can be reduced to an effective $1+1$ dimensional problem by exploiting spherical symmetry. The reduction of a higher-dimensional field theory to fields along the radial direction has a well-established precedent. In flat-space quantum field theory, Srednicki's classic computation of entanglement entropy exploited a decomposition into angular-momentum sectors and a radial lattice, demonstrating that spherical reduction can turn a genuinely higher-dimensional problem into a numerically tractable collection of one-dimensional systems \cite{Srednicki:1993im}. More generally, spherical reduction to an effective two-dimensional description is a standard tool in curved-spacetime physics, where it underlies the treatment of spherically symmetric systems in dilaton gravity and related models \cite{Grumiller:2002nm}. \textcolor{black}{These examples motivate organizing the calculation by angular momentum, with a radial chain for each retained field. Truncating the coupled set of fields simplifies the calculation. Whether the retained fields capture the most relevant part of the full interacting dynamics remains to be established.}

We consider the metric on the Poincar\'e patch
\begin{equation}
ds^2=\Omega^2(\eta)\left(-d\eta^2+dr^2+r^2 d\Omega_2^2\right)\;,
\qquad
\Omega(\eta)=-\frac{1}{H\eta}\;.
\end{equation}
For a scalar field with interaction
\begin{equation}
S=\int d^4x \sqrt{-g}\left[
-\frac12(\nabla\phi)^2-\frac12 m^2\phi^2-\frac{\lambda}{4!}\phi^4
\right]\;,
\end{equation}
we expand the field in a real orthonormal basis of spherical harmonics,
\begin{equation}
\phi(\eta,r,\theta,\varphi)=\sum_{\ell m}\frac{\chi_{\ell m}(\eta,r)}{r}Y_{\ell m}(\theta,\varphi)\;.
\end{equation}

After integrating over the sphere, the quadratic action becomes a sum over $(\ell,m)$ sectors,
\begin{equation}
S_2=
\sum_{\ell m}\int d\eta\,dr
\left[
\frac12\Omega^2
\left(
(\partial_\eta \chi_{\ell m})^2
-(\partial_r \chi_{\ell m})^2
\right)
-\frac12
\left(
\Omega^2\frac{\ell(\ell+1)}{r^2}
+\Omega^4 m^2
\right)
\chi_{\ell m}^2
\right].
\end{equation}
Each spherical harmonic mode therefore behaves as an effective field propagating in a $1+1$ dimensional spacetime $(\eta,r)$ with an angular-momentum barrier $\ell(\ell+1)/r^2$.

The quartic interaction couples different angular momentum sectors,
\begin{equation}
S_{\rm int}=-\frac{\lambda}{4!}\int d\eta\,dr\,\frac{\Omega^4}{r^2}\sum_{\ell_i m_i}C_{\ell_1 m_1 \ell_2 m_2 \ell_3 m_3 \ell_4 m_4}\,\chi_{\ell_1 m_1}\chi_{\ell_2 m_2}\chi_{\ell_3 m_3}\chi_{\ell_4 m_4}\;,
\end{equation}
where the coefficients
\begin{equation}
C_{\ell_1 m_1 \ell_2 m_2 \ell_3 m_3 \ell_4 m_4}=\int d\Omega\,Y_{\ell_1 m_1}Y_{\ell_2 m_2}Y_{\ell_3 m_3}Y_{\ell_4 m_4}\;,
\end{equation}
are Gaunt coefficients that can be written in terms of Wigner $3j$ symbols.

In practice one may truncate the spherical harmonic expansion to $\ell\le \ell_{\rm max}$, yielding a finite set of interacting scalar fields propagating along the radial direction. The resulting theory is a $1+1$ dimensional multi-field quantum field theory with position-dependent couplings. Since tensor-network methods naturally operate on one-dimensional spatial lattices, this formulation opens the possibility of studying sectors of $3+1$ dimensional de Sitter quantum field theory using MPS techniques.

A particularly simple truncation keeps only the s-wave mode $(\ell=0)$, for which the theory reduces to a single scalar field with an effective coupling proportional to $1/r^2$. \textcolor{black}{In this tractable radial model of the interacting theory, the omitted modes are discarded without being integrated out. A controlled approximation to the full four-dimensional theory needs convergence with $\ell_{\max}$ and consistent renormalization, which is left for a future detailed study.}

\subsection{Low angular-momentum truncations}

A practical implementation of the above reduction is obtained by truncating the spherical harmonic expansion to a finite set of angular momentum modes. Since the number of fields grows as $(\ell_{\rm max}+1)^2$, the simplest nontrivial truncations correspond to $\ell_{\rm max}=0$ and $\ell_{\rm max}=1$.

\paragraph{S-wave truncation ($\ell_{\rm max}=0$).}

Keeping only the s-wave mode,
\begin{equation}
\phi(\eta,r,\theta,\varphi)=\frac{\chi(\eta,r)}{r\sqrt{4\pi}}\;,
\end{equation}
the angular integration becomes trivial. The action reduces to a single scalar field propagating in the $(\eta,r)$ plane,
\begin{equation}
\label{eq:swave_action_chi}
S=\int d\eta\,dr\left[\frac12\Omega^2\left((\partial_\eta \chi)^2-(\partial_r \chi)^2\right)-\frac12\Omega^4 m^2\chi^2-\frac{\lambda}{4!}\frac{\Omega^4}{4\pi r^2}\chi^4\right]\;.
\end{equation}
The kinetic term carries a factor of $\Omega^2(\eta)$, so the conjugate
momentum of $\chi$ is $\pi_\chi=\Omega^2\partial_\eta\chi$ and the Hamiltonian has a time-dependent kinetic term $\pi_\chi^2/(2\Omega^2)$. Its coefficient differs from the constant kinetic coefficient in the $1+1$ dimensional Hamiltonian used earlier.

\paragraph{Canonical rescaling.}
To bring the kinetic term into canonical form we define
\begin{equation}
\label{eq:rescaling}
\tilde\varphi(\eta,r)\equiv\Omega(\eta)\,\chi(\eta,r)\;.
\end{equation}
Substituting $\chi=\tilde\varphi/\Omega$ into
\eqref{eq:swave_action_chi} and integrating the cross-term
$\partial_\eta\tilde\varphi\cdot\tilde\varphi\,(\dot\Omega/\Omega)$ by
parts (the boundary term at $\eta_i$ and $\eta_f$ is set to zero by the
adiabatic switching prescription), we obtain the canonically normalised
action
\begin{equation}
\label{eq:swave_action_tilde}
S_{\rm s\text{-}wave}=
\int d\eta\,dr
\left[
\frac{1}{2}(\partial_\eta\tilde\varphi)^2
-\frac{1}{2}(\partial_r\tilde\varphi)^2
-\frac{m_{\rm eff}^2(\eta)}{2}\,\tilde\varphi^2
-\frac{\lambda}{4!\cdot 4\pi r^2}\,\tilde\varphi^4
\right]\;,
\end{equation}
where the effective time-dependent mass is
\begin{equation}
\label{eq:meff}
m_{\rm eff}^2(\eta)\equiv\Omega^2 m^2 - \frac{\ddot\Omega}{\Omega}\;.
\end{equation}
For the regulated conformal factor
$\Omega(\eta)=\bigl[H\sqrt{\eta^2+\eta_0^2}\bigr]^{-1}$ one finds
\begin{equation}
\label{eq:meff_reg}
m_{\rm eff}^2(\eta)=\frac{m^2}{H^2(\eta^2+\eta_0^2)}-\frac{2\eta^2-\eta_0^2}{(\eta^2+\eta_0^2)^2}\;.
\end{equation}
In the limit $\eta_0\to0$ this reduces to
$m_{\rm eff}^2=(m^2/H^2-2)/\eta^2$, the standard Mukhanov--Sasaki effective mass.  The quartic coupling in \eqref{eq:swave_action_tilde} is \emph{time-independent} because the factor of $\Omega^4$ in eq.~\eqref{eq:swave_action_chi} is exactly canceled by the four powers of $1/\Omega$ from $\chi=\tilde\varphi/\Omega$.  This is in contrast to the $1+1$ dimensional action, where the quartic carries $\Omega^2(\eta)$ and must be updated at every time step.

\paragraph{Boundary conditions.}
Regularity of $\phi$ at the origin requires $\chi(\eta,0)=0$, which
implies $\tilde\varphi(\eta,0)=0$.  We impose Dirichlet conditions also at $r=r_{\rm max}$, which plays the role of an infrared cutoff. These boundary conditions allow us to represent the radial state by an MPS with open ends, as elsewhere in this work.

\paragraph{Lattice Hamiltonian.}
We discretize $r\in[0,r_{\rm max}]$ with $N$ sites at $r_j=ja$
($j=1,\ldots,N$), $r_{\rm max}=(N+1)a$, and
$\tilde\varphi_0=\tilde\varphi_{N+1}=0$.
As in~\eqref{eq:lattice_operators}, we use the rescaled lattice momentum
$\pi_j=a\,\partial_\eta\tilde\varphi_j$, so that
$[\tilde\varphi_j,\pi_k]=i\delta_{jk}$. The lattice Hamiltonian is
\begin{equation}
\label{eq:Hlat_swave}
\boxed{
H_{\rm s\text{-}wave}(\eta)=\sum_{j=1}^{N}\!\left[\frac{\pi_j^2}{2a}+\frac{m_{\rm eff}^2(\eta)\,a}{2}\,\tilde\varphi_j^2+\frac{\lambda\,a}{4!\cdot 4\pi\, j^2 a^2}\,\tilde\varphi_j^4\right]+\frac{1}{a}\sum_{j=1}^{N}\tilde\varphi_j^2-\frac{1}{a}\sum_{j=1}^{N-1}\tilde\varphi_j\tilde\varphi_{j+1}}
\end{equation}
with $m_{\rm eff}^2(\eta)$ given by eq.~\eqref{eq:meff_reg}.
In dimensionless units $a=1$ this reads
\begin{equation}
\label{eq:Hlat_swave_dimless}
H_{\rm s\text{-}wave}(\eta)=\sum_{j=1}^{N}\!\left[\frac{\pi_j^2}{2}+\frac{m_{\rm eff}^2(\eta)+2}{2}\,\tilde\varphi_j^2+\frac{\lambda}{4!\cdot 4\pi j^2}\,\tilde\varphi_j^4\right]-\sum_{j=1}^{N-1}\tilde\varphi_j\tilde\varphi_{j+1}\;.
\end{equation}
Comparing with the $1+1$ dimensional Hamiltonian (\ref{eq:lattice_H_final}), the structural differences are:
\begin{enumerate}
\item The kinetic term $\pi_j^2/2$ is identical and canonical.
\item The mass coefficient is $m_{\rm eff}^2(\eta)$ rather than $\Omega^2(\eta)m^2$; the additional term $-\ddot\Omega/\Omega$ arises from the canonical field rescaling.
\item The effective quartic coupling $\lambda_j=\lambda/(4\pi j^2)$, which multiplies $\tilde\varphi_j^4/4!$, is time independent and decreases as $1/j^2$. At $j=1$, $\lambda_j=\lambda/(4\pi)\approx0.08\lambda$; at large $j$ it falls to zero, reflecting the factor of $4\pi r^2$ from the angular integration.
\end{enumerate}
Because only the mass coefficient is time-dependent, only the onsite mass term in the matrix product operator (MPO) needs to be updated at each time step. The quartic term is assembled once, using the coupling $\lambda_j=\lambda/(4\pi j^2)$ at site $j$. All other aspects of the implementation---DMRG ground-state preparation, TDVP/TEBD evolution, and all observables---carry over without modification.

\paragraph{Free-theory check.}
For $\lambda=0$ eq.~\eqref{eq:Hlat_swave_dimless} reduces to a chain of harmonic oscillators.  The mode equation for the rescaled field $\tilde\varphi$ is the standard Mukhanov--Sasaki equation,
\begin{equation}
\tilde f_n''(\eta)+\!\left(\hat k_n^2+m_{\rm eff}^2(\eta)\right)\tilde f_n(\eta)=0\;,\quad\hat k_n=2\sin\!\left(\frac{k_n}{2}\right)\;,\quad
k_n=\frac{n\pi}{N+1}\;,
\end{equation}
which differs from the $1+1$D mode equation~(\ref{eq:mode_fn_eq}) only through the replacement $\Omega^2 m^2\to m_{\rm eff}^2(\eta)$.  The ODE calculation of the rescaled power spectrum $\tilde P(k_n,\eta)=|\tilde f_n(\eta)|^2$ therefore requires a minor modification of the existing solver.  The Bunch--Davies initial conditions at $\eta_i$ are unchanged in form:
\begin{equation}
\tilde f_n(\eta_i)=\frac{1}{\sqrt{2\tilde\omega_n(\eta_i)}}\;,\qquad \tilde f_n'(\eta_i)=-i\,\tilde\omega_n(\eta_i)\,\tilde f_n(\eta_i)\;,\qquad
\tilde\omega_n^2=\hat k_n^2+m_{\rm eff}^2(\eta_i)\;.
\end{equation}

We present the free-theory comparison ($\lambda = 0$) using masses $m^2_{1+1} = 0.1$ for the $1+1$D model and $m^2_s = 2.1$ for the canonical s-wave model. We choose these masses by equating the two effective masses in the unregulated limit ($\eta_0\to0$). The s-wave Mukhanov--Sasaki mass in Eq.~\eqref{eq:meff} reduces to
$(m^2_s/H^2 - 2)/\eta^2$, while the $1+1$D effective mass goes as 
$m^2_{1+1}/\eta^2$. Equating the two gives
\begin{equation}
  m^2_s = m^2_{1+1} + 2H^2,
  \label{eq:mass-matching}
\end{equation}
which at $H=1$ yields $m^2_s = m^2_{1+1} + 2 = 0.1 + 2 = 2.1$. Equation~\eqref{eq:mass-matching} equates the \(1/\eta^2\) terms in the two free mode equations after canonical rescaling when $\eta_0=0$. It does not imply equivalence of the two bare theories.

We used the same numerical setup in both runs:
\[
N=39,\quad a=1,\quad \eta_i=-20,\quad \eta_*=-0.2,\quad \eta_0=0.05,\quad N_{\max}=10,\quad \chi=32,\quad \Delta\eta_{\max}=0.1.
\]
For each model we extracted the sine-projected spectrum \(P(k_n,\eta_*)\) from the connected equal-time correlator and compared \(k_nP_{\mathrm{MPS}}\) against the corresponding free mode-equation result \(k_nP_{\mathrm{ODE}}\).

\begin{figure}[hbt]
    \centering
    \includegraphics[width=0.85\linewidth]{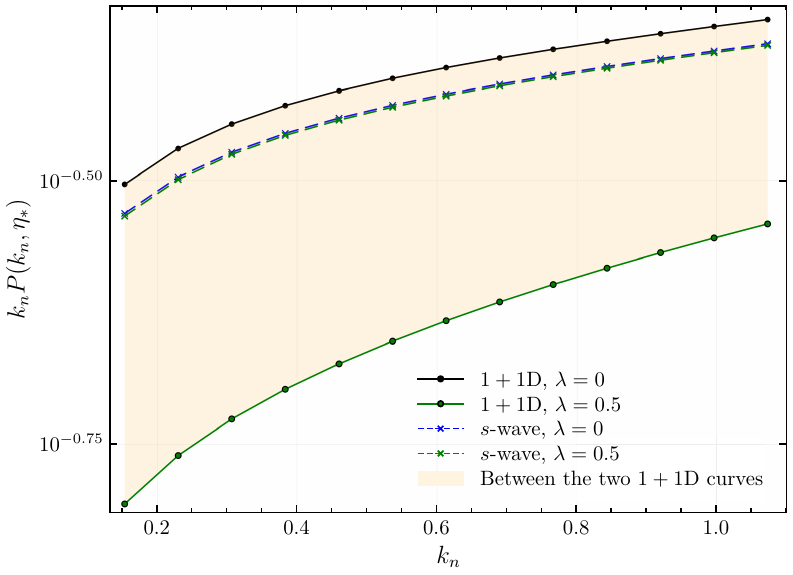}
   \caption{Equal-time spectra at $\eta_*=-0.2$ for $m_{1+1}^2=0.1$ and
   $m_{\mathrm{s\text{-}wave}}^2=2.1$. Solid lines and circles show the $1+1$D model;
   dashed lines and crosses show the canonical $s$-wave model.
   Black and blue denote $\lambda=0$, and green denotes $\lambda=0.5$.
   At $\lambda=0$, the lines are obtained from the free mode equations and the
   markers from MPS evolution. The shading marks the region between the two
   $1+1$D curves. The smaller change in the $s$-wave spectrum at $\lambda=0.5$
   is consistent with its site-dependent coupling $\lambda_j\propto 1/j^2$.}

    \label{fig:free_match_1p1d_swave}
\end{figure}

The agreement is quantitatively strong in both cases. Using
\[
\textcolor{black}{r_n(\eta)\equiv \log\!\left[\frac{P_{\mathrm{MPS}}(k_n,\eta)}{P_{\mathrm{ODE}}(k_n,\eta)}\right],}
\]
we find in the mode range \(n=4,\dots,12\):
\[
\max|r_n(\eta_i)|\simeq 1.56\times 10^{-4}\ \text{(both models)},
\]
\[
\max|r_n(\eta_*)|=1.33\times 10^{-4}\ \text{(1+1D)},\qquad
\max|r_n(\eta_*)|=1.78\times 10^{-4}\ \text{(s-wave)}.
\]
For the masses chosen above, both MPS implementations agree with the solutions of their respective free mode equations.

Figure~\ref{fig:free_match_1p1d_swave} also shows that the free spectra of the two models agree closely for these masses, reproducing the expected linear dynamics. At $\lambda=0.5$, the s-wave spectrum stays close to its free result and lies between the $1+1$ dimensional spectra at $\lambda=0$ and $\lambda=0.5$. The s-wave interaction is weaker at fixed bare $\lambda$ because the radial coupling $\lambda_j \propto 1/j^2$ decreases away from the origin. This explains the smaller shift in its spectrum.

\subsection{Toward a quantum-circuit implementation}
\label{sec:quantum_implementation}

It is natural to ask whether the same dS$_2$ correlators and the s-wave sector in 3+1d studied here with tensor-network methods
can also be formulated on gate-based quantum hardware\footnote{For a recent summary of implementations of low dimensional quantum field theories on a quantum computer see \cite{Sinha:2025hyq}.}.
We outline the state encoding, time evolution and overlap measurements needed for a proof-of-principle calculation.

As a minimal test case, consider the truncated bosonic lattice Hamiltonian on a small system with $N=4$ sites and local cutoff $N_{\max}=4$. Using a binary encoding, each site is represented by $n_q=\lceil \log_2(N_{\max}+1)\rceil=3$ qubits, so the full register contains $12$ system qubits. The real-time evolution may then be implemented by a second-order Suzuki--Trotter decomposition, with the time-dependent Hamiltonian evaluated at the midpoint of each step. This gives a direct implementation with quantum gates of the same real-time evolution problem studied
classically.

The equal-time in--in correlator,
\begin{equation}
B_{\rm in\text{-}in}(\eta_*)
=
\frac{\langle \psi(\eta_*)|\phi_j^2|\psi(\eta_*)\rangle}
     {\langle \psi(\eta_*)|\psi(\eta_*)\rangle},
\end{equation}
is then an ordinary expectation value and can be measured after expressing $\phi_j^2$
in the qubit basis. By contrast, the in--out quantity involves an overlap between two differently evolved branches. A natural hardware implementation therefore requires ancilla-assisted interferometric methods of Hadamard-test type in order to extract the real and imaginary parts of the relevant matrix element.

A useful simplification follows from the structure of the nearest-neighbour coupling.
Rather than synthesizing the bond evolution as a completely generic multiqubit unitary,
one can exploit the factorized form proportional to $\phi_j\phi_{j+1}$.
This significantly reduces the number of two-qubit gates needed for the bond step.
Even so, the total circuit depth remains substantial for present-day hardware once the full
time evolution is assembled.

The main conceptual obstruction is that selecting the vacuum in the in--out analysis of \cite{Donath:2024}
requires an operation that is not strictly unitary. The contour tilt employed in the classical treatment corresponds to a
non-Hermitian deformation and therefore does not admit a direct realization on standard
gate-based hardware. \textcolor{black}{The Approach-A real-time calculation uses adiabatic switching and is more directly compatible with unitary gates. We carry out the nonunitary contour evolution of Section~\ref{sec:asymptotic-projection} classically with MPS. Implementing its endpoint projection with the circuit construction described here needs additional resources.}

\begin{table}[H]
\centering
\begin{tabular}{|l|l|}
\hline
System size & $N=4$, $N_{\max}=4$ \\
Qubit cost & $12$ system qubits; $+1$ ancilla for in--out overlaps \\
Time evolution & second-order Suzuki--Trotter decomposition \\
$B_{\rm in\text{-}in}$ measurement & standard expectation-value estimation \\
$B_{\rm in\text{-}out}$ measurement & ancilla-assisted overlap measurement \\
Bond-gate optimization & exploit factorized $\phi_j\phi_{j+1}$ structure \\
Main limitation & contour tilt requires nonunitary operations \\
Current status & proof of principle \\
\hline
\end{tabular}
\caption{Minimal resource summary for a gate-based formulation of the regulated dS$_2$ lattice problem on a small system.}
\label{tab:quantum_summary}
\end{table}

Small systems may already be used to test encodings, Trotterization, overlap measurements and error mitigation against exact-diagonalization and MPS results. Quantum simulations are particularly relevant in the light-field regime, for example at $m^2=0.1$, where the entanglement growth is visibly stronger than in the perturbatively safe $m^2=1$ case and the classical tensor-network cost is driven by the need for much larger bond dimension. In that regime, the nonperturbative in--out test improves only when $\chi$ is increased from $30$ to $200$, and since the leading MPS cost scales as $\mathcal{O}(N d \chi^3)$, where $d=N_{\max}+1$, this corresponds to an increase in classical cost by roughly a factor of $(200/30)^3 \simeq 3\times 10^2$. By contrast, a gate-based encoding does not introduce a bond-dimension truncation. For the same light-field parameters used in the MPS runs, namely $N=10$ and $N_{\max}=10$, a binary encoding would require $n_q=\lceil \log_2(N_{\max}+1)\rceil = 4$ qubits per site, i.e.\ $40$ system qubits, plus one ancilla for the in--out overlap measurement. The regime where entanglement makes classical MPS evolution expensive is therefore the most interesting long-term target for quantum simulations. Their practical limits are then circuit depth, coherence time and the number of measurements. Meeting these requirements remains a medium-term goal rather than a near-term NISQ application.

\section{Conclusions and Outlook}
\label{sec:conclusion}

We used MPS to compute cosmological observables in a controlled lattice realization of interacting quantum field theory in dS$_2$. Our main focus was the comparison between the standard in--in correlator and the normalized in--out ratio proposed in \cite{Donath:2024}. At finite volume and finite initial time, we checked the regulated Bunch--Davies vacuum using mode covariances, solutions of the free mode equations and checks of the local oscillator cutoff. These checks were essential for distinguishing physical effects from finite-volume and finite-cutoff effects.

\begingroup\color{black}
We compared two main time contour implementations: the first being the real time contour motivated by simplicity while the second being a complex contour that carefully accounts for the extra pole that the choice of the regulator needed for zero conformal time. The real contour led to a residual imaginary part as well as increasing entanglement through time evolution across the patch. These were shown to be artifacts which were removed by the complex contour. The complex contour needs imaginary time evolution at portions along the contour. This will make implementation of the in-out approach on a futuristic quantum computer challenging. 
\endgroup

The bra-side bipartite entanglement entropy grows approximately linearly during the in-out evolution in figs.\ref{fig:ee_inout}, \ref{fig:ee-halfchain-inout}. This is qualitatively consistent with the Lieb--Robinson picture \cite{liebrobinson} for local lattice dynamics. In a one-dimensional short-range system, Lieb--Robinson bounds imply that the entanglement across a fixed cut can grow at most linearly in time, up to model-dependent prefactors. Since our simulations evolve a regulated nearest-neighbor bosonic chain with a finite on-site truncation, we view the observed linear trend as a natural feature of the truncated lattice Hamiltonian used in the MPS calculation, rather than as a statement about the untruncated continuum theory itself. In this sense, the difficulty of evolving the in-out state to late times is plausibly tied to the entanglement on the bra branch continuing to grow ballistically.

Here we have focused primarily on equal-time observables, but the same formalism can be used to study more general real-time probes, including unequal-time correlators and out-of-time-ordered correlators, which may reveal scrambling and emergent thermal behavior in curved spacetime. It would also be interesting to apply the same methods to theories beyond the $\phi^4$ model considered here. The dS$_2$ Schwinger model would allow us to test these methods in another interacting theory by comparing with recent exact results \cite{Anninos:2024fty}.

More broadly, many ingredients of our construction should carry over to other conformally flat backgrounds. With appropriate modifications of the conformal factor, boundary conditions, and observables, one can hope to adapt the same numerical methods to AdS$_2$ and to two-dimensional flat space\footnote{In a rather different direction, tensor-network constructions have also recently been developed \cite{cummings} for constrained gravitational systems themselves, including diffeomorphism-invariant states in three-dimensional gravity. While this is conceptually distinct from our use of MPS for real-time quantum field dynamics on a fixed de Sitter background, it points to a broader interface between tensor-network methods and gravitational physics.}. MPS calculations of interacting $\phi^4$ theory in AdS$_2$ have already extracted low-lying boundary scaling dimensions from the global-AdS energy spectrum~\cite{Bhat:2026epg}. In particular, it would be interesting to explore whether suitable curved-spacetime analogues of LSZ-type constructions can be formulated, thereby extending the present tensor-network approach from correlators to more directly scattering-like observables. In the AdS$_2$ setting, one may also ask whether such methods can be used to compute bulk correlators with sufficient precision to extract useful holographic data, perhaps in settings related to Jackiw--Teitelboim gravity and the Sachdev--Ye--Kitaev model.

\begingroup\color{black}
\paragraph{Scale-adapted tensor networks.}
\label{par:ref1-mera}
The scale covariance of de Sitter correlators motivates the multiscale entanglement renormalization ansatz (MERA), which organizes entanglement by length scale and can efficiently represent critical ground-state correlations~\cite{Vidal2008MERA,PfeiferEvenblyVidal2009ScaleMERA}. Its continuous counterpart, cMERA, offers a related continuum construction~\cite{HaegemanEtAl2013cMERA}. Applying these representations to our evolving state needs further study because de Sitter scale covariance acts on space and time together, whereas the scale-invariant ground states described above refer to a fixed spatial Hamiltonian. Finite volume, initial-time preparation and the geometric regulator also restrict the available scaling regime. We chose MPS because the discretized Hamiltonian is a local chain with established ground-state and time-evolution algorithms, and because the required expectation values and independently evolved bra--ket overlaps can be computed directly. A multiscale representation may improve the treatment of correlations over widely separated scales. Its cost relative to MPS for these real- or complex-time calculations remains an open question, to be assessed by comparing the same observables at the same accuracy.

\paragraph{Infrared dynamics and stochastic inflation.}
\label{par:ref1-stochastic}
For light scalar fields, stochastic inflation provides an effective description of the long-wavelength sector that can resum infrared effects missed by a finite perturbative expansion. The interacting equilibrium distribution and its correlation functions provide useful comparisons for tensor-network calculations~\cite{StarobinskyYokoyama1994,Gorbenko:2019rza}. Tensor networks could complement this description by evolving the underlying quantum state and measuring a field averaged over a specified physical scale. Its probability distribution, second and connected fourth moments, and relaxation rates could then be compared with the corresponding stochastic predictions at the same parameters. Such a comparison could test where the long-wavelength approximation applies and assess quantum or gradient corrections, beyond checking whether interactions make a moment nonzero. The coarse-graining prescription, accessible volume, evolution time and ultraviolet renormalization must be controlled for that purpose. 

\paragraph{Superhorizon simplification and angular truncation.}
\label{par:ref1-angular}
Suppression of spatial gradients on superhorizon scales motivates an effectively local long-wavelength description. Spatial correlations, horizon-crossing initial data and the effects of discarded modes still need to be determined. Even in a light-field regime with logarithmic correlations, relating the full quantum state's simulation cost to that of a two-dimensional conformal field theory needs a separate assessment. We retain low angular momentum $\ell$ to obtain a one-dimensional radial chain. The physical angular-gradient scale is $\ell(\ell+1)/[\Omega^2(\eta)r^2]$: at fixed comoving radius, expansion can make a mode superhorizon without making its angular index small. Interactions couple the retained and omitted sectors, so even observables constructed from the angularly averaged field require an angular-cutoff check. We describe the s-wave implementation in Section~\ref{3plus1dS} and the proposed $\ell=1$ extension in Appendix~\ref{sec:spin1trunc}. Convergence in $\ell_{\max}$ remains to be tested for selected observables. Higher-dimensional networks such as projected entangled-pair states offer another direction for extending the calculation.

\paragraph{Even and odd spacetime dimensions.}
\label{par:ref1-parity}
The odd-dimensional transparency result discussed in \cite{LagogiannisMaloneyWang2011Transparent} concerns free fields in \emph{global} de Sitter and specified asymptotic in/out particle states, distinct from the Euclidean (Bunch--Davies) vacuum~. For scalars, its oscillatory asymptotic interpretation applies when $m^2/H^2>(D-1)^2/4$, where $D$ is the spacetime dimension. This result concerns transparency for those states while allowing other quantum effects in odd $D$. Our regulated patched-Poincar\'e calculations use interactions and different boundary prescriptions, and the reported dS$_2$ and s-wave dS$_4$ examples are both even-dimensional. Testing transparency and any dependence of the entanglement cost on dimensional parity will be interesting to examine. 
\endgroup

Finally, our results help clarify where quantum-computing approaches may eventually become useful. In the present paper we only considered a proof-of-principle quantum-circuit formulation, but the tensor-network analysis already indicates which regimes are likely to be classically difficult, namely those in which the entanglement growth drives the required bond dimension rapidly upward. Tensor networks provide controlled classical comparisons in the weakly and moderately entangled regimes, while more strongly entangled states may be suitable for future quantum simulations. A worthwhile next step would be to obtain explicit circuit-depth and measurement estimates, including the effect of noise and finite-depth approximations in platforms such as Qiskit.

\section*{Acknowledgments} We thank Faizan Bhat, Diptarka Das, Hiranmay Das , Arnab Saha and Vijay Shenoy for valuable discussions. AS is supported by an ANRF grant ANRF/ARG/2025/001338/PS and a Quantum Horizons Alberta senior fellowship. XZ is supported by NSFC Grant No. 12275273, funds from Chinese Academy of Sciences, University of Chinese Academy of Sciences, and the Kavli Institute for Theoretical Sciences. We have used generative AI tools for programming, literature survey, and stylistic work. All outputs from AI have been cross-checked by us and we have also written our own code for several of the numerical simulations reported here. UB acknowledges support from the National Quantum Mission of the DST, Govt of India, through a research assistant position.

\appendix
\section{Interaction-picture identities for time-dependent Hamiltonians}
\label{app:pictures}

We collect the standard relations between the Schr\"{o}dinger, Heisenberg,
and interaction pictures when the Hamiltonian depends explicitly on conformal time~$\eta$
and is split as $H(\eta)=H_0(\eta)+V(\eta)$.

\subsection{Time-evolution operators and interaction picture}
\label{app:timeevol}

For any (possibly time-dependent) Hamiltonian, the unitary evolution operator is
\begin{equation}
U(\eta_2,\eta_1)\equiv\mathcal{T}\exp\!\left(-i\int_{\eta_1}^{\eta_2} d\eta\, H(\eta)\right)\;,\qquad U(\eta,\eta)=\mathbf{1}\;.
\label{eq:Udef}
\end{equation}
The free part $H_0$ generates $U_0(\eta_2,\eta_1)$ analogously.
In the Schr\"{o}dinger picture, states evolve with $U$ while operators are time independent;
in the Heisenberg picture, operators evolve with $U$ while states remain fixed at~$\eta_i$.
In the interaction picture, states evolve with
\begin{equation}
U_I(\eta,\eta_i)\equiv U_0^\dagger(\eta,\eta_i)\,U(\eta,\eta_i)\;,
\label{eq:UIdef_simple}
\end{equation}
while operators carry only the free evolution,
$O_I(\eta)\equiv U_0^\dagger(\eta,\eta_i)\,O_S\,U_0(\eta,\eta_i)$.
Differentiating~\eqref{eq:UIdef_simple} gives $i\partial_\eta U_I = V_I(\eta)\,U_I$
where $V_I\equiv U_0^\dagger V U_0$, so the formal solution is
\begin{equation}
U_I(\eta_2,\eta_1)=\mathcal{T}\exp\!\left(-i\int_{\eta_1}^{\eta_2} d\eta\, V_I(\eta)\right)\;.
\label{eq:UIsolution}
\end{equation}
When the interaction picture is defined with reference time $\eta_1$, the factorization identity, exact even when $H_0(\eta)$ is time-dependent, is
\begin{equation}
\boxed{U(\eta_2,\eta_1)=U_0(\eta_2,\eta_1)\,U_I(\eta_2,\eta_1)\;.}
\label{eq:Ufactor}
\end{equation}

\subsection{Useful rewritings for in--out quantities}
\label{app:inout_rewrite}

The in--out ratio proposed by Donath and Pajer is naturally expressed in the interaction picture.
For a fixed reference time $\eta_i$ and arbitrary endpoints $\eta_1,\eta_2$, the corresponding identity is
\begin{equation}
U_I(\eta_2,\eta_1)=U_0^\dagger(\eta_2,\eta_i)\,U(\eta_2,\eta_1)\,U_0(\eta_1,\eta_i)\;,
\label{eq:UI_general}
\end{equation}
valid for any reference time~$\eta_i$\;.

\paragraph{Remark (``out bra'' in Schr\"{o}dinger picture).}
If the interaction-picture bra is $\langle 0|$, the corresponding
Schr\"{o}dinger-picture free bra at time~$\eta_f$ is
\begin{equation}
\langle 0,\eta_f|\equiv \langle 0|\,U_0^\dagger(\eta_f,\eta_i)\;,
\end{equation}
which generally differs from $\langle 0|$ when $H_0(\eta)$ is time dependent.
The identity test $O=\mathbf{1}$ in the in--out ratio requires using the \emph{same}
free-evolved bra on both numerator and denominator.

\section{Tensor-network algorithms}
\label{app:MPS}

The numerical simulations use Matrix Product States (MPS) as the variational ansatz~\cite{Schollwock:2011nx,Orus:2013kga}. Physically, an MPS factorizes the exponentially large many-body wavefunction of $N$ sites into a one-dimensional chain of local tensors. The state is written as
\begin{equation}
  |\psi\rangle  = \sum_{n_1,\ldots,n_N}    A^{n_1}[1]\,A^{n_2}[2]\cdots A^{n_N}[N] \;|n_1,\ldots,n_N\rangle\;,
\end{equation}
where each $A^{n_j}[j]$ in the bulk is a $\chi\times\chi$ complex matrix, and $A^{n_1}[1]$ and $A^{n_N}[N]$ at the boundaries are $\chi$ dimensional vectors. The {\it bond dimension} $\chi$ limits the Schmidt rank across each bond and determines the cost of contracting the MPS. Computing expectation values and reduced density matrices scales as $\mathcal{O}(N\chi^3 d)$, where $d=N_{\max}+1$ is the local physical dimension. The maximum entanglement entropy across a single cut is therefore $S\leq\log\chi$. Because the ground states of 1D gapped Hamiltonians strictly obey an entanglement area law, a finite $\chi$ is mathematically guaranteed to capture the true ground state to any desired precision. Finally, after each numerical update, the state is explicitly restored to a {\it canonical form}. In this gauge choice, the singular values on the virtual bonds (denoted as $\{\sigma_\alpha\}$) are exactly equal to the system's Schmidt coefficients. Maintaining this canonical gauge stabilizes the algorithm against truncation errors and gives the entanglement entropy directly from the Schmidt coefficients via $S=-\sum_\alpha\sigma_\alpha^2\log\sigma_\alpha^2$.

\paragraph{Ground-state preparation (DMRG).} We prepare the instantaneous ground state at $\eta_i$ using the Density Matrix Renormalization Group (DMRG) \cite{White:1992zz}. In Approach A the interaction is suppressed at $\eta_i$, whereas Approach B uses the instantaneous interacting Hamiltonian. DMRG optimizes the tensor network via a variational sweep across the 1D chain. At each step, the algorithm isolates one or two adjacent sites and projects the global many-body Hamiltonian into an effective local basis defined by the surrounding left and right environments. The energy is then minimized by solving a local eigenvalue problem, typically with an iterative method such as Lanczos. For gapped free theories, the spatial correlations decay exponentially. Consequently, the MPS ansatz captures the relevant physics, allowing the algorithm to converge to the true ground state in a small number of sweeps.

\paragraph{Real-time evolution (TDVP and TEBD).}
 We evolve the state under the time-dependent Hamiltonian $H(\eta)$ with TDVP and TEBD. We control the numerical errors by comparing the two methods for selected runs.

 \begin{itemize}
     \item {\bf Time-Dependent Variational Principle
(TDVP) \cite{Haegeman:2016}.} The TDVP method evolves the system by projecting the Schr\"{o}dinger equation
$i\partial_\eta|\psi\rangle = H(\eta)|\psi\rangle$ onto the tangent space of the MPS manifold at each time step. The Julia calculations use one-site TDVP, which evolves within the MPS manifold at fixed bond dimension. The TenPy calculations also use two-site TDVP, which updates adjacent tensors and allows the bond dimension to increase up to the imposed cutoff. The projected real-time equation preserves the norm for a Hermitian Hamiltonian, even when the Hamiltonian depends on time. Numerical integration errors and, in the two-site method, truncation of Schmidt coefficients can produce a small norm drift. We monitor this drift continuously as a check on numerical convergence. \textcolor{black}{In the original real-time runs this deviation was below $10^{-3}$. Norm conservation provides a check on the evolution, while observable errors need separate convergence checks. For the nonunitary complex-contour evolution in Section~\ref{sec:asymptotic-projection}, we instead track the changing normalization.}
\item {\bf Time-Evolving Block Decimation (TEBD) \cite{Vidal:2004,HauschildPollmann2018}.} The TEBD algorithm takes a different approach, approximating the time-evolution operator $e^{-iH\,d\eta}$ by a Trotter-Suzuki decomposition. By partitioning the Hamiltonian into sums of commuting even and odd nearest-neighbor bond terms $H = H_{\rm even}+H_{\rm odd}$, the second-order scheme expands the evolution operator as
\begin{equation}
  e^{-iH\,d\eta}\approx e^{-iH_{\rm even}\,d\eta/2}\, e^{-iH_{\rm odd}\,d\eta}\, e^{-iH_{\rm even}\,d\eta/2}  + \mathcal{O}(d\eta^3)\;.
\end{equation}
For the time-dependent Hamiltonian, all factors use $H(\eta+d\eta/2)$ during a step from $\eta$ to $\eta+d\eta$.
Because the local terms within $H_{\rm even}$ (and similarly $H_{\rm odd}$) commute with one another, each exponential factor breaks down into a product of independent two-site gates that are applied sequentially across the chain. After applying each gate, the local bond is truncated by discarding the smallest singular values to limit the bond dimension. For equal time steps, the accumulated Trotter error is $\mathcal{O}(N_{\rm steps}\,d\eta^3)$, or $\mathcal{O}(d\eta^2)$ at a fixed total evolution time. We control this by choosing a time step $d\eta$ sufficiently small so that the dynamics generated by TEBD agree with the TDVP results to our quoted precision.
 \end{itemize}
For the local nearest-neighbor Hamiltonians studied here, TEBD is in practice usually faster per time step than two-site TDVP, making it well suited for exploratory scans over parameters. TDVP, however, avoids Trotterization and therefore serves as an independent comparison for controlling systematic errors in the real-time evolution.

\section{The regulated in--out comparison}
\label{app:cosmo}

\subsection{Free-theory and lattice formulas}
\label{app:cosmo:formulas}
We use the mode normalization of Section~\ref{subsec:continuum_theory}.
In the expanding patch ($\eta<0$), the continuum Bunch--Davies (BD) modes are
\begin{equation}
\label{eq:BDmodeschoice}
    f_k^{\rm BD}(\eta)=\frac{\sqrt{-\pi\eta}}{2}\,
    e^{i\pi(2\nu+1)/4}\,H^{(1)}_\nu(-|k|\eta),
    \qquad
    \nu=\sqrt{\tfrac{1}{4}-\tfrac{m^2}{H^2}},
\end{equation}
where $H^{(1)}_\nu$ is the Hankel function of the first kind.
They obey $f_k f_k^{*\prime}-f_k'f_k^*=i$ and are selected by
$f_k\to e^{-i|k|\eta}/\sqrt{2|k|}$ as $\eta\to-\infty$.
The prefactor also fixes the normalization when $\nu$ is imaginary.

On the lattice of $N$ sites (spacing $a$, Dirichlet boundary conditions
$\phi_0=\phi_{N+1}=0$), we use the sine basis $u_j(n)$ in~\eqref{eq:sine_modes}.
The integer $j$ labels a lattice mode, while $k_j=\pi j/[a(N+1)]$ denotes
its momentum. Each mode function $v_j(\eta)$ satisfies
\begin{equation}
\label{eq:latticemodeq}
    v_j'' + \omega_j(\eta)^2\,v_j = 0,
    \qquad
    \omega_j(\eta)^2
    = \frac{4}{a^2}\sin^2\!\Bigl(\frac{j\pi}{2(N+1)}\Bigr)
      + \Omega(\eta)^2\,m^2,
    \quad j=1,\ldots,N.
\end{equation}
For the finite-time calculations, we impose the free instantaneous-vacuum
initial data~\eqref{eq:mode_initial_data}, with the Wronskian
in~\eqref{eq:lattice_mode_wronskian}. This finite-time choice approaches
the BD prescription as $\eta_i\to-\infty$. The field expansion is
\begin{equation}
\label{eq:discexpansion}
    \phi_n(\eta)
    = \sqrt{\frac{2}{a(N+1)}}\sum_{j=1}^{N}
      \bigl(b_j v_j(\eta)+b_j^{\dagger}v_j^*(\eta)\bigr)
      \sin\!\Bigl(\tfrac{nj\pi}{N+1}\Bigr),
\end{equation}
where $[b_j,b_\ell^\dagger]=\delta_{j\ell}$ and
$b_j|0_{\eta_i}\rangle=0$. Contracting these operators gives
\begin{equation}
\label{eq:Wightman}
    W_{nm}(\eta_1,\eta_2)
    \equiv \langle 0_{\eta_i}|\phi_n(\eta_1)\phi_m(\eta_2)|0_{\eta_i}\rangle
    = \frac{2}{a(N+1)}\sum_{j=1}^{N}
      v_j(\eta_1)\,v_j^*(\eta_2)
      \sin\!\Bigl(\tfrac{nj\pi}{N+1}\Bigr)
      \sin\!\Bigl(\tfrac{mj\pi}{N+1}\Bigr),
\end{equation}
in agreement with~\eqref{eq:wightmanfree}.

The instantaneous vacuum $|0_\eta\rangle$ is the ground state of the free
Hamiltonian held fixed at $\eta$. Its equal-time covariance is
\begin{equation}
\label{eq:instcorr}
    \langle 0_\eta|\phi_n(\eta)\phi_m(\eta)|0_\eta\rangle
    = \frac{1}{a(N+1)}\sum_{j=1}^{N}
      \frac{\sin\!\bigl(\frac{nj\pi}{N+1}\bigr)
            \sin\!\bigl(\frac{mj\pi}{N+1}\bigr)}
           {\omega_j(\eta)}\;.
\end{equation}
The factor $1/\omega_j$ follows from the vacuum variance
$\langle Q_j^2\rangle=1/(2\omega_j)$ and the normalization of the sine basis.
For exact de Sitter space at $H=1$, $\omega_j^2=\hat k_j^2+m^2/\eta^2$;
with the metric regulator, it is
$\omega_j^2=\hat k_j^2+m^2/[H^2(\eta^2+\eta_0^2)]$.
At $\eta=\eta_i$, Eq.~\eqref{eq:instcorr} equals
$W_{nm}(\eta_i,\eta_i)$. At later times, the covariance of the evolved
initial vacuum must instead be computed from~\eqref{eq:Wightman}.

\subsection{First-order perturbative criterion for the DP equality}
\label{app:cosmo:pert}
We now examine the Donath--Pajer (DP) equality at first order in perturbation theory by comparing the leading corrections to the in-in and in-out correlators.

\paragraph{In-in formula.} We expand the evolution operator to first order in the interaction
$H_{\rm int}(\eta) = (\lambda/4!)\sum_m a\,\Omega^2(\eta)\,\phi_m^4$
via the Keldysh closed-time-path formalism. Performing Wick-contractions, we get 
\begin{equation}
  B^{(1)}_{\rm in\text{-}in}  = -\int_{\eta_i}^{\eta_*}\!d\eta\;    \Omega^2(\eta)    \sum_{m} a\;    \mathrm{Im}\bigl[W_{mi}(\eta,\eta_*)\,W_{mk}(\eta,\eta_*)\,W_{mm}(\eta,\eta)\bigr]\;,
  \label{eq:B1inin}
\end{equation}
where $W_{ij}(\eta_1,\eta_2)=\langle 0_{\eta_i}|\phi_i(\eta_1)\phi_j(\eta_2)|0_{\eta_i}\rangle$
is the free Wightman function \eqref{eq:Wightman}.

\paragraph{In-out formula.}
Expanding the single forward contour of the DP ratio~\eqref{eq:Binout_DP_def} to first order
in $\lambda$, the Feynman propagator connecting external leg $i$ (at $\eta_*$) to
the vertex at time $\eta$ changes structure across $\eta=\eta_*$:
\begin{itemize}
  \item For $\eta<\eta_*$ (ket region), time-ordering puts the vertex to the right
    of the operator insertion, giving $W_{mi}^*(\eta,\eta_*)W_{mk}^*(\eta,\eta_*)W_{mm}(\eta,\eta)$.
    Using $\mathrm{Im}[z^*]=-\mathrm{Im}[z]$, the contribution is $-I_{\rm ket}/2$.
  \item For $\eta>\eta_*$ (bra region), the vertex is to the left, giving
    $W_{mi}(\eta,\eta_*)W_{mk}(\eta,\eta_*)W_{mm}(\eta,\eta)$, contributing $+I_{\rm bra}/2$.
\end{itemize}
Let us define the partial integrals
\begin{equation}
  I_{\rm ket} \equiv \int_{\eta_i}^{\eta_*}\!d\eta\;\Omega^2\sum_m a\,    \mathrm{Im}[W_{mi}W_{mk}W_{mm}]\;,\quad  I_{\rm bra} \equiv \int_{\eta_*}^{\eta_f}\!d\eta\;(\cdots)\;.
  \label{eq:IketIbra}
\end{equation}
Then the first order correction to the in-in correlator is 
\begin{equation}
    B^{(1)}_{\rm in\text{-}in} = -I_{\rm ket}\;,
\end{equation}
and the real part of the correction to the in-out correlator is
\begin{equation}
 \mathrm{Re}\bigl[B^{(1)}_{\rm in\text{-}out}\bigr]  = \frac{I_{\rm bra} - I_{\rm ket}}{2}\;.  \label{eq:B1inout}
\end{equation}
\textcolor{black}{The real-part condition $B^{(1)}_{\rm in\text{-}in}=\mathrm{Re}[B^{(1)}_{\rm in\text{-}out}]$, necessary for the full complex equality, gives}
\begin{equation}
  -I_{\rm ket} = \frac{I_{\rm bra}-I_{\rm ket}}{2}\;,
  \quad\Longrightarrow\quad
  I_{\rm ket}+I_{\rm bra} = 0\;.
  \label{eq:DP_algebra}
\end{equation}

\paragraph{\textcolor{black}{A necessary real-part condition at first order.}}
\textcolor{black}{Equation~\eqref{eq:DP_algebra} tests the real-part condition at $O(\lambda)$ through the following total vertex integral. Vanishing of the imaginary part of the in--out correction is a separate requirement:}
\begin{equation}
 \int_{\eta_i}^{+\infty}d\eta\;\Omega^2(\eta)\sum_{m} a\;  \mathrm{Im}\bigl[W_{mi}(\eta,\eta_*)\,W_{mk}(\eta,\eta_*)\,W_{mm}(\eta,\eta)\bigr]  = 0\;,  \label{eq:DP_condition}
\end{equation}
where we have further extended $\eta_f$ to infinity. Note that $I_{\rm ket}$ and $I_{\rm bra}$ individually need not vanish.
In the flat-space limit where $\Omega\to\mathrm{const}$, the product of Wightman functions contains oscillatory factors
$\propto e^{-i(\omega_j+\omega_\ell)\eta}$ with $\omega_j+\omega_\ell>0$; \textcolor{black}{the endpoint prescription must include adiabatic damping before taking the infinite-time limit. Positive-frequency Fourier components then vanish in the distributional sense. At finite endpoints, an undamped oscillatory integral can remain nonzero.}
In exact Poincar\'e-patch de~Sitter, the Bunch--Davies mode functions
(Hankel functions $H_\nu^{(1)}$) are analytic in the upper half of the
complex $z$-plane ($z=-\hat k_j\eta$), which means the integrand is
analytic in the lower half $\eta$-plane. \textcolor{black}{With sufficient falloff and no late-time singular contribution, the appropriate continued contour gives the perturbative relation under the hypotheses of \cite{Donath:2024}. Light-field interactions with late-time divergences need additional control at the gluing point.}

\paragraph{Effect of the metric regulator.}
To regulate the singular behavior of the conformal factor at $\eta=0$, we introduced
$\Omega_{\rm reg}(\eta)=1/\sqrt{\eta^2+\eta_0^2}$ in the numerical implementation,
using $H=1$. The square root has branch points at $\eta=\pm i\eta_0$, whereas
the coefficient $\Omega_{\rm reg}^2$ in the mode equation is single-valued, with
simple poles at these points. Near the lower pole, put $w=\eta+i\eta_0$ and
$c=m^2/(-2i\eta_0)$. The mode equation becomes
\begin{equation}
  \frac{d^2v_j}{dw^2}+\left[\frac{c}{w}+O(1)\right]v_j=0.
\end{equation}
Two independent local solutions can be chosen as
\begin{equation}
  v^{(1)}(w)=w-\frac{c}{2}w^2+O(w^3),\qquad
  v^{(2)}(w)=h(w)-c\,v^{(1)}(w)\log w,
\end{equation}
where $h$ is analytic and $h(0)=1$. Thus a generic massive mode has a
logarithmic branch point. A counterclockwise circuit changes $v^{(2)}$ by
$-2\pi i c\,v^{(1)}$; repeated circuits do not in general return the original
solution. The upper pole has the same local structure, with $c=m^2/(2i\eta_0)$.
\begingroup\color{black}
For $m=0$, $c=0$: the free mode equation reduces to $v_j''+\hat k_j^2v_j=0$, so the modes have no regulator-induced logarithmic branch point. The regulated $\phi^4$ vertex retains $\Omega_{\rm reg}^2$ and its poles at $\pm i\eta_0$. The massless covariant derivative interaction instead has a first-order coefficient proportional to $\Omega_{\rm reg}^{-2}=H^2(\eta^2+\eta_0^2)$, an entire polynomial, and we evaluate it without the geometric regulator. For the generic massive case considered here we choose the lower cut to extend downward along the imaginary axis.
\endgroup

The contour argument must account for both the poles of $\Omega_{\rm reg}^2$
and the continuation of the mode functions in the Wightman products; the
local expansion alone does not determine $I_{\rm ket}+I_{\rm bra}$.
Numerically, we find that the relative difference $\Delta_{\rm rel}=\lambda|I_{\rm ket}+I_{\rm bra}|/(2|B^{(0)}|)$ is
$6.0\%$ at $\eta_0=0.3$ and grows non-monotonically as $\eta_0\to0$,
\textcolor{black}{demonstrating regulator sensitivity at the stated finite box and endpoints. The separate finite-time, finite-volume and continuation effects remain to be determined.} A numerical scan over $\eta_0$ is shown in Table~\ref{tab:eta0_pert}.

\begin{table}[htbp]
\centering
\renewcommand{\arraystretch}{1.3}
\begin{tabular}{|c|r|r|r|r|c|}
\hline
$\eta_0$ & $B^{(0)}$ & $I_{\rm ket}$ & $I_{\rm bra}$ & $I_{\rm total}$ & $\Delta_{\rm rel}$ \\
\hline
$\infty$ (flat) & $+0.127$ & $-2.29\times10^{-4}$ & $+2.29\times10^{-4}$ & $-4.7\times10^{-7}$ & $0.00\%$ \\
$1.0$ & $-5.5\times10^{-4}$ & $-4.96\times10^{-4}$ & $+5.16\times10^{-4}$ & $+2.0\times10^{-5}$ & $0.18\%$ \\
$0.5$ & $-1.53\times10^{-3}$ & $-2.70\times10^{-4}$ & $-9.74\times10^{-4}$ & $-1.24\times10^{-3}$ & $4.1\%$ \\
$0.3$ & $-1.71\times10^{-3}$ & $-2.19\times10^{-4}$ & $-1.84\times10^{-3}$ & $-2.06\times10^{-3}$ & $6.0\%$ \\
$0.1$ & $-1.80\times10^{-3}$ & $-1.93\times10^{-4}$ & $-1.52\times10^{-3}$ & $-1.72\times10^{-3}$ & $4.8\%$ \\
\hline
\end{tabular}
\caption{Absolute difference between the real parts of the in-in and in-out correlators at first order, divided by the magnitude of the free correlator, as a function of the regulator $\eta_0$.
The parameters are $N=39$, $\eta_i=-20$, $\eta_f=+20$, $\eta_*=-1$, $O=\phi(20)\phi(25)$, $\lambda=0.1$.
\textcolor{black}{The branch and finite-time contributions to this nonmonotone dependence remain undetermined. The small value at $\eta_0=1.0$ is a finite-regulator result, with the limiting equality still open.}}
\label{tab:eta0_pert}
\end{table}

Let us also make two comments regarding the mismatch for different operator insertion locations. 
\begin{itemize}
    \item {\bf Coincident operators.} For coincident operator insertions at the central lattice site, we often observe much smaller $\Delta_{\rm rel}$. For instance, for $\eta_0=0.3$ the relative difference $\Delta_{\rm rel}$ for $O=\phi^2(20)$ is only $0.16\%$, compared to $6.0\%$ for $O=\phi(20)\phi(25)$ as shown in Table \ref{tab:eta0_pert}. The reduction comes from the normalization. The absolute difference $\lambda|I_{\rm total}|/2$ is 2.7 times larger for coincident operators, but $B^{(0)}=\langle\phi^2\rangle$, the UV-divergent coincidence limit, is roughly 100 times larger than the separated correlator. The fractional correction is therefore small regardless of whether the Donath--Pajer relation holds, and its smallness does not imply a smaller absolute difference.
    \item \textcolor{black}{\textbf{Unequal-time operators.} For $O=\phi_i(\eta_1)\phi_k(\eta_2)$ with $\eta_1<\eta_2$, the in--in Wightman function is generally complex. Testing equality then entails retaining the contour ordering and all time intervals relative to the two insertions. We therefore compare the full complex difference $|B_{\rm in\text{-}out}-B_{\rm in\text{-}in}|/|B_{\rm in\text{-}in}|$, with the same specified operator ordering. For a projected prescription, equality remains to be tested with a consistent analytic continuation of the regulated modes and vertex.}
\end{itemize}

\subsection{Regulator-induced branch point and contour deformation}
\label{app:cosmo:contour}

The branch-cut obstruction identified above can be isolated by deforming the
integration contour.
We examine the contour geometry and the resulting order-of-magnitude reduction in the difference between the real parts of the correlators at first order.

\begingroup\color{black}
\paragraph{The integrand and its analytic continuation.}
Before applying a contour argument, define the \emph{complex} vertex function
\begin{equation}
\mathcal F(z)=\Omega_{\rm reg}^2(z)\sum_m a\,
 \mathcal W_{mi}(z,\eta_*)\mathcal W_{mk}(z,\eta_*)
 \mathcal W_{mm}(z,z).
\label{eq:complex-vertex-integrand}
\end{equation}
Here $\mathcal W$ is the analytic continuation of the real-axis Wightman product on a specified branch. More explicitly, a real-axis factor $v_j(\eta)v_j^*(\eta)$ continues to $v_j(z)\widetilde v_j(z)$, where $\widetilde v_j$ is the independently continued solution whose real-axis values are $v_j^*(\eta)$. Complex-conjugating $v_j(z)$ pointwise at complex $z$ would change this continuation. On the real axis the integrand of \eqref{eq:IketIbra} is $\operatorname{Im}\mathcal F(\eta)$, so
\[
I_{\rm ket}=\operatorname{Im}\int_{\eta_i}^{\eta_*}\mathcal F(\eta)\,d\eta,
\qquad
I_{\rm bra}=\operatorname{Im}\int_{\eta_*}^{\eta_f}\mathcal F(\eta)\,d\eta.
\]
For a complex contour we first form
$\mathcal J_C=\int_C\mathcal F(z)\,dz$ and only then take the required real or imaginary part. We apply Cauchy's theorem to the analytic function $\mathcal F$ at this stage because $\operatorname{Im}\mathcal F(z)$ is nonholomorphic.
\endgroup

\begin{figure}[hbt]
\centering
\includegraphics[width=\textwidth]{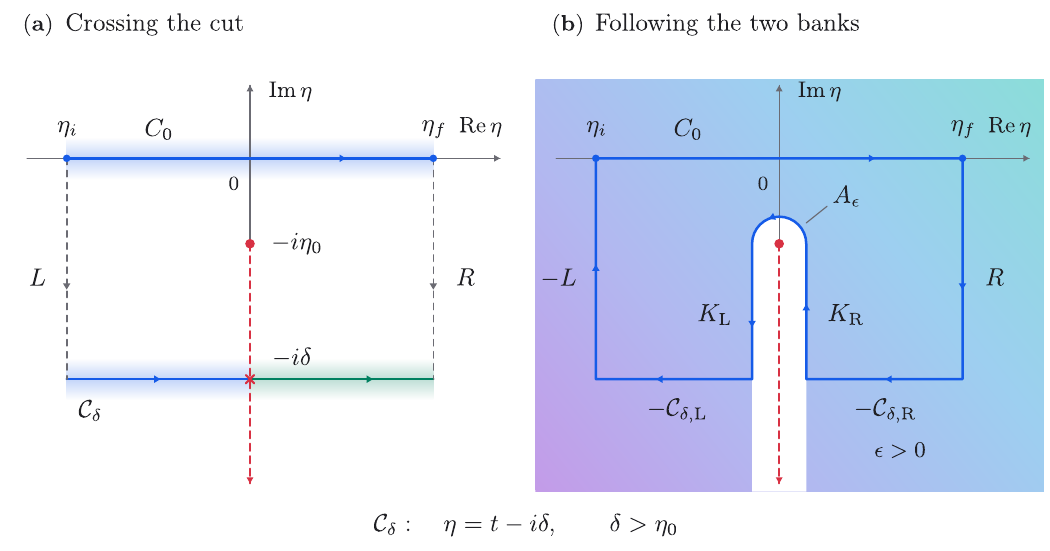}
\caption{Contour geometry in the complex $\eta$-plane, with $\delta>\eta_0$.
The red dashed line is the cut beginning at $\eta=-i\eta_0$.
Blue and green distinguish values along the path before and after continuation
through the cut. In (a), the fading colour bands follow these path segments.
(a) The shifted path $\mathcal{C}_\delta:\eta=t-i\delta$ crosses the cut
at $\eta=-i\delta$, marked $\times$.
The dashed segments $L$ and $R$ show the finite endpoint connectors
projected onto the $\eta$-plane.
(b) A closed contour on the chosen branch follows the right bank $K_{\mathrm R}$
upward, the upper semicircle $A_\epsilon$ of radius $\epsilon>0$ around the
branch point, and the left bank $K_{\mathrm L}$ downward.
The tinted domain in (b) represents a single branch, including both banks;
the white slit and its rounded end mark the excluded neighbourhood of the cut.
The lower horizontal segments are traversed from right to left, and $-L$
returns to $\eta_i$. Both banks, the small arc and both endpoint connectors
are retained at finite cutoffs.
The green segment in (a) and the lower-right blue segment in (b) use
different branch prescriptions.}
\label{fig:contour}
\end{figure}

\paragraph{Riemann sheet structure.}
For continuation into the lower half-plane, we choose the cut of the multivalued
integrand \textcolor{black}{$\mathcal F(z)$ of \eqref{eq:complex-vertex-integrand}} to run downward from $\eta=-i\eta_0$ along the imaginary axis
($\mathrm{Re}(\eta)=0$, $\mathrm{Im}(\eta)<-\eta_0$).
We label the branches encountered in this deformation as follows:
\begin{itemize}
  \item \textbf{Sheet I} is reached from the real axis without crossing the cut.
  The real-axis contour $C_0$ lies entirely on Sheet~I.
  \item \textbf{Sheet II} is reached by continuing across the lower cut once from left to right.
\end{itemize}
Each branch is defined on a copy of the cut plane, with both banks of the cut.
The labels identify the continuation along the path, rather than its position
in the complex plane; they do not imply that there are only two branches.
We consider the shifted contour $\mathcal{C}_\delta: \eta(t)=t-i\delta$,
$t\in[\eta_i,\eta_f]$, with $\delta>\eta_0$, and we should eventually take the $|\eta_i|, |\eta_f|\to \infty$ limit.
As $t$ increases from $-\infty$ to $+\infty$, this contour crosses the branch cut
at the single point $\eta=-i\delta$ (where $\mathrm{Re}(\eta)=0$).
The left portion of the shifted contour ($t<0$) is on Sheet~I while the right portion ($t>0$) is on Sheet~II. Figure~\ref{fig:contour}(a) shows this crossing; panel~(b) shows a closed contour that stays on the chosen branch by following both banks of the cut.
\textcolor{black}{Analytic continuation below the lower singularity removes the local pole obstruction from the deformed path. Applying Jordan's lemma to close the contour also entails checking the continued integrand and its large-time falloff. }

\begingroup\color{black}
\paragraph{Cauchy relation and endpoint terms.}
A closed-contour identity must use a single specified branch throughout its boundary. In Figure~\ref{fig:contour}(b), let $\widehat C_\delta$ denote the two lower horizontal segments evaluated on that chosen branch and oriented from left to right. Orient both endpoint connectors downward, $L:\eta_i\to\eta_i-i\delta$ and $R:\eta_f\to\eta_f-i\delta$. Let $K$ comprise the right bank, the small arc about the singularity and the left bank, with the orientation drawn in (b). For the cut-out domain at finite arc radius, Cauchy's theorem gives
\begin{equation}
\mathcal J_{C_0}-\mathcal J_{\widehat C_\delta}
 =\mathcal J_L-\mathcal J_R-\mathcal J_K .
\label{eq:cauchy}
\end{equation}
At finite endpoints we retain both connectors together with the bank and small-arc contributions. Along the path $C_\delta$ in panel (a), continuous analytic continuation takes the right portion onto Sheet II. This differs from $\widehat C_\delta$ on the branch used in panel (b), so relating their integrals entails including the corresponding difference of branch values. Identifying the real-axis integral with a cut discontinuity then depends on control of the connector limits, the small-arc limit, and the branch difference, even when the shifted integral or the difference between real parts is small.
\endgroup

\paragraph{Numerical demonstration.}
Table~\ref{tab:contour_deform} shows the relative difference between the real parts of the correlators for the shifted bra integral
as a function of $\delta$, for the same parameters as Table~\ref{tab:eta0_pert}
at $\eta_0=0.3$. Note that crossing the branch cut at $\delta=\eta_0$ reduces $|I_{\rm bra}|$ by a factor
of $\sim17$.
The mismatch drops from $6.01\%$ (real axis) to $0.35\%$ at $\delta=0.35$.
\textcolor{black}{The decomposition of the residual $0.35$--$0.8\%$ into endpoint, branch and quadrature contributions remains open. The left connector has yet to be measured separately, so its relative contribution is undetermined.}
\textcolor{black}{The reported real-valued bra contribution changes sign between $\delta=0.5$ and $1.0$ and is nonmonotone after continuation. A topological interpretation of these sign changes needs further evidence.}

\begin{table}[htbp]
\centering
\renewcommand{\arraystretch}{1.3}
\begin{tabular}{|c|c|r|r|r|r|}
\hline
$\delta$ & status & $I_{\rm ket}$ & $I_{\rm bra}^{(\delta)}$ & $I_{\rm total}^{(\delta)}$ & $\Delta_{\rm rel}$ \\
\hline
$0.000$ & real axis & $-2.190\times10^{-4}$ & $-1.838\times10^{-3}$ & $-2.057\times10^{-3}$ & $6.01\%$ \\
$0.100$ & above cut & $-2.190\times10^{-4}$ & $-1.297\times10^{-3}$ & $-1.516\times10^{-3}$ & $4.43\%$ \\
$0.250$ & above cut & $-2.190\times10^{-4}$ & $-7.868\times10^{-4}$ & $-1.006\times10^{-3}$ & $2.94\%$ \\
\hline
$0.3=\eta_0$ & \textit{branch point} & \multicolumn{4}{c|}{\textit{singular}} \\
\hline
$0.350$ & \textbf{Sheet II} & $-2.190\times10^{-4}$ & $+9.99\times10^{-5}$ & $-1.19\times10^{-4}$ & $\mathbf{0.35\%}$ \\
$0.500$ & Sheet II & $-2.190\times10^{-4}$ & $+6.73\times10^{-5}$ & $-1.52\times10^{-4}$ & $0.44\%$ \\
$1.000$ & Sheet II & $-2.190\times10^{-4}$ & $-5.16\times10^{-5}$ & $-2.71\times10^{-4}$ & $0.79\%$ \\
\hline
\end{tabular}
\caption{Absolute difference between the real parts of the in-in and in-out correlators at first order, divided by the magnitude of the free correlator, with the bra integral evaluated on the shifted contour $\mathcal{C}_\delta$
at $\eta_0=0.3$, $N=39$, $O=\phi(20)\phi(25)$.
\textcolor{black}{These archived values compare the real parts with the bra contribution shifted to $\eta=t-i\delta$ and the ket contribution unchanged. The complex $\mathcal J_C$ integrals and separate connectors in \eqref{eq:cauchy}, together with the full complex correlators, remain to be evaluated for this comparison.}}
\label{tab:contour_deform}
\end{table}

\begingroup\color{black}
The contour deformation in this appendix concerns first-order integrals. The original nonperturbative MPS results use real conformal time, whereas in Section~\ref{sec:asymptotic-projection} we evolve along a complex contour and use different finite-time boundary data. Here we examine the difference between the original real-contour correlators. The perturbative coefficients of the projected prescription need a separate calculation with its contour and boundary data.
\endgroup

\subsection{Reality properties and the relative difference between correlators}
\label{app:cosmo:metric}

\begingroup\color{black}
For a Hermitian insertion $O$, the in--in expectation is real. Write the in--out ratio as $N/D=\langle\chi|O|\psi\rangle/\langle\chi|\psi\rangle$, with both states evaluated at the insertion and normalized when discussing their overlap. If $\langle\chi|=e^{i\alpha}\langle\psi|$, their common phase cancels and the full correlators agree. This condition concerns the evolved states at the insertion and must be checked even when the boundary vacua coincide.

The geometric regulator and the real-contour continuation can produce free Bogoliubov mixing and change the relation between the independently evolved states. The reported $|\beta_k|^2\simeq0.03$--$0.06$ and the less-than-$0.5\%$ variation of $|\operatorname{Im}B_{\rm in\text{-}out}|$ when $\eta_f$ is varied from $6$ to $10$ refer to the original real-contour setup. The sensitivity to asymptotic vacuum preparation remains open beyond this short endpoint scan. Section~\ref{sec:asymptotic-projection} tests a different contour and boundary prescription and suppresses the full complex mismatch. The behavior of the old real-contour imaginary part as $\eta_0\to0$ remains to be determined.
\endgroup

\paragraph{The relative difference between correlators.}
For a Hermitian equal-time operator $O=\phi_i(\eta_*)\phi_k(\eta_*)$,
$O^\dagger=O$ (since equal-time fields commute for $i\neq k$), so
\begin{equation}
  B_{\rm in\text{-}out}^* = \frac{\langle\psi|O|\chi\rangle}{\langle\psi|\chi\rangle}.
\end{equation}
\begingroup\color{black}
Taking the real part gives $(B_{\rm in\text{-}out}+B_{\rm in\text{-}out}^*)/2$, evaluated with the same out-state and contour prescription as the original ratio. Full equality requires both matching real parts and vanishing imaginary part for this Hermitian insertion, with the out-state and contour specified consistently.
\endgroup
We therefore define the complex difference
\begin{equation}
  \Delta \equiv B_{\rm in\text{-}out} - B_{\rm in\text{-}in},
  \label{eq:Delta}
\end{equation}
and the relative difference of the real parts
\begin{equation}
  \Delta_{\rm DP} \equiv
  \frac{|\mathrm{Re}\,\Delta|}{|B_{\rm in\text{-}in}|}
  = \frac{|\mathrm{Re}[B_{\rm in\text{-}out}] - B_{\rm in\text{-}in}|}{|B_{\rm in\text{-}in}|},
  \label{eq:DeltaDP}
\end{equation}
\textcolor{black}{The historical label $\Delta_{\rm DP}$ denotes this comparison of real parts. To test the full Donath--Pajer equality we use the relative complex difference $|\Delta|/|B_{\rm in\text{-}in}|$, with $\mathrm{Im}[B_{\rm in\text{-}out}]$ providing an additional check for Hermitian insertions. These quantities depend on the specified contour, boundary states and cutoffs as well as the geometric regulator.}
For unequal-time operators, where $B_{\rm in\text{-}in}$ is complex, we compare
the full complex correlators using $|B_{\rm in\text{-}out}-B_{\rm in\text{-}in}|/|B_{\rm in\text{-}in}|$.
No algebraic manipulation of $B_{\rm in\text{-}out}$ alone can fix the imaginary part;
\textcolor{black}{the contour, endpoint states and normalization must be specified consistently. The endpoint $i\varepsilon$ prescription fixes asymptotic preparation, while removal of the geometric regulator $\eta_0$ needs a separate check.}

\subsection{Nonperturbative \texorpdfstring{$\lambda$}{lambda} scan}
\label{app:cosmo:scan}

\begin{figure}[hbt]
\centering
\includegraphics[width=\textwidth]{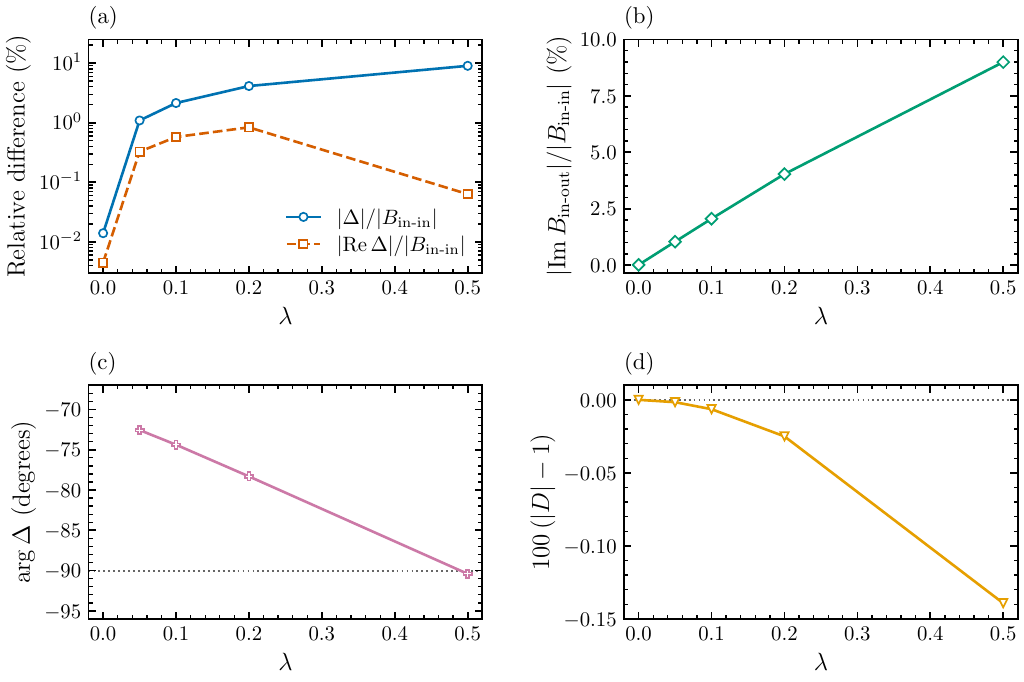}
\caption{Dependence on $\lambda$ at $\eta_0=1$, with $N=6$, $d=4$, $\chi=64$,
$O=\phi(3)\phi(5)$, $\eta_i=-6$, $\eta_f=6$ and $\eta_*=-1$.
(a)~Relative differences $|\Delta|/|B_{\rm in\text{-}in}|$ (blue circles)
and $|\mathrm{Re}\,\Delta|/|B_{\rm in\text{-}in}|$ (orange-red squares), expressed as percentages,
where $\Delta=B_{\rm in\text{-}out}-B_{\rm in\text{-}in}$.
(b)~$|\mathrm{Im}\,B_{\rm in\text{-}out}|/|B_{\rm in\text{-}in}|$, also expressed as a percentage.
(c)~Phase of $\Delta$; the dotted line marks $-90^\circ$.
(d)~Percentage change of the overlap magnitude from unity, $100(|D|-1)$,
where $D=\langle\chi|\psi\rangle$; the dotted line marks $|D|=1$.}
\label{fig:lambda_scan}
\end{figure}

Table~\ref{tab:lambda_mps} shows the nonperturbative MPS results
for $O=\phi(3)\phi(5)$ on the $N=6$, $d=4$, $\chi=64$ lattice.
All MPS runs were cross-checked against exact diagonalisation (ED); the ED-MPS agreement
is at the $4\times10^{-9}\%$ level. \textcolor{black}{For $N=6,d=4$, $\chi=64=d^3$ permits the full finite-system Schmidt rank, so this checks the MPS representation at that local cutoff. Common time-discretization errors, oscillator-cutoff errors and endpoint effects need separate checks, as does convergence on larger lattices.} Figure~\ref{fig:lambda_scan} displays the $\lambda$ scan graphically.

\begin{table}[!htb]
\centering
\renewcommand{\arraystretch}{1.3}
\begin{tabular}{|c|c|c|c|c|c|c|}
\hline
$\lambda$ & $|D|$ & $\mathrm{Re}\,B_{\rm in\text{-}in}$
  & $\mathrm{Re}\,B_{\rm in\text{-}out}$ & $\mathrm{Im}\,B_{\rm in\text{-}out}$
  & $\Delta_{\rm DP}$ & $\arg(\Delta)$ \\
\hline
$0.0$  & $1.000000$ & $0.06725$ & $0.06725$ & $-8.9\times10^{-6}$  & $0.004\%$ & $-109^\circ$ \\
$0.05$ & $0.999984$ & $0.06676$ & $0.06697$ & $-6.9\times10^{-4}$  & $0.326\%$ & $-73^\circ$ \\
$0.1$  & $0.999936$ & $0.06627$ & $0.06665$ & $-1.36\times10^{-3}$ & $0.576\%$ & $-74^\circ$ \\
$0.2$  & $0.999750$ & $0.06533$ & $0.06587$ & $-2.63\times10^{-3}$ & $0.835\%$ & $-78^\circ$ \\
$0.5$  & $0.998608$ & $0.06267$ & $0.06263$ & $-5.64\times10^{-3}$ & $0.063\%$ & $-90^\circ$ \\
\hline
\end{tabular}
\caption{$\lambda$ scan at $\eta_0=1.0$ ($\lambda_{\rm eff}=\lambda/\eta_0^4=\lambda$,
perturbative regime), $N=6$, $d=4$, $\chi=64$, $O=\phi(3)\phi(5)$,
$\eta_i=-6$, $\eta_f=+6$, $\eta_*=-1$.
$\Delta_{\rm DP} = |\mathrm{Re}\,\Delta|/|B_{\rm in\text{-}in}|$, 
with $\Delta \equiv B_{\rm in\text{-}out} - B_{\rm in\text{-}in}$ 
as defined in eq.~\eqref{eq:Delta}. The relative magnitude of the complex difference
$|\Delta|/|B_{\rm in\text{-}in}|$ (not shown) grows monotonically 
from $0.014\%$ to $8.38\%$.
The relative difference of the real parts $\Delta_{\rm DP}$ is non-monotone because the phase of
$\Delta$ rotates toward $-90^\circ$ at large $\lambda$,
making $\mathrm{Re}(\Delta)\approx0$ at $\lambda=0.5$ even though $|\Delta|$ is large.
The $\lambda=0$ check passes to $0.004\%$ (residual from finite $\eta_f$).}
\label{tab:lambda_mps}
\end{table}

\begingroup\color{black}
The archived scan at $\lambda=0.1$ gives $\Delta_{\rm DP}=0.58\%$ at $\eta_0=1.0$ and $1.10\%$ at $\eta_0=0.3$, with reported ED--MPS agreement below $10^{-8}\%$ for this small system. This demonstrates sensitivity to the regulated finite-time prescription and agreement between implementations. The separate branch-cut and endpoint contributions remain undetermined. The relative difference of the real parts and the relative magnitude of the imaginary part increase over the sampled values as $\eta_0$ is reduced. Their limits at $\eta_0=0$ remain open. Determining whether either residual disappears entails testing the full complex difference with the specified asymptotic contour and state prescription, as in Section~\ref{sec:asymptotic-projection}.
\endgroup

\section{The \texorpdfstring{$\ell=1$}{ell=1} sector beyond the s-wave} \label{sec:spin1trunc}

The 3+1d analysis in this work is restricted to the s-wave sector ($\ell=0$), for which the scalar field reduces to a single radial field. The next natural truncation is to include the $\ell=1$ multiplet, which captures the leading anisotropic fluctuations while remaining within reach of MPS methods.

Following the conventions of Section \ref{3plus1dS}, we expand
\begin{equation}
\phi(\eta,r,\theta,\varphi)=\sum_{\ell,m}\frac{\chi_{\ell m}(\eta,r)}{r}\,Y_{\ell m}(\theta,\varphi)\;.
\end{equation}
The factor of $1/r$ removes first-derivative terms from the radial Laplacian and yields a standard radial kinetic term. As in the s-wave case, it is convenient to introduce canonically normalized fields
\begin{equation}
\tilde{\chi}_{\ell m}(\eta,r)=\Omega(\eta)\chi_{\ell m}(\eta,r)\;.
\end{equation}

For the $\ell=1$ sector the quadratic Hamiltonian becomes
\begin{equation}
H_{\ell=1}
=
\sum_{m=-1}^{1}
\int dr\,
\left[
\frac{1}{2}\tilde{\pi}_{1m}^{\,2}
+
\frac{1}{2}(\partial_r \tilde{\chi}_{1m})^2
+
\frac{1}{2}
\left(
m_{\rm eff}^2(\eta)+\frac{2}{r^2}
\right)
\tilde{\chi}_{1m}^2
\right],
\end{equation}
where
\begin{equation}
m_{\rm eff}^2(\eta)=\Omega^2(\eta)m^2-\frac{\ddot{\Omega}(\eta)}{\Omega(\eta)}
\end{equation}
is the Mukhanov--Sasaki effective mass. The centrifugal barrier $2/r^2$ is time independent after the canonical rescaling, so it can be built once into the matrix-product-operator (MPO) representation rather than updated at every time step. The three $\ell=1$ modes form the vector representation of the spatial rotation group $SO(3)$. Using Cartesian labels for the real $\ell=1$ triplet, we may write
\begin{equation}
\vec{\tilde{\chi}}_{1}=(\tilde{\chi}_{1x},\tilde{\chi}_{1y},\tilde{\chi}_{1z}),
\end{equation}
so that the quadratic Hamiltonian is manifestly $SO(3)$ invariant.

The interaction truncated to $\ell\leq 1$ is determined by Gaunt integrals. Parity eliminates all terms with an odd number of $\ell=1$ insertions, and the remaining structures are fixed by
\begin{equation}
\int d\Omega\,Y_{00}^4=\frac{1}{4\pi},\qquad\int d\Omega\,Y_{00}^2Y_{1m}Y_{1m'}=\frac{1}{4\pi}\delta_{mm'},\qquad\int d\Omega\,Y_{10}^4=\frac{9}{20\pi}.
\end{equation}
In canonical variables this gives
\begin{equation}
H_{\rm int}^{(\ell\leq 1)}=\int dr\,\frac{\lambda}{4!\,4\pi r^2}\left(\tilde{\chi}_{00}^4+6\,\tilde{\chi}_{00}^2\,\vec{\tilde{\chi}}_1^{\,2}+\frac{9}{5}\big(\vec{\tilde{\chi}}_1^{\,2}\big)^2\right),\qquad\vec{\tilde{\chi}}_1^{\,2}\equiv\tilde{\chi}_{1x}^2+\tilde{\chi}_{1y}^2+\tilde{\chi}_{1z}^2.
\end{equation}
As in the s-wave sector, all conformal factors cancel in the quartic term after the canonical rescaling, so the interaction is time independent and can be assembled once in the MPO.

We next compare the local Hilbert-space dimensions required by two occupation-number truncations. A truncation of angular momentum at $\ell_{\max}$ introduces
\begin{equation}
n_{\rm fields}=(\ell_{\max}+1)^2
\end{equation}
radial fields. If each field is truncated independently to $N_{\max}$ excitations, the local Hilbert-space dimension is
\begin{equation}
d_{\rm indep}=(N_{\max}+1)^{(\ell_{\max}+1)^2}\;.
\end{equation}
For example, for $\ell_{\max}=1$ and $N_{\max}=6$, this gives
\begin{equation}
d_{\rm indep}=7^4=2401\;.
\end{equation}
A more efficient truncation is obtained by restricting the total occupation number per site,
\begin{equation}
n_0+n_{1,-1}+n_{1,0}+n_{1,1}\le N_{\max}\;,
\end{equation}
which yields
\begin{equation}
d_{\rm joint}=\binom{N_{\max}+4}{4}\;.
\end{equation}
For $N_{\max}=6$, this gives
\begin{equation}
d_{\rm joint}=210\;,
\end{equation}
which is an order-of-magnitude reduction relative to the naive tensor-product truncation. Additional savings arise from organizing the local basis into irreducible $SO(3)$ sectors, which block-diagonalizes the MPS tensors. For a radial lattice with $N$ sites and bond dimension $\chi$, the dominant cost of two-site TDVP scales as
\begin{equation}
\mathcal{O}(N\,d^2\,\chi^3)\;.
\end{equation}
Replacing $d_{\rm indep}=2401$ by $d_{\rm joint}=210$ therefore lowers the leading cost by
\begin{equation}
\left(\frac{2401}{210}\right)^2 \approx 130\;,
\end{equation}
i.e. by more than two orders of magnitude.

\begingroup\color{black}
Retaining the $\ell=1$ triplet adds the first nonspherical field modes to the s-wave model while allowing rotationally invariant states with isotropic one-point functions. The angular interaction coefficients couple the retained fields and conserve total angular momentum. Omitted higher-$\ell$ fluctuations can still contribute interaction corrections even to correlators of the angularly averaged field. The centrifugal term $\ell(\ell+1)/r^2$ motivates an angular expansion at fixed radius. Uniform control of truncation errors across the radial domain and through cosmological evolution remains to be established. The construction above is therefore a proposed extension. Its accuracy must be checked by increasing $\ell_{\max}$ at the same physical parameters and controlling the other cutoffs, to assess the reliability of $\ell\leq1$ for the full interacting theory. This angular-cutoff study remains open. The estimates of the local Hilbert-space dimension suggest possible computational savings, with the accuracy and cost of a full simulation still to be determined.
\endgroup

\section{\texorpdfstring{{Connected fourth moment at weak coupling}}{Connected fourth moment at weak coupling}}
\label{app:kurtosis}

\subsection{\texorpdfstring{{First-order connected fourth moment}}{First-order connected fourth moment}}
\label{app:kurtosis-first-order}
\begingroup\color{black}
Let $\kappa_j$ denote the local connected moment whose spatial average
gives $\kappa$ in Section~\ref{sec:kurtosis_weak_coupling}. We write
$\kappa_j(\lambda,\eta)=\lambda\kappa_j^{(1)}(\eta)+O(\lambda^2)$
and denote its averaged coefficient by
$\kappa^{(1)}=\overline{\kappa_j^{(1)}}$. Let $|0\rangle$ be the
free instantaneous vacuum at $\eta_i$ and
$W_{\ell j}(\tau,\eta)=\langle0|\phi_{I,\ell}(\tau)
\phi_{I,j}(\eta)|0\rangle$, where the interaction-picture fields evolve
with the free Hamiltonian. For a constant quartic coupling, the first-order
contribution generated during the evolution is
\begin{equation}
\kappa_{j,\mathrm{evol}}^{(1)}(\eta)
=-2a\int_{\eta_i}^{\eta}d\tau\,\Omega^2(\tau)
\sum_{\ell=1}^{N}\operatorname{Im}
\bigl[W_{\ell j}(\tau,\eta)^4\bigr].
\label{eq:kurtosis_contact}
\end{equation}
All four fields at the quartic vertex contract with fields at the measured
site. Their $4!$ contractions cancel the factorial in the interaction,
while the disconnected contractions cancel in
$\langle\phi_j^4\rangle-3\langle\phi_j^2\rangle^2$.

The interacting initial vacuum contributes at the same order. Writing
$|0;\lambda\rangle=|0\rangle+\lambda|\delta0\rangle+O(\lambda^2)$,
ordinary ground-state perturbation theory gives
\begin{equation}
|\delta0\rangle=-\sum_{n\ne0}
\frac{|n\rangle\langle n|V_i|0\rangle}{E_n-E_0},
\qquad
V_i=\left.\frac{\partial H(\eta_i;\lambda)}{\partial\lambda}
\right|_{\lambda=0},
\label{eq:kurtosis_initial_vacuum}
\end{equation}
where $|n\rangle$ and $E_n$ are the many-body eigenstates and energies of
$H_0(\eta_i)$, and $\langle0|\delta0\rangle=0$. With
$G_j(\eta)=W_{jj}(\eta,\eta)$, the full coefficient is
\begin{equation}
\kappa_j^{(1)}(\eta)=\kappa_{j,\mathrm{evol}}^{(1)}(\eta)
+2\operatorname{Re}\langle\delta0|
\bigl[\phi_{I,j}^4(\eta)-6G_j(\eta)\phi_{I,j}^2(\eta)\bigr]|0\rangle.
\label{eq:kurtosis_first_order}
\end{equation}
The second term accounts for the connected moment already present in the
initial state and its subsequent free evolution. We include it when the
initial vacuum is prepared with the interaction, as in Figure~\ref{fig:non_gaussianity_and_Sint}.
Equation~\eqref{eq:kurtosis_first_order} predicts both the magnitude and
the time dependence of the departure from Gaussianity.
\endgroup

\subsection{\texorpdfstring{{Connected fourth moment on three sites}}{Connected fourth moment on three sites}}
\label{app:kurtosis_small_chain}
\begingroup\color{black}
On a short chain, we can also compare TDVP and TEBD with direct
Schr\"odinger evolution in the same oscillator basis. The chosen bond
dimension retains every Schmidt amplitude, so this comparison tests the
time evolution within the finite Hilbert space. As shown in
Table~\ref{tab:kurtosis_small_chain}, both methods reproduce the connected
fourth moment, and halving the time step reduces their differences from
the direct evolution by a factor of four. For the comparison with
perturbation theory, the local oscillator basis is enlarged to approach
the free-mode result. At the final time and smallest coupling, the MPS
values of $q$ agree with the first-order predictions within $0.2\%$ for both
free and interacting initial vacua. In these small-chain calculations,
the remainder $\kappa(\lambda)-\kappa(0)-\lambda\kappa^{(1)}$ is
approximately proportional to $\lambda^2$. The evolution therefore
reproduces the leading interaction correction for both initial states.
The parameters of each comparison are specified in the caption.

\begin{table}[hbt]
    \centering
    \color{black}
    \captionsetup{font+={color=black},labelfont+={color=black}}
    \begin{tabular}{|c|c|c|c|}
        \hline
        \multicolumn{4}{|c|}{Largest difference in $\kappa_{\rm rel}^{\rm int}$} \\
        \hline
        $\Delta\eta$ & TDVP--Schr\"odinger & TEBD--Schr\"odinger & TDVP--TEBD \\
        \hline
        $0.025$ & $8.77\times10^{-8}$ & $8.80\times10^{-8}$ & $6.59\times10^{-9}$ \\
        $0.0125$ & $2.19\times10^{-8}$ & $2.20\times10^{-8}$ & $1.65\times10^{-9}$ \\
        \hline
    \end{tabular}
    \par\medskip
    \begin{tabular}{|l|c|c|c|}
        \hline
        Initial vacuum & $\kappa^{(1)}(-1)$ & $q(0.005,-1)$ & Difference (\%) \\
        \hline
        Free & $-0.00673032$ & $-0.00671778$ & $0.186$ \\
        Interacting & $-0.00682737$ & $-0.00681796$ & $0.138$ \\
        \hline
    \end{tabular}
    \caption{Comparisons on three Dirichlet sites with $a=H=m^2=1$,
    $\eta_0=0$, $\eta_i=-2$, $\eta_*=-1$, and measurement at the central
    site. The upper table gives the largest absolute differences over
    five observation times and $\lambda=0.005,0.01,0.02$, starting from
    the free instantaneous vacuum. It uses $N_{\max}=7$, $\chi=8$,
    and the MPS time steps shown. Direct Schr\"odinger evolution uses
    a step of $0.001$. The lower table compares with first-order
    perturbation theory at $\lambda=0.005$ and $\eta=-1$, using
    $N_{\max}=15$, $\chi=16$, and a maximum step of $0.00625$.
    Its last column is $100|q-\kappa^{(1)}|/|\kappa^{(1)}|$.}
    \label{tab:kurtosis_small_chain}
\end{table}
\endgroup

\newpage
\printbibliography
\end{document}